%% file: P02b_LFI_calibration.tex
\documentclass[longauth,traditabstract]{aa}
\pdfoutput=1

\input Planck.tex

\usepackage{amsmath}
\usepackage{natbib}
\usepackage{graphicx}
\usepackage{txfonts}
\usepackage{natbib}
\usepackage{fixltx2e}
\usepackage{url}
\usepackage[draft,breaklinks,colorlinks,citecolor=blue]{hyperref}

\bibpunct{(}{)}{;}{a}{}{,}

\newcommand{\ud}{\mathrm{d}}
\newcommand{\timeder}{\partial_t}
\newcommand{\cc}{\mathcal{C}} % "cc" stands for "colour correction"
\def\xversor{\vec{x}}
\def\Svector{\vec{S}}

\newenvironment{DIFnomarkup}{}{}

\begin{document}

\all2013resultspapers

\title{\Planck{} 2013 results. V. LFI calibration}

\input AuthorList_P02b_LFI_Calibration_authors_and_institutes.tex

\authorrunning{Planck Collaboration}

\abstract{We discuss the methods employed to photometrically calibrate the data acquired by the Low Frequency Instrument on \Planck. Our calibration is based on a combination of the Orbital Dipole plus the Solar Dipole, caused respectively by the motion of the \Planck{} spacecraft with respect to the Sun and by motion of the Solar System with respect to the CMB rest frame. The latter provides a signal of a few mK with the same spectrum as the CMB anisotropies and is visible throughout the mission. In this data release we rely on the characterization of the Solar Dipole as measured by {\it WMAP}. We also present preliminary results (at 44\,GHz only) on the study of the Orbital Dipole, which agree with the {\it WMAP} value of the Solar System speed within our uncertainties. We compute the calibration constant for each radiometer roughly once per hour, in order to keep track of changes in the detectors' gain. Since non-idealities in the optical response of the beams proved to be important, we implemented a fast convolution algorithm which considers the full beam response in estimating the signal generated by the dipole. Moreover, in order to further reduce the impact of residual systematics due to sidelobes, we estimated time variations in the calibration constant of the 30\,GHz radiometers (the ones with the largest sidelobes) using the signal of an internal reference load at 4\,K instead of the CMB dipole. We have estimated the accuracy of the LFI calibration following two strategies: (1) we have run a set of simulations to assess the impact of statistical errors and systematic effects in the instrument and in the calibration procedure, and (2) we have performed a number of internal consistency checks on the data and on the brightness temperature of Jupiter. Errors in the calibration of this \Planck/LFI data release are expected to be about 0.6\,\% at 44 and 70 GHz, and 0.8\,\% at 30 GHz. Both these preliminary results at low and high $\ell$ are consistent with WMAP results within uncertainties and comparison of power spectra indicates good consistency in the absolute calibration with HFI (0.3\,\%) and a $1.4\sigma$ discrepancy with WMAP (0.9\,\%). }


\maketitle

%%%%%%%%%%%%%%%%%%%%%%%%%%%%%%%%%%%%%%%%%%%%%%%%%%%%%%%%%%%%%%%%%%%%%%

\section{Introduction}

\input P02b_1_introduction.tex

%%%%%%%%%%%%%%%%%%%%%%%%%%%%%%%%%%%%%%%%%%%%%%%%%%%%%%%%%%%%%%%%%%%%%%

\section{Calibration philosophy and conventions}
\label{sec:calibrationPhilosophyAndConventions}

	\subsection{Time scale of gain variations}
	\label{sec:timeScale}
	\input P02b_2_1_time_scale.tex

	\subsection{Handling beam efficiency}
	\label{sec:beamEfficiencyAndWindowFunctions}
	\input P02b_2_2_beam_efficiency_window_functions.tex
	
	\subsection{Colour corrections}
	\label{sec:colourCorrections}
	\input P02b_2_3_colour_corrections.tex

%%%%%%%%%%%%%%%%%%%%%%%%%%%%%%%%%%%%%%%%%%%%%%%%%%%%%%%%%%%%%%%%%%%%%%

\section{Calibration techniques for LFI}
\label{sec:calibrationMethods}

\input P02b_3_0_introduction.tex

\subsection{OSG dipole calibration}
\label{sec:OSGCalibration}

\input P02b_3_1_kinematic_dipole.tex

\subsection{4\,K calibration}
\label{sec:dVV}

\input P02b_3_2_4K_calibration.tex

\subsection{Orbital dipole calibration}
\label{sec:orbitalDipole}

\input P02b_3_3_orbital_dipole.tex

\subsection{Setting the zero levels in the maps}
\label{sec:zeroLevel}

\input P02b_3_4_zero_level.tex

%%%%%%%%%%%%%%%%%%%%%%%%%%%%%%%%%%%%%%%%%%%%%%%%%%%%%%%%%%%%%%%%%%%%%%

\section{Systematic effects that affect calibration}
\label{sec:systematics}

Raw gains from the nominal pipeline are sensitive to various effects related both to previous data processing steps and to instrument characteristics. A full treatment of systematic effects in the current release of LFI data is provided by \citet{planck2013-p02a}; here we concentrate only on those systematics that have a significant impact on the calibration of the radiometers.

\subsection{Impact of the optics on calibration estimation}
\label{sec:optics}

\input P02b_4_1_impact_of_optics_on_calibration.tex

\subsection{Efficiently convolving the dipole with a realistic beam}
\label{sec:fourPiConvolver}
\input P02b_4_2_beam_convolution.tex

\subsection{Imperfect behaviour of the ADC}
\label{sec:ADCNonIdealities}

\input P02b_4_3_ADC_systematic_effects.tex

%%%%%%%%%%%%%%%%%%%%%%%%%%%%%%%%%%%%%%%%%%%%%%%%%%%%%%%%%%%%%%%%%%%%%%

\section{Accuracy of the calibration}
\label{sec:accuracy}

\input P02b_5_accuracy_introduction.tex

\subsection{Statistical uncertainty}
\label{sec:noiseAndGain}
\input P02b_5_1_noise_and_gain.tex

\subsection{Impact of known systematics}
\label{sec:impactOfKnownSystematics}

\input P02b_5_2_adc_effect.tex

\subsection{Impact of unknown systematics}
\label{sec:impactOfUnknownSystematics}

To estimate the impact of all those systematics which affect the calibration, but for which we do not have simulations or direct measurements at hand, we relied on internal consistency checks, namely inter-channel comparisons and null tests. We explain the logic behind each analysis as well as the results in the next paragraphs.

\subsubsection{Inter-channel calibration consistency}
\label{sec:interchannelCalConsistency}
\input P02b_5_3_1_interchannel_calibration.tex

\subsubsection{Null tests}
\label{sec:nullTests}
\input P02b_5_3_2_null_tests.tex

\subsection{Checks with external calibrations}
\label{sec:checksWithExternalCalibrations}

\subsubsection{Using bright point sources for relative calibration}
\label{sec:sourceFluxes}
\input P02b_5_4_1_source_fluxes.tex

\subsubsection{Planet flux densities}
\label{sec:planets}
\input P02b_5_4_2_planets.tex

\subsection{Consistency with HFI and WMAP}
\input P02b_5_5_consistency.tex

%%%%%%%%%%%%%%%%%%%%%%%%%%%%%%%%%%%%%%%%%%%%%%%%%%%%%%%%%%%%%%%%%%%%%%

\section{Conclusions and next steps}
\label{sec:conclusions}
\input P02b_6_conclusions.tex

\begin{acknowledgements}

The development of \Planck\ has been supported by: ESA; CNES and CNRS/INSU-IN2P3-INP (France); ASI, CNR, and INAF (Italy); NASA and DoE (USA); STFC and UKSA (UK); CSIC, MICINN, JA and RES (Spain); Tekes, AoF and CSC (Finland); DLR and MPG (Germany); CSA (Canada); DTU Space (Denmark); SER/SSO (Switzerland); RCN (Norway); SFI (Ireland); FCT/MCTES (Portugal); and PRACE (EU). A description of the Planck Collaboration and a list of its members, including the technical or scientific activities in which they have been involved, can be found at \url{http://www.sciops.esa.int/index.php?project=planck&page=Planck_Collaboration}.  

\end{acknowledgements}

\bibliographystyle{aa}
\bibliography{Planck_bib,custom_bibliography}

\appendix

\section{Fast convolution of the beam with the dipole}
\label{sec:computingDipoleConvolutionParams}
\input P02b_appendix_A_beam_convolution_calculation.tex

\raggedright
\end{document}

%% file: Planck.tex
\def\setsymbol#1#2{\expandafter\def\csname #1\endcsname{#2}}
\def\getsymbol#1{\csname #1\endcsname}

\def\Planck{\textit{Planck}}

\def\all2013resultspapers{\nocite{planck2013-p01, planck2013-p02, planck2013-p02a, planck2013-p02d, planck2013-p02b, planck2013-p03, planck2013-p03c, planck2013-p03f, planck2013-p03d, planck2013-p03e, planck2013-p01a, planck2013-p06, planck2013-p03a, planck2013-pip88, planck2013-p08, planck2013-p11, planck2013-p12, planck2013-p13, planck2013-p14, planck2013-p15, planck2013-p05b, planck2013-p17, planck2013-p09, planck2013-p09a, planck2013-p20, planck2013-p19, planck2013-pipaberration, planck2013-p05, planck2013-p05a, planck2013-pip56, planck2013-p06b}}

\newbox\tablebox    \newdimen\tablewidth
\def\leaderfil{\leaders\hbox to 5pt{\hss.\hss}\hfil}
\def\endPlancktable{\tablewidth=\columnwidth 
    $$\hss\copy\tablebox\hss$$
    \vskip-\lastskip\vskip -2pt}
\def\endPlancktablewide{\tablewidth=\textwidth 
    $$\hss\copy\tablebox\hss$$
    \vskip-\lastskip\vskip -2pt}
\def\tablenote#1 #2\par{\begingroup \parindent=0.8em
    \abovedisplayshortskip=0pt\belowdisplayshortskip=0pt
    \noindent
    $$\hss\vbox{\hsize\tablewidth \hangindent=\parindent \hangafter=1 \noindent
    \hbox to \parindent{$^#1$\hss}\strut#2\strut\par}\hss$$
    \endgroup}
\def\doubleline{\vskip 3pt\hrule \vskip 1.5pt \hrule \vskip 5pt}

\def\L2{\ifmmode L_2\else $L_2$\fi}

\def\DeltaT{\ifmmode \Delta T\else $\Delta T$\fi}
\def\deltat{\ifmmode \Delta t\else $\Delta t$\fi}
\def\fknee{\ifmmode f_{\rm knee}\else $f_{\rm knee}$\fi}
\def\Fmax{\ifmmode F_{\rm max}\else $F_{\rm max}$\fi}
\def\solar{\ifmmode{\rm M}_{\mathord\odot}\else${\rm M}_{\mathord\odot}$\fi}
\def\Msolar{\ifmmode{\rm M}_{\mathord\odot}\else${\rm M}_{\mathord\odot}$\fi}
\def\Lsolar{\ifmmode{\rm L}_{\mathord\odot}\else${\rm L}_{\mathord\odot}$\fi}

\def\inv{\ifmmode^{-1}\else$^{-1}$\fi}
\def\mo{\ifmmode^{-1}\else$^{-1}$\fi}
\def\sup#1{\ifmmode ^{\rm #1}\else $^{\rm #1}$\fi}
\def\expo#1{\ifmmode \times 10^{#1}\else $\times 10^{#1}$\fi}
\def\,{\thinspace}
\def\lsim{\mathrel{\raise .4ex\hbox{\rlap{$<$}\lower 1.2ex\hbox{$\sim$}}}}
\def\gsim{\mathrel{\raise .4ex\hbox{\rlap{$>$}\lower 1.2ex\hbox{$\sim$}}}}

\def\simprop{\mathrel{\raise .4ex\hbox{\rlap{$\propto$}\lower 1.2ex\hbox{$\sim$}}}}
\def\deg{\ifmmode^\circ\else$^\circ$\fi}
\def\pdeg{\ifmmode $\setbox0=\hbox{$^{\circ}$}\rlap{\hskip.11\wd0 .}$^{\circ}
          \else \setbox0=\hbox{$^{\circ}$}\rlap{\hskip.11\wd0 .}$^{\circ}$\fi}
\def\arcs{\ifmmode {^{\scriptstyle\prime\prime}}
          \else $^{\scriptstyle\prime\prime}$\fi}
\def\arcm{\ifmmode {^{\scriptstyle\prime}}
          \else $^{\scriptstyle\prime}$\fi}
\newdimen\sa  \newdimen\sb
\def\parcs{\sa=.07em \sb=.03em
     \ifmmode \hbox{\rlap{.}}^{\scriptstyle\prime\kern -\sb\prime}\hbox{\kern -\sa}
     \else \rlap{.}$^{\scriptstyle\prime\kern -\sb\prime}$\kern -\sa\fi}
\def\parcm{\sa=.08em \sb=.03em
     \ifmmode \hbox{\rlap{.}\kern\sa}^{\scriptstyle\prime}\hbox{\kern-\sb}
     \else \rlap{.}\kern\sa$^{\scriptstyle\prime}$\kern-\sb\fi}
\def\ra[#1 #2 #3.#4]{#1\sup{h}#2\sup{m}#3\sup{s}\llap.#4}
\def\dec[#1 #2 #3.#4]{#1\deg#2\arcm#3\arcs\llap.#4}
\def\deco[#1 #2 #3]{#1\deg#2\arcm#3\arcs}
\def\rra[#1 #2]{#1\sup{h}#2\sup{m}}

\def\dots{\relax\ifmmode \ldots\else $\ldots$\fi}
\def\WHzsr{\ifmmode $W\,Hz\mo\,sr\mo$\else W\,Hz\mo\,sr\mo\fi}
\def\mHz{\ifmmode $\,mHz$\else \,mHz\fi}
\def\GHz{\ifmmode $\,GHz$\else \,GHz\fi}
\def\mKs{\ifmmode $\,mK\,s$^{1/2}\else \,mK\,s$^{1/2}$\fi}
\def\muKs{\ifmmode \,\mu$K\,s$^{1/2}\else \,$\mu$K\,s$^{1/2}$\fi}
\def\muKRJs{\ifmmode \,\mu$K$_{\rm RJ}$\,s$^{1/2}\else \,$\mu$K$_{\rm RJ}$\,s$^{1/2}$\fi}
\def\muKHz{\ifmmode \,\mu$K\,Hz$^{-1/2}\else \,$\mu$K\,Hz$^{-1/2}$\fi}
\def\MJysr{\ifmmode \,$MJy\,sr\mo$\else \,MJy\,sr\mo\fi}
\def\MJysrmK{\ifmmode \,$MJy\,sr\mo$\,mK$_{\rm CMB}\mo\else \,MJy\,sr\mo\,mK$_{\rm CMB}\mo$\fi}
\def\microns{\ifmmode \,\mu$m$\else \,$\mu$m\fi}

\def\muK{\ifmmode \,\mu$K$\else \,$\mu$\hbox{K}\fi}
\def\microK{\ifmmode \,\mu$K$\else \,$\mu$\hbox{K}\fi}
\def\muW{\ifmmode \,\mu$W$\else \,$\mu$\hbox{W}\fi}
\def\kms{\ifmmode $\,km\,s$^{-1}\else \,km\,s$^{-1}$\fi}
\def\kmsMpc{\ifmmode $\,\kms\,Mpc\mo$\else \,\kms\,Mpc\mo\fi}
\setsymbol{LFI:center:frequency:70GHz:units}{70.3\,GHz}
\setsymbol{LFI:center:frequency:44GHz:units}{44.1\,GHz}
\setsymbol{LFI:center:frequency:30GHz:units}{28.5\,GHz}

\setsymbol{LFI:center:frequency:70GHz}{70.3}
\setsymbol{LFI:center:frequency:44GHz}{44.1}
\setsymbol{LFI:center:frequency:30GHz}{28.5}

\setsymbol{LFI:center:frequency:LFI18:Rad:M:units}{71.7\GHz}
\setsymbol{LFI:center:frequency:LFI19:Rad:M:units}{67.5\GHz}
\setsymbol{LFI:center:frequency:LFI20:Rad:M:units}{69.2\GHz}
\setsymbol{LFI:center:frequency:LFI21:Rad:M:units}{70.4\GHz}
\setsymbol{LFI:center:frequency:LFI22:Rad:M:units}{71.5\GHz}
\setsymbol{LFI:center:frequency:LFI23:Rad:M:units}{70.8\GHz}
\setsymbol{LFI:center:frequency:LFI24:Rad:M:units}{44.4\GHz}
\setsymbol{LFI:center:frequency:LFI25:Rad:M:units}{44.0\GHz}
\setsymbol{LFI:center:frequency:LFI26:Rad:M:units}{43.9\GHz}
\setsymbol{LFI:center:frequency:LFI27:Rad:M:units}{28.3\GHz}
\setsymbol{LFI:center:frequency:LFI28:Rad:M:units}{28.8\GHz}
\setsymbol{LFI:center:frequency:LFI18:Rad:S:units}{70.1\GHz}
\setsymbol{LFI:center:frequency:LFI19:Rad:S:units}{69.6\GHz}
\setsymbol{LFI:center:frequency:LFI20:Rad:S:units}{69.5\GHz}
\setsymbol{LFI:center:frequency:LFI21:Rad:S:units}{69.5\GHz}
\setsymbol{LFI:center:frequency:LFI22:Rad:S:units}{72.8\GHz}
\setsymbol{LFI:center:frequency:LFI23:Rad:S:units}{71.3\GHz}
\setsymbol{LFI:center:frequency:LFI24:Rad:S:units}{44.1\GHz}
\setsymbol{LFI:center:frequency:LFI25:Rad:S:units}{44.1\GHz}
\setsymbol{LFI:center:frequency:LFI26:Rad:S:units}{44.1\GHz}
\setsymbol{LFI:center:frequency:LFI27:Rad:S:units}{28.5\GHz}
\setsymbol{LFI:center:frequency:LFI28:Rad:S:units}{28.2\GHz}

\setsymbol{LFI:center:frequency:LFI18:Rad:M}{71.7}
\setsymbol{LFI:center:frequency:LFI19:Rad:M}{67.5}
\setsymbol{LFI:center:frequency:LFI20:Rad:M}{69.2}
\setsymbol{LFI:center:frequency:LFI21:Rad:M}{70.4}
\setsymbol{LFI:center:frequency:LFI22:Rad:M}{71.5}
\setsymbol{LFI:center:frequency:LFI23:Rad:M}{70.8}
\setsymbol{LFI:center:frequency:LFI24:Rad:M}{44.4}
\setsymbol{LFI:center:frequency:LFI25:Rad:M}{44.0}
\setsymbol{LFI:center:frequency:LFI26:Rad:M}{43.9}
\setsymbol{LFI:center:frequency:LFI27:Rad:M}{28.3}
\setsymbol{LFI:center:frequency:LFI28:Rad:M}{28.8}
\setsymbol{LFI:center:frequency:LFI18:Rad:S}{70.1}
\setsymbol{LFI:center:frequency:LFI19:Rad:S}{69.6}
\setsymbol{LFI:center:frequency:LFI20:Rad:S}{69.5}
\setsymbol{LFI:center:frequency:LFI21:Rad:S}{69.5}
\setsymbol{LFI:center:frequency:LFI22:Rad:S}{72.8}
\setsymbol{LFI:center:frequency:LFI23:Rad:S}{71.3}
\setsymbol{LFI:center:frequency:LFI24:Rad:S}{44.1}
\setsymbol{LFI:center:frequency:LFI25:Rad:S}{44.1}
\setsymbol{LFI:center:frequency:LFI26:Rad:S}{44.1}
\setsymbol{LFI:center:frequency:LFI27:Rad:S}{28.5}
\setsymbol{LFI:center:frequency:LFI28:Rad:S}{28.2}

\setsymbol{LFI:white:noise:sensitivity:70GHz:units}{134.7\muKs}
\setsymbol{LFI:white:noise:sensitivity:44GHz:units}{164.7\muKs}
\setsymbol{LFI:white:noise:sensitivity:30GHz:units}{143.4\muKs}

\setsymbol{LFI:white:noise:sensitivity:70GHz}{134.7}
\setsymbol{LFI:white:noise:sensitivity:44GHz}{164.7}
\setsymbol{LFI:white:noise:sensitivity:30GHz}{143.4}

\setsymbol{LFI:white:noise:sensitivity:LFI18:Rad:M:units}{512.0\muKs}
\setsymbol{LFI:white:noise:sensitivity:LFI19:Rad:M:units}{581.4\muKs}
\setsymbol{LFI:white:noise:sensitivity:LFI20:Rad:M:units}{590.8\muKs}
\setsymbol{LFI:white:noise:sensitivity:LFI21:Rad:M:units}{455.2\muKs}
\setsymbol{LFI:white:noise:sensitivity:LFI22:Rad:M:units}{492.0\muKs}
\setsymbol{LFI:white:noise:sensitivity:LFI23:Rad:M:units}{507.7\muKs}
\setsymbol{LFI:white:noise:sensitivity:LFI24:Rad:M:units}{462.2\muKs}
\setsymbol{LFI:white:noise:sensitivity:LFI25:Rad:M:units}{413.6\muKs}
\setsymbol{LFI:white:noise:sensitivity:LFI26:Rad:M:units}{478.6\muKs}
\setsymbol{LFI:white:noise:sensitivity:LFI27:Rad:M:units}{277.7\muKs}
\setsymbol{LFI:white:noise:sensitivity:LFI28:Rad:M:units}{312.3\muKs}
\setsymbol{LFI:white:noise:sensitivity:LFI18:Rad:S:units}{465.7\muKs}
\setsymbol{LFI:white:noise:sensitivity:LFI19:Rad:S:units}{555.6\muKs}
\setsymbol{LFI:white:noise:sensitivity:LFI20:Rad:S:units}{623.2\muKs}
\setsymbol{LFI:white:noise:sensitivity:LFI21:Rad:S:units}{564.1\muKs}
\setsymbol{LFI:white:noise:sensitivity:LFI22:Rad:S:units}{534.4\muKs}
\setsymbol{LFI:white:noise:sensitivity:LFI23:Rad:S:units}{542.4\muKs}
\setsymbol{LFI:white:noise:sensitivity:LFI24:Rad:S:units}{399.2\muKs}
\setsymbol{LFI:white:noise:sensitivity:LFI25:Rad:S:units}{392.6\muKs}
\setsymbol{LFI:white:noise:sensitivity:LFI26:Rad:S:units}{418.6\muKs}
\setsymbol{LFI:white:noise:sensitivity:LFI27:Rad:S:units}{302.9\muKs}
\setsymbol{LFI:white:noise:sensitivity:LFI28:Rad:S:units}{285.3\muKs}

\setsymbol{LFI:white:noise:sensitivity:LFI18:Rad:M}{512.0}
\setsymbol{LFI:white:noise:sensitivity:LFI19:Rad:M}{581.4}
\setsymbol{LFI:white:noise:sensitivity:LFI20:Rad:M}{590.8}
\setsymbol{LFI:white:noise:sensitivity:LFI21:Rad:M}{455.2}
\setsymbol{LFI:white:noise:sensitivity:LFI22:Rad:M}{492.0}
\setsymbol{LFI:white:noise:sensitivity:LFI23:Rad:M}{507.7}
\setsymbol{LFI:white:noise:sensitivity:LFI24:Rad:M}{462.2}
\setsymbol{LFI:white:noise:sensitivity:LFI25:Rad:M}{413.6}
\setsymbol{LFI:white:noise:sensitivity:LFI26:Rad:M}{478.6}
\setsymbol{LFI:white:noise:sensitivity:LFI27:Rad:M}{277.7}
\setsymbol{LFI:white:noise:sensitivity:LFI28:Rad:M}{312.3}
\setsymbol{LFI:white:noise:sensitivity:LFI18:Rad:S}{465.7}
\setsymbol{LFI:white:noise:sensitivity:LFI19:Rad:S}{555.6}
\setsymbol{LFI:white:noise:sensitivity:LFI20:Rad:S}{623.2}
\setsymbol{LFI:white:noise:sensitivity:LFI21:Rad:S}{564.1}
\setsymbol{LFI:white:noise:sensitivity:LFI22:Rad:S}{534.4}
\setsymbol{LFI:white:noise:sensitivity:LFI23:Rad:S}{542.4}
\setsymbol{LFI:white:noise:sensitivity:LFI24:Rad:S}{399.2}
\setsymbol{LFI:white:noise:sensitivity:LFI25:Rad:S}{392.6}
\setsymbol{LFI:white:noise:sensitivity:LFI26:Rad:S}{418.6}
\setsymbol{LFI:white:noise:sensitivity:LFI27:Rad:S}{302.9}
\setsymbol{LFI:white:noise:sensitivity:LFI28:Rad:S}{285.3}

\setsymbol{LFI:knee:frequency:70GHz:units}{29.5\mHz}
\setsymbol{LFI:knee:frequency:44GHz:units}{56.2\mHz}
\setsymbol{LFI:knee:frequency:30GHz:units}{113.7\mHz}

\setsymbol{LFI:knee:frequency:70GHz}{29.5}
\setsymbol{LFI:knee:frequency:44GHz}{56.2}
\setsymbol{LFI:knee:frequency:30GHz}{113.7}

\setsymbol{LFI:knee:frequency:LFI18:Rad:M:units}{16.3\mHz}
\setsymbol{LFI:knee:frequency:LFI19:Rad:M:units}{15.1\mHz}
\setsymbol{LFI:knee:frequency:LFI20:Rad:M:units}{18.7\mHz}
\setsymbol{LFI:knee:frequency:LFI21:Rad:M:units}{37.2\mHz}
\setsymbol{LFI:knee:frequency:LFI22:Rad:M:units}{12.7\mHz}
\setsymbol{LFI:knee:frequency:LFI23:Rad:M:units}{34.6\mHz}
\setsymbol{LFI:knee:frequency:LFI24:Rad:M:units}{46.2\mHz}
\setsymbol{LFI:knee:frequency:LFI25:Rad:M:units}{24.9\mHz}
\setsymbol{LFI:knee:frequency:LFI26:Rad:M:units}{67.6\mHz}
\setsymbol{LFI:knee:frequency:LFI27:Rad:M:units}{187.4\mHz}
\setsymbol{LFI:knee:frequency:LFI28:Rad:M:units}{122.2\mHz}
\setsymbol{LFI:knee:frequency:LFI18:Rad:S:units}{17.7\mHz}
\setsymbol{LFI:knee:frequency:LFI19:Rad:S:units}{22.0\mHz}
\setsymbol{LFI:knee:frequency:LFI20:Rad:S:units}{8.7\mHz}
\setsymbol{LFI:knee:frequency:LFI21:Rad:S:units}{25.9\mHz}
\setsymbol{LFI:knee:frequency:LFI22:Rad:S:units}{15.8\mHz}
\setsymbol{LFI:knee:frequency:LFI23:Rad:S:units}{129.8\mHz}
\setsymbol{LFI:knee:frequency:LFI24:Rad:S:units}{100.9\mHz}
\setsymbol{LFI:knee:frequency:LFI25:Rad:S:units}{38.9\mHz}
\setsymbol{LFI:knee:frequency:LFI26:Rad:S:units}{58.9\mHz}
\setsymbol{LFI:knee:frequency:LFI27:Rad:S:units}{104.4\mHz}
\setsymbol{LFI:knee:frequency:LFI28:Rad:S:units}{40.7\mHz}

\setsymbol{LFI:knee:frequency:LFI18:Rad:M}{16.3}
\setsymbol{LFI:knee:frequency:LFI19:Rad:M}{15.1}
\setsymbol{LFI:knee:frequency:LFI20:Rad:M}{18.7}
\setsymbol{LFI:knee:frequency:LFI21:Rad:M}{37.2}
\setsymbol{LFI:knee:frequency:LFI22:Rad:M}{12.7}
\setsymbol{LFI:knee:frequency:LFI23:Rad:M}{34.6}
\setsymbol{LFI:knee:frequency:LFI24:Rad:M}{46.2}
\setsymbol{LFI:knee:frequency:LFI25:Rad:M}{24.9}
\setsymbol{LFI:knee:frequency:LFI26:Rad:M}{67.6}
\setsymbol{LFI:knee:frequency:LFI27:Rad:M}{187.4}
\setsymbol{LFI:knee:frequency:LFI28:Rad:M}{122.2}
\setsymbol{LFI:knee:frequency:LFI18:Rad:S}{17.7}
\setsymbol{LFI:knee:frequency:LFI19:Rad:S}{22.0}
\setsymbol{LFI:knee:frequency:LFI20:Rad:S}{8.7}
\setsymbol{LFI:knee:frequency:LFI21:Rad:S}{25.9}
\setsymbol{LFI:knee:frequency:LFI22:Rad:S}{15.8}
\setsymbol{LFI:knee:frequency:LFI23:Rad:S}{129.8}
\setsymbol{LFI:knee:frequency:LFI24:Rad:S}{100.9}
\setsymbol{LFI:knee:frequency:LFI25:Rad:S}{38.9}
\setsymbol{LFI:knee:frequency:LFI26:Rad:S}{58.9}
\setsymbol{LFI:knee:frequency:LFI27:Rad:S}{104.4}
\setsymbol{LFI:knee:frequency:LFI28:Rad:S}{40.7}

\setsymbol{LFI:slope:70GHz:units}{$-1.03$\mHz}
\setsymbol{LFI:slope:44GHz:units}{$-0.89$\mHz}
\setsymbol{LFI:slope:30GHz:units}{$-0.87$\mHz}

\setsymbol{LFI:slope:70GHz}{$-1.03$}
\setsymbol{LFI:slope:44GHz}{$-0.89$}
\setsymbol{LFI:slope:30GHz}{$-0.87$}

\setsymbol{LFI:slope:LFI18:Rad:M:units}{$-1.04$\mHz}
\setsymbol{LFI:slope:LFI19:Rad:M:units}{$-1.09$\mHz}
\setsymbol{LFI:slope:LFI20:Rad:M:units}{$-0.69$\mHz}
\setsymbol{LFI:slope:LFI21:Rad:M:units}{$-1.56$\mHz}
\setsymbol{LFI:slope:LFI22:Rad:M:units}{$-1.01$\mHz}
\setsymbol{LFI:slope:LFI23:Rad:M:units}{$-0.96$\mHz}
\setsymbol{LFI:slope:LFI24:Rad:M:units}{$-0.83$\mHz}
\setsymbol{LFI:slope:LFI25:Rad:M:units}{$-0.91$\mHz}
\setsymbol{LFI:slope:LFI26:Rad:M:units}{$-0.95$\mHz}
\setsymbol{LFI:slope:LFI27:Rad:M:units}{$-0.87$\mHz}
\setsymbol{LFI:slope:LFI28:Rad:M:units}{$-0.88$\mHz}
\setsymbol{LFI:slope:LFI18:Rad:S:units}{$-1.15$\mHz}
\setsymbol{LFI:slope:LFI19:Rad:S:units}{$-1.00$\mHz}
\setsymbol{LFI:slope:LFI20:Rad:S:units}{$-0.95$\mHz}
\setsymbol{LFI:slope:LFI21:Rad:S:units}{$-0.92$\mHz}
\setsymbol{LFI:slope:LFI22:Rad:S:units}{$-1.01$\mHz}
\setsymbol{LFI:slope:LFI23:Rad:S:units}{$-0.95$\mHz}
\setsymbol{LFI:slope:LFI24:Rad:S:units}{$-0.73$\mHz}
\setsymbol{LFI:slope:LFI25:Rad:S:units}{$-1.16$\mHz}
\setsymbol{LFI:slope:LFI26:Rad:S:units}{$-0.79$\mHz}
\setsymbol{LFI:slope:LFI27:Rad:S:units}{$-0.82$\mHz}
\setsymbol{LFI:slope:LFI28:Rad:S:units}{$-0.91$\mHz}

\setsymbol{LFI:slope:LFI18:Rad:M}{$-1.04$}
\setsymbol{LFI:slope:LFI19:Rad:M}{$-1.09$}
\setsymbol{LFI:slope:LFI20:Rad:M}{$-0.69$}
\setsymbol{LFI:slope:LFI21:Rad:M}{$-1.56$}
\setsymbol{LFI:slope:LFI22:Rad:M}{$-1.01$}
\setsymbol{LFI:slope:LFI23:Rad:M}{$-0.96$}
\setsymbol{LFI:slope:LFI24:Rad:M}{$-0.83$}
\setsymbol{LFI:slope:LFI25:Rad:M}{$-0.91$}
\setsymbol{LFI:slope:LFI26:Rad:M}{$-0.95$}
\setsymbol{LFI:slope:LFI27:Rad:M}{$-0.87$}
\setsymbol{LFI:slope:LFI28:Rad:M}{$-0.88$}
\setsymbol{LFI:slope:LFI18:Rad:S}{$-1.15$}
\setsymbol{LFI:slope:LFI19:Rad:S}{$-1.00$}
\setsymbol{LFI:slope:LFI20:Rad:S}{$-0.95$}
\setsymbol{LFI:slope:LFI21:Rad:S}{$-0.92$}
\setsymbol{LFI:slope:LFI22:Rad:S}{$-1.01$}
\setsymbol{LFI:slope:LFI23:Rad:S}{$-0.95$}
\setsymbol{LFI:slope:LFI24:Rad:S}{$-0.73$}
\setsymbol{LFI:slope:LFI25:Rad:S}{$-1.16$}
\setsymbol{LFI:slope:LFI26:Rad:S}{$-0.79$}
\setsymbol{LFI:slope:LFI27:Rad:S}{$-0.82$}
\setsymbol{LFI:slope:LFI28:Rad:S}{$-0.91$}

\setsymbol{LFI:FWHM:70GHz:units}{13\parcm01}
\setsymbol{LFI:FWHM:44GHz:units}{27\parcm92}
\setsymbol{LFI:FWHM:30GHz:units}{32\parcm65}

\setsymbol{LFI:FWHM:70GHz}{13.01}
\setsymbol{LFI:FWHM:44GHz}{27.92}
\setsymbol{LFI:FWHM:30GHz}{32.65}

\setsymbol{LFI:FWHM:LFI18:units}{13\parcm39}
\setsymbol{LFI:FWHM:LFI19:units}{13\parcm01}
\setsymbol{LFI:FWHM:LFI20:units}{12\parcm75}
\setsymbol{LFI:FWHM:LFI21:units}{12\parcm74}
\setsymbol{LFI:FWHM:LFI22:units}{12\parcm87}
\setsymbol{LFI:FWHM:LFI23:units}{13\parcm27}
\setsymbol{LFI:FWHM:LFI24:units}{22\parcm98}
\setsymbol{LFI:FWHM:LFI25:units}{30\parcm46}
\setsymbol{LFI:FWHM:LFI26:units}{30\parcm31}
\setsymbol{LFI:FWHM:LFI27:units}{32\parcm65}
\setsymbol{LFI:FWHM:LFI28:units}{32\parcm66}

\setsymbol{LFI:FWHM:LFI18}{13.39}
\setsymbol{LFI:FWHM:LFI19}{13.01}
\setsymbol{LFI:FWHM:LFI20}{12.75}
\setsymbol{LFI:FWHM:LFI21}{12.74}
\setsymbol{LFI:FWHM:LFI22}{12.87}
\setsymbol{LFI:FWHM:LFI23}{13.27}
\setsymbol{LFI:FWHM:LFI24}{22.98}
\setsymbol{LFI:FWHM:LFI25}{30.46}
\setsymbol{LFI:FWHM:LFI26}{30.31}
\setsymbol{LFI:FWHM:LFI27}{32.65}
\setsymbol{LFI:FWHM:LFI28}{32.66}

\setsymbol{LFI:FWHM:uncertainty:LFI18:units}{0.170\arcm}
\setsymbol{LFI:FWHM:uncertainty:LFI19:units}{0.174\arcm}
\setsymbol{LFI:FWHM:uncertainty:LFI20:units}{0.170\arcm}
\setsymbol{LFI:FWHM:uncertainty:LFI21:units}{0.156\arcm}
\setsymbol{LFI:FWHM:uncertainty:LFI22:units}{0.164\arcm}
\setsymbol{LFI:FWHM:uncertainty:LFI23:units}{0.171\arcm}
\setsymbol{LFI:FWHM:uncertainty:LFI24:units}{0.652\arcm}
\setsymbol{LFI:FWHM:uncertainty:LFI25:units}{1.075\arcm}
\setsymbol{LFI:FWHM:uncertainty:LFI26:units}{1.131\arcm}
\setsymbol{LFI:FWHM:uncertainty:LFI27:units}{1.266\arcm}
\setsymbol{LFI:FWHM:uncertainty:LFI28:units}{1.287\arcm}

\setsymbol{LFI:FWHM:uncertainty:LFI18}{0.170}
\setsymbol{LFI:FWHM:uncertainty:LFI19}{0.174}
\setsymbol{LFI:FWHM:uncertainty:LFI20}{0.170}
\setsymbol{LFI:FWHM:uncertainty:LFI21}{0.156}
\setsymbol{LFI:FWHM:uncertainty:LFI22}{0.164}
\setsymbol{LFI:FWHM:uncertainty:LFI23}{0.171}
\setsymbol{LFI:FWHM:uncertainty:LFI24}{0.652}
\setsymbol{LFI:FWHM:uncertainty:LFI25}{1.075}
\setsymbol{LFI:FWHM:uncertainty:LFI26}{1.131}
\setsymbol{LFI:FWHM:uncertainty:LFI27}{1.266}
\setsymbol{LFI:FWHM:uncertainty:LFI28}{1.287}

\setsymbol{HFI:center:frequency:100GHz:units}{100\,GHz}
\setsymbol{HFI:center:frequency:143GHz:units}{143\,GHz}
\setsymbol{HFI:center:frequency:217GHz:units}{217\,GHz}
\setsymbol{HFI:center:frequency:353GHz:units}{353\,GHz}
\setsymbol{HFI:center:frequency:545GHz:units}{545\,GHz}
\setsymbol{HFI:center:frequency:857GHz:units}{857\,GHz}

\setsymbol{HFI:center:frequency:100GHz}{100}
\setsymbol{HFI:center:frequency:143GHz}{143}
\setsymbol{HFI:center:frequency:217GHz}{217}
\setsymbol{HFI:center:frequency:353GHz}{353}
\setsymbol{HFI:center:frequency:545GHz}{545}
\setsymbol{HFI:center:frequency:857GHz}{857}

\setsymbol{HFI:Ndetectors:100GHz}{8}
\setsymbol{HFI:Ndetectors:143GHz}{11}
\setsymbol{HFI:Ndetectors:217GHz}{12}
\setsymbol{HFI:Ndetectors:353GHz}{12}
\setsymbol{HFI:Ndetectors:545GHz}{3}
\setsymbol{HFI:Ndetectors:857GHz}{4}

\setsymbol{HFI:FWHM:Maps:100GHz:units}{9\parcm88}
\setsymbol{HFI:FWHM:Maps:143GHz:units}{7\parcm18}
\setsymbol{HFI:FWHM:Maps:217GHz:units}{4\parcm87}
\setsymbol{HFI:FWHM:Maps:353GHz:units}{4\parcm65}
\setsymbol{HFI:FWHM:Maps:545GHz:units}{4\parcm72}
\setsymbol{HFI:FWHM:Maps:857GHz:units}{4\parcm39}
\setsymbol{HFI:FWHM:Maps:100GHz}{9.88}
\setsymbol{HFI:FWHM:Maps:143GHz}{7.18}
\setsymbol{HFI:FWHM:Maps:217GHz}{4.87}
\setsymbol{HFI:FWHM:Maps:353GHz}{4.65}
\setsymbol{HFI:FWHM:Maps:545GHz}{4.72}
\setsymbol{HFI:FWHM:Maps:857GHz}{4.39}

\setsymbol{HFI:beam:ellipticity:Maps:100GHz}{1.15}
\setsymbol{HFI:beam:ellipticity:Maps:143GHz}{1.01}
\setsymbol{HFI:beam:ellipticity:Maps:217GHz}{1.06}
\setsymbol{HFI:beam:ellipticity:Maps:353GHz}{1.05}
\setsymbol{HFI:beam:ellipticity:Maps:545GHz}{1.14}
\setsymbol{HFI:beam:ellipticity:Maps:857GHz}{1.19}

\setsymbol{HFI:FWHM:Mars:100GHz:units}{9\parcm37}
\setsymbol{HFI:FWHM:Mars:143GHz:units}{7\parcm04}
\setsymbol{HFI:FWHM:Mars:217GHz:units}{4\parcm68}
\setsymbol{HFI:FWHM:Mars:353GHz:units}{4\parcm43}
\setsymbol{HFI:FWHM:Mars:545GHz:units}{3\parcm80}
\setsymbol{HFI:FWHM:Mars:857GHz:units}{3\parcm67}

\setsymbol{HFI:FWHM:Mars:100GHz}{9.37}
\setsymbol{HFI:FWHM:Mars:143GHz}{7.04}
\setsymbol{HFI:FWHM:Mars:217GHz}{4.68}
\setsymbol{HFI:FWHM:Mars:353GHz}{4.43}
\setsymbol{HFI:FWHM:Mars:545GHz}{3.80}
\setsymbol{HFI:FWHM:Mars:857GHz}{3.67}

\setsymbol{HFI:beam:ellipticity:Mars:100GHz}{1.18}
\setsymbol{HFI:beam:ellipticity:Mars:143GHz}{1.03}
\setsymbol{HFI:beam:ellipticity:Mars:217GHz}{1.14}
\setsymbol{HFI:beam:ellipticity:Mars:353GHz}{1.09}
\setsymbol{HFI:beam:ellipticity:Mars:545GHz}{1.25}
\setsymbol{HFI:beam:ellipticity:Mars:857GHz}{1.03}

\setsymbol{HFI:CMB:relative:calibration:100GHz}{$\lsim 1\%$}
\setsymbol{HFI:CMB:relative:calibration:143GHz}{$\lsim 1\%$}
\setsymbol{HFI:CMB:relative:calibration:217GHz}{$\lsim 1\%$}
\setsymbol{HFI:CMB:relative:calibration:353GHz}{$\lsim 1\%$}
\setsymbol{HFI:CMB:relative:calibration:545GHz}{}
\setsymbol{HFI:CMB:relative:calibration:857GHz}{}

\setsymbol{HFI:CMB:absolute:calibration:100GHz}{$\lsim 2\%$}
\setsymbol{HFI:CMB:absolute:calibration:143GHz}{$\lsim 2\%$}
\setsymbol{HFI:CMB:absolute:calibration:217GHz}{$\lsim 2\%$}
\setsymbol{HFI:CMB:absolute:calibration:353GHz}{$\lsim 2\%$}
\setsymbol{HFI:CMB:absolute:calibration:545GHz}{}
\setsymbol{HFI:CMB:absolute:calibration:857GHz}{}

\setsymbol{HFI:FIRAS:gain:calibration:accuracy:statistical:100GHz}{}
\setsymbol{HFI:FIRAS:gain:calibration:accuracy:statistical:143GHz}{}
\setsymbol{HFI:FIRAS:gain:calibration:accuracy:statistical:217GHz}{}
\setsymbol{HFI:FIRAS:gain:calibration:accuracy:statistical:353GHz}{2.5\%}
\setsymbol{HFI:FIRAS:gain:calibration:accuracy:statistical:545GHz}{1\%}
\setsymbol{HFI:FIRAS:gain:calibration:accuracy:statistical:857GHz}{0.5\%}

\setsymbol{HFI:FIRAS:gain:calibration:accuracy:systematic:100GHz}{}
\setsymbol{HFI:FIRAS:gain:calibration:accuracy:systematic:143GHz}{}
\setsymbol{HFI:FIRAS:gain:calibration:accuracy:systematic:217GHz}{}
\setsymbol{HFI:FIRAS:gain:calibration:accuracy:systematic:353GHz}{}
\setsymbol{HFI:FIRAS:gain:calibration:accuracy:systematic:545GHz}{7\%}
\setsymbol{HFI:FIRAS:gain:calibration:accuracy:systematic:857GHz}{7\%}

\setsymbol{HFI:FIRAS:zero:point:accuracy:100GHz:units}{0.8\MJysr}
\setsymbol{HFI:FIRAS:zero:point:accuracy:143GHz:units}{}
\setsymbol{HFI:FIRAS:zero:point:accuracy:217GHz:units}{}
\setsymbol{HFI:FIRAS:zero:point:accuracy:353GHz:units}{1.4\MJysr}
\setsymbol{HFI:FIRAS:zero:point:accuracy:545GHz:units}{2.2\MJysr}
\setsymbol{HFI:FIRAS:zero:point:accuracy:857GHz:units}{1.7\MJysr}

\setsymbol{HFI:FIRAS:zero:point:accuracy:100GHz}{0.8}
\setsymbol{HFI:FIRAS:zero:point:accuracy:143GHz}{}
\setsymbol{HFI:FIRAS:zero:point:accuracy:217GHz}{}
\setsymbol{HFI:FIRAS:zero:point:accuracy:353GHz}{1.4}
\setsymbol{HFI:FIRAS:zero:point:accuracy:545GHz}{2.2}
\setsymbol{HFI:FIRAS:zero:point:accuracy:857GHz}{1.7}

\setsymbol{HFI:unit:conversion:100GHz:units}{0.2415\MJysrmK}
\setsymbol{HFI:unit:conversion:143GHz:units}{0.3694\MJysrmK}
\setsymbol{HFI:unit:conversion:217GHz:units}{0.4811\MJysrmK}
\setsymbol{HFI:unit:conversion:353GHz:units}{0.2883\MJysrmK}
\setsymbol{HFI:unit:conversion:545GHz:units}{0.05826\MJysrmK}
\setsymbol{HFI:unit:conversion:857GHz:units}{0.002238\MJysrmK}

\setsymbol{HFI:unit:conversion:100GHz}{0.2415}
\setsymbol{HFI:unit:conversion:143GHz}{0.3694}
\setsymbol{HFI:unit:conversion:217GHz}{0.4811}
\setsymbol{HFI:unit:conversion:353GHz}{0.2883}
\setsymbol{HFI:unit:conversion:545GHz}{0.05826}
\setsymbol{HFI:unit:conversion:857GHz}{0.002238}

\setsymbol{HFI:colour:correction:alpha=-2:V1.01:100GHz}{0.9893}
\setsymbol{HFI:colour:correction:alpha=-2:V1.01:143GHz}{0.9759}
\setsymbol{HFI:colour:correction:alpha=-2:V1.01:217GHz}{1.0007}
\setsymbol{HFI:colour:correction:alpha=-2:V1.01:353GHz}{1.0028}
\setsymbol{HFI:colour:correction:alpha=-2:V1.01:545GHz}{1.0019}
\setsymbol{HFI:colour:correction:alpha=-2:V1.01:857GHz}{0.9889}

\setsymbol{HFI:colour:correction:alpha=0:V1.01:100GHz}{1.0008}
\setsymbol{HFI:colour:correction:alpha=0:V1.01:143GHz}{1.0148}
\setsymbol{HFI:colour:correction:alpha=0:V1.01:217GHz}{0.9909}
\setsymbol{HFI:colour:correction:alpha=0:V1.01:353GHz}{0.9888}
\setsymbol{HFI:colour:correction:alpha=0:V1.01:545GHz}{0.9878}
\setsymbol{HFI:colour:correction:alpha=0:V1.01:857GHz}{1.0014}

\providecommand{\sorthelp}[1]{}

%% file: AuthorList_P02b_LFI_Calibration_authors_and_institutes.tex
%This author list corresponds to \title{Author list for SVN P02b\_LFI\_Calibration, Proj. Ref. 1\_4: LFI Calibration}
%Prepared by R. Leonardi (rleonardi@sciops.esa.int), ESAC/ESA
%This version is from Thu Sep 05 13:37:08 2013 CET
%\subtitle{There are 219 co-authors in this list}
\author{\small
Planck Collaboration:
N.~Aghanim\inst{59}
\and
C.~Armitage-Caplan\inst{89}
\and
M.~Arnaud\inst{72}
\and
M.~Ashdown\inst{69, 6}
\and
F.~Atrio-Barandela\inst{17}
\and
J.~Aumont\inst{59}
\and
C.~Baccigalupi\inst{83}
\and
A.~J.~Banday\inst{92, 8}
\and
R.~B.~Barreiro\inst{66}
\and
E.~Battaner\inst{93}
% ...
}
\institute{\small
APC, AstroParticule et Cosmologie, Universit\'{e} Paris Diderot, CNRS/IN2P3, CEA/lrfu, Observatoire de Paris, Sorbonne Paris Cit\'{e}, 10, rue Alice Domon et L\'{e}onie Duquet, 75205 Paris Cedex 13, France\\
\and
Aalto University Mets\"{a}hovi Radio Observatory, Mets\"{a}hovintie 114, FIN-02540 Kylm\"{a}l\"{a}, Finland\\
\and
African Institute for Mathematical Sciences, 6-8 Melrose Road, Muizenberg, Cape Town, South Africa\\
\and
Agenzia Spaziale Italiana Science Data Center, Via del Politecnico snc, 00133, Roma, Italy\\
\and
Agenzia Spaziale Italiana, Viale Liegi 26, Roma, Italy\\
\and
Astrophysics Group, Cavendish Laboratory, University of Cambridge, J J Thomson Avenue, Cambridge CB3 0HE, U.K.\\
\and
CITA, University of Toronto, 60 St. George St., Toronto, ON M5S 3H8, Canada\\
\and
CNRS, IRAP, 9 Av. colonel Roche, BP 44346, F-31028 Toulouse cedex 4, France\\
\and
California Institute of Technology, Pasadena, California, U.S.A.\\
\and
Centre for Theoretical Cosmology, DAMTP, University of Cambridge, Wilberforce Road, Cambridge CB3 0WA, U.K.\\
\and
Centro de Estudios de F\'{i}sica del Cosmos de Arag\'{o}n (CEFCA), Plaza San Juan, 1, planta 2, E-44001, Teruel, Spain\\
\and
Computational Cosmology Center, Lawrence Berkeley National Laboratory, Berkeley, California, U.S.A.\\
\and
Consejo Superior de Investigaciones Cient\'{\i}ficas (CSIC), Madrid, Spain\\
\and
DSM/Irfu/SPP, CEA-Saclay, F-91191 Gif-sur-Yvette Cedex, France\\
\and
DTU Space, National Space Institute, Technical University of Denmark, Elektrovej 327, DK-2800 Kgs. Lyngby, Denmark\\
\and
D\'{e}partement de Physique Th\'{e}orique, Universit\'{e} de Gen\`{e}ve, 24, Quai E. Ansermet,1211 Gen\`{e}ve 4, Switzerland\\
\and
Departamento de F\'{\i}sica Fundamental, Facultad de Ciencias, Universidad de Salamanca, 37008 Salamanca, Spain\\
\and
Departamento de F\'{\i}sica, Universidad de Oviedo, Avda. Calvo Sotelo s/n, Oviedo, Spain\\
\and
Department of Astronomy and Astrophysics, University of Toronto, 50 Saint George Street, Toronto, Ontario, Canada\\
\and
Department of Astrophysics/IMAPP, Radboud University Nijmegen, P.O. Box 9010, 6500 GL Nijmegen, The Netherlands\\
\and
Department of Electrical Engineering and Computer Sciences, University of California, Berkeley, California, U.S.A.\\
\and
Department of Physics \& Astronomy, University of British Columbia, 6224 Agricultural Road, Vancouver, British Columbia, Canada\\
\and
Department of Physics and Astronomy, Dana and David Dornsife College of Letter, Arts and Sciences, University of Southern California, Los Angeles, CA 90089, U.S.A.\\
\and
Department of Physics and Astronomy, University College London, London WC1E 6BT, U.K.\\
\and
Department of Physics, Florida State University, Keen Physics Building, 77 Chieftan Way, Tallahassee, Florida, U.S.A.\\
\and
Department of Physics, Gustaf H\"{a}llstr\"{o}min katu 2a, University of Helsinki, Helsinki, Finland\\
\and
Department of Physics, Princeton University, Princeton, New Jersey, U.S.A.\\
\and
Department of Physics, University of California, One Shields Avenue, Davis, California, U.S.A.\\
\and
Department of Physics, University of California, Santa Barbara, California, U.S.A.\\
\and
Department of Physics, University of Illinois at Urbana-Champaign, 1110 West Green Street, Urbana, Illinois, U.S.A.\\
\and
Dipartimento di Fisica e Astronomia G. Galilei, Universit\`{a} degli Studi di Padova, via Marzolo 8, 35131 Padova, Italy\\
\and
Dipartimento di Fisica e Scienze della Terra, Universit\`{a} di Ferrara, Via Saragat 1, 44122 Ferrara, Italy\\
\and
Dipartimento di Fisica, Universit\`{a} La Sapienza, P. le A. Moro 2, Roma, Italy\\
\and
Dipartimento di Fisica, Universit\`{a} degli Studi di Milano, Via Celoria, 16, Milano, Italy\\
\and
Dipartimento di Fisica, Universit\`{a} degli Studi di Trieste, via A. Valerio 2, Trieste, Italy\\
\and
Dipartimento di Fisica, Universit\`{a} di Roma Tor Vergata, Via della Ricerca Scientifica, 1, Roma, Italy\\
\and
Discovery Center, Niels Bohr Institute, Blegdamsvej 17, Copenhagen, Denmark\\
\and
Dpto. Astrof\'{i}sica, Universidad de La Laguna (ULL), E-38206 La Laguna, Tenerife, Spain\\
\and
European Space Agency, ESAC, Planck Science Office, Camino bajo del Castillo, s/n, Urbanizaci\'{o}n Villafranca del Castillo, Villanueva de la Ca\~{n}ada, Madrid, Spain\\
\and
European Space Agency, ESTEC, Keplerlaan 1, 2201 AZ Noordwijk, The Netherlands\\
\and
Haverford College Astronomy Department, 370 Lancaster Avenue, Haverford, Pennsylvania, U.S.A.\\
\and
Helsinki Institute of Physics, Gustaf H\"{a}llstr\"{o}min katu 2, University of Helsinki, Helsinki, Finland\\
\and
INAF - Osservatorio Astrofisico di Catania, Via S. Sofia 78, Catania, Italy\\
\and
INAF - Osservatorio Astronomico di Padova, Vicolo dell'Osservatorio 5, Padova, Italy\\
\and
INAF - Osservatorio Astronomico di Roma, via di Frascati 33, Monte Porzio Catone, Italy\\
\and
INAF - Osservatorio Astronomico di Trieste, Via G.B. Tiepolo 11, Trieste, Italy\\
\and
INAF Istituto di Radioastronomia, Via P. Gobetti 101, 40129 Bologna, Italy\\
\and
INAF/IASF Bologna, Via Gobetti 101, Bologna, Italy\\
\and
INAF/IASF Milano, Via E. Bassini 15, Milano, Italy\\
\and
INFN, Sezione di Bologna, Via Irnerio 46, I-40126, Bologna, Italy\\
\and
INFN, Sezione di Roma 1, Universit\`{a} di Roma Sapienza, Piazzale Aldo Moro 2, 00185, Roma, Italy\\
\and
IPAG: Institut de Plan\'{e}tologie et d'Astrophysique de Grenoble, Universit\'{e} Joseph Fourier, Grenoble 1 / CNRS-INSU, UMR 5274, Grenoble, F-38041, France\\
\and
ISDC Data Centre for Astrophysics, University of Geneva, ch. d'Ecogia 16, Versoix, Switzerland\\
\and
IUCAA, Post Bag 4, Ganeshkhind, Pune University Campus, Pune 411 007, India\\
\and
Imperial College London, Astrophysics group, Blackett Laboratory, Prince Consort Road, London, SW7 2AZ, U.K.\\
\and
Infrared Processing and Analysis Center, California Institute of Technology, Pasadena, CA 91125, U.S.A.\\
\and
Institut N\'{e}el, CNRS, Universit\'{e} Joseph Fourier Grenoble I, 25 rue des Martyrs, Grenoble, France\\
\and
Institut Universitaire de France, 103, bd Saint-Michel, 75005, Paris, France\\
\and
Institut d'Astrophysique Spatiale, CNRS (UMR8617) Universit\'{e} Paris-Sud 11, B\^{a}timent 121, Orsay, France\\
\and
Institut d'Astrophysique de Paris, CNRS (UMR7095), 98 bis Boulevard Arago, F-75014, Paris, France\\
\and
Institute for Space Sciences, Bucharest-Magurale, Romania\\
\and
Institute of Astronomy and Astrophysics, Academia Sinica, Taipei, Taiwan\\
\and
Institute of Astronomy, University of Cambridge, Madingley Road, Cambridge CB3 0HA, U.K.\\
\and
Institute of Theoretical Astrophysics, University of Oslo, Blindern, Oslo, Norway\\
\and
Instituto de Astrof\'{\i}sica de Canarias, C/V\'{\i}a L\'{a}ctea s/n, La Laguna, Tenerife, Spain\\
\and
Instituto de F\'{\i}sica de Cantabria (CSIC-Universidad de Cantabria), Avda. de los Castros s/n, Santander, Spain\\
\and
Jet Propulsion Laboratory, California Institute of Technology, 4800 Oak Grove Drive, Pasadena, California, U.S.A.\\
\and
Jodrell Bank Centre for Astrophysics, Alan Turing Building, School of Physics and Astronomy, The University of Manchester, Oxford Road, Manchester, M13 9PL, U.K.\\
\and
Kavli Institute for Cosmology Cambridge, Madingley Road, Cambridge, CB3 0HA, U.K.\\
\and
LAL, Universit\'{e} Paris-Sud, CNRS/IN2P3, Orsay, France\\
\and
LERMA, CNRS, Observatoire de Paris, 61 Avenue de l'Observatoire, Paris, France\\
\and
Laboratoire AIM, IRFU/Service d'Astrophysique - CEA/DSM - CNRS - Universit\'{e} Paris Diderot, B\^{a}t. 709, CEA-Saclay, F-91191 Gif-sur-Yvette Cedex, France\\
\and
Laboratoire Traitement et Communication de l'Information, CNRS (UMR 5141) and T\'{e}l\'{e}com ParisTech, 46 rue Barrault F-75634 Paris Cedex 13, France\\
\and
Laboratoire de Physique Subatomique et de Cosmologie, Universit\'{e} Joseph Fourier Grenoble I, CNRS/IN2P3, Institut National Polytechnique de Grenoble, 53 rue des Martyrs, 38026 Grenoble cedex, France\\
\and
Laboratoire de Physique Th\'{e}orique, Universit\'{e} Paris-Sud 11 \& CNRS, B\^{a}timent 210, 91405 Orsay, France\\
\and
Lawrence Berkeley National Laboratory, Berkeley, California, U.S.A.\\
\and
Max-Planck-Institut f\"{u}r Astrophysik, Karl-Schwarzschild-Str. 1, 85741 Garching, Germany\\
\and
McGill Physics, Ernest Rutherford Physics Building, McGill University, 3600 rue University, Montr\'{e}al, QC, H3A 2T8, Canada\\
\and
MilliLab, VTT Technical Research Centre of Finland, Tietotie 3, Espoo, Finland\\
\and
Niels Bohr Institute, Blegdamsvej 17, Copenhagen, Denmark\\
\and
Observational Cosmology, Mail Stop 367-17, California Institute of Technology, Pasadena, CA, 91125, U.S.A.\\
\and
SB-ITP-LPPC, EPFL, CH-1015, Lausanne, Switzerland\\
\and
SISSA, Astrophysics Sector, via Bonomea 265, 34136, Trieste, Italy\\
\and
School of Physics and Astronomy, Cardiff University, Queens Buildings, The Parade, Cardiff, CF24 3AA, U.K.\\
\and
School of Physics and Astronomy, University of Nottingham, Nottingham NG7 2RD, U.K.\\
\and
Space Sciences Laboratory, University of California, Berkeley, California, U.S.A.\\
\and
Special Astrophysical Observatory, Russian Academy of Sciences, Nizhnij Arkhyz, Zelenchukskiy region, Karachai-Cherkessian Republic, 369167, Russia\\
\and
Stanford University, Dept of Physics, Varian Physics Bldg, 382 Via Pueblo Mall, Stanford, California, U.S.A.\\
\and
Sub-Department of Astrophysics, University of Oxford, Keble Road, Oxford OX1 3RH, U.K.\\
\and
Theory Division, PH-TH, CERN, CH-1211, Geneva 23, Switzerland\\
\and
UPMC Univ Paris 06, UMR7095, 98 bis Boulevard Arago, F-75014, Paris, France\\
\and
Universit\'{e} de Toulouse, UPS-OMP, IRAP, F-31028 Toulouse cedex 4, France\\
\and
University of Granada, Departamento de F\'{\i}sica Te\'{o}rica y del Cosmos, Facultad de Ciencias, Granada, Spain\\
% ...
}

%% file: P02b_1_introduction.tex
This paper, one of a set associated with the 2013 release of data from the \Planck\footnote{\Planck\ (\url{http://www.esa.int/Planck}) is a project of the European Space Agency (ESA) with instruments provided by two scientific consortia funded by ESA member states (in particular the lead countries France and Italy), with contributions from NASA (USA) and telescope reflectors provided by a collaboration between ESA and a scientific consortium led and funded by Denmark.} mission \citep{planck2013-p01}, describes the techniques we employed to calibrate the voltages measured by the LFI radiometers into a set of thermodynamic temperatures (\emph{photometric calibration}).  We also discuss the quality of our calibration in terms of the required accuracy needed to achieve \Planck{}'s final science goals. This paper is part of a larger set of articles \citep{planck2013-p02,planck2013-p02a,planck2013-p02d} which explain the methodology used to produce maps from raw LFI data which have been issued in the \Planck{} 2013 data release. A similar paper, \citet{planck2013-p03f}, describes the approach used by HFI. \citet{planck2013-p01a}, in preparation, will  contain a comparison of the LFI/HFI approaches and assess the consistency of \Planck's maps and power spectra.

By ``calibration'' here we mean the process that converts each voltage measured by an analogue-to-digital converter (ADC) into a thermodynamic temperature. (We describe the incoming flux as a thermodynamic temperature because the CMB signal is a nearly perfect blackbody, and thus the temperature is a more physically significant quantity to measure.) The process can be modelled by the following equation:
\begin{equation}
\label{eq:calibrationEquation}
V_\mathrm{out}(t) = G(t)\times\Bigl[B * \bigl(T_\mathrm{sky} + D\bigr)\Bigr]_{\xversor(t), t} + M,
\end{equation}
where $\xversor(t)$ is the direction of the beam axis at time $t$. This relates the voltage $V_\mathrm{out}$ measured by the ADC with the sum of three terms: (1) the convolution\footnote{In this work we use the following notation for convolution:
\[
(A * B)(\theta, \phi) = \int_{4\pi} A(\theta', \phi') B(\theta - \theta', \phi - \phi')\,\ud\Omega'.
\]} between the brightness temperature $T_\mathrm{sky}$ of the sky (CMB and galactic/extragalactic foregrounds) and the beam response $B$ of the instrument ($\int_{4\pi} B\,\ud\Omega = 1$) at a given time, (2) the convolution between $B$ and the CMB dipole $D$ (including the solar and orbital terms, as well as their associated kinematic quadrupoles), and (3) an offset term $M$ (monopole, including instrumental offsets), which is of little importance for differential instruments\footnote{Although LFI directly measures $V_\text{out}$, it can be considered a differential instrument \protect\citep{bersanelli2010} as the analysis is performed on the value $V_\text{out} - r V_\text{ref}$, where $V_\text{ref}$ is the measurement of the temperature of a stable 4\,K heat load (Sect.~\protect\ref{sec:dVV}) and $r$ is a coefficient which removes the contribution of $M$ from both $V_\text{out}$ and $V_\text{ref}$, so that the time average of $V_\text{out} - r V_\text{ref}$ is zero. Refer to \protect\citet{planck2013-p02} for further details.} like LFI. (The dependence of the $B * (T_\mathrm{sky} + D)$ term on both $\xversor$ and $t$, expressed in Eq.~\ref{eq:calibrationEquation} by a subscript, is due to the fact that $D$ depends on the velocity of \Planck{}, which is a time-dependent quantity.) The transfer function $G$ represents the overall ``gain'' of the instrument. We are primarily interested in $K = G^{-1}$, as the purpose of the calibration is to convert $V$ back into a temperature. Since a number of environmental factors influence the value of $K$, we expect it to change with time. The output of a calibration procedure is therefore a time series of values $K_i$, which sample the unknown function $K(t)$ at a reasonable frequency (i.e., higher than the frequency of the expected instrumental fluctuations in $K$) and which allow us to reconstruct the value of $B * T_\mathrm{sky}$ in Eq.~\ref{eq:calibrationEquation} with good accuracy (for LFI, this accuracy is between 0.6 and 0.8\,\%, with 0.25\,\% of uncertainty coming from the error bars on the characterization of the CMB dipole provided by \textit{WMAP}).

Proper \emph{relative calibration} (i.e., precise tracking of the variations of each radiometer's gain throughout the mission) is a necessary condition for a self-consistent map-making. In addition, accurate determination of the \emph{absolute calibration} (i.e., translation of the observed voltages into physical units, in terms of antenna temperature), together with proper reconstruction of the beam window function, is crucial for any scientific exploitation of the maps and power spectra. The LFI calibrated maps are used extensively in the \Planck{} data analysis. They are a fundamental input to the component separation process \citep{planck2013-p06}, which leads to \Planck's full-sky CMB map. This map is the basis for the extraction of the Planck power spectrum in the low-multipole regime \citep{planck2013-p08}, and for all analyses on non-Gaussianity \citep{planck2013-p09a}, isotropy, and second-order statistics \citep{planck2013-p09}. Furthermore, the LFI power spectrum provides a unique consistency check internal to \Planck, particularly in the comparison between the LFI 70 GHz and the HFI 100 GHz channels \citep{planck2013-p01a}. The LFI beams and window functions are discussed in detail in \citet{planck2013-p02d}. Here we give a detailed account of the LFI absolute and relative calibration. Earlier accounts of the calibration procedure for LFI were given by \citet{villa2010} and \citet{mennella2010}, which present the results of the LFI on-ground calibration campaign, and by \citet{planck2011-1.4}, which describes the LFI calibration procedure used for producing the \Planck{} Early Results \citep{planck2011-1.1}.

The structure of this paper is the following. In Sect.~\ref{sec:calibrationPhilosophyAndConventions} we introduce a number of important ideas that are going to be used in this work, namely the time scales of variations in $K$ (Sect.~\ref{sec:timeScale}), the treatment of beam sidelobes in the calibration and their impact on subsequent analyses of LFI's calibrated data (Sect.~\ref{sec:beamEfficiencyAndWindowFunctions}), and an updated list of colour corrections (Sect~\ref{sec:colourCorrections}). Then, in Sect.~\ref{sec:calibrationMethods}, we explain the methods we have developed to calibrate the data acquired by the LFI radiometers. We discuss the type of systematic effects affecting the calibration procedure in Sect.~\ref{sec:systematics}. In Sect.~\ref{sec:accuracy} we estimate the accuracy of our calibration, and we include the most relevant results from \citet{planck2013-p01a}, which compares LFI results to HFI as well as WMAP. Finally, in Sect.~\ref{sec:conclusions} we summarize our results and propose a number of improvements to be implemented for the future releases of LFI data.

%% file: P02b_2_1_time_scale.tex
In this section we establish the time scale over which we expect significant variations in the gain of the LFI radiometers. This quantity drives the design of the calibration algorithms we then discuss in the next sections.

Changes in the gain of the LFI radiometers are mainly triggered by changes in the thermal environment of the LFI instrument, particularly in the front-end and back-end modules \citep{bersanelli2010}. The time scale of gain changes can therefore be estimated either by considering the rate of change of temperature sampled near the radiometer amplifiers, or by using the radiometer to continuously measure the temperature of a load kept at a stable temperature. The latter solution is viable for LFI, because each radiometer continuously observes a stable 4\,K load mounted on the external shield of HFI \citep{valenziano2009,lamarre2010}; the temperature of each 4\,K load drifts by less than one mK per year (so that the ratio of a yearly drift to the system temperature, $T_\text{sys} \approx 20\,\text{K}$, is $\lesssim 0.01\,\%$), and the 4\,K signal entering LFI radiometers goes through the same chain as the signal coming from the telescope. It can therefore be used to assess the rate of change in the gain with good accuracy.

\begin{figure}
	\centering
	\includegraphics{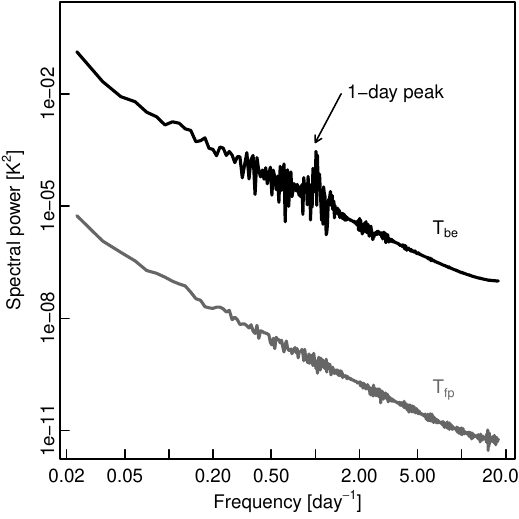}
	\caption{\label{fig:HKFourierTransforms}Fourier transforms of the temperatures of the focal plane ($T_\text{fp}$, sensor \texttt{TS5R}) and the back-end ($T_\text{be}$, sensor \DIFnomarkup{\texttt{R\_BEM1}}) measured during the first three months of data acquisition. Apart from the daily fluctuation induced by the transponder, there are no significant peaks at high frequencies.}
\end{figure}

Despite the fact that \Planck{} orbits around the Sun-Earth $L_2$ point, which grants a stable thermal environment \citep{tauber2010a}, there are, however, some phenomena that can induce variations in the temperature near the amplifiers:
\begin{enumerate}
\item Fluctuations in the temperature of the 20\,K sorption cooler cold end \citep{planck2011-1.3}. The cycle period of each cooler's bed is of the order of tens of minutes, but the induced change of temperature in the focal plane is minimal (less than 0.01\,\%). We must move to much longer time scales (i.e., weeks) to see significant variations in the temperature.
\item Fluctuations in the temperature of the warm back-end of the radiometers. The biggest variations we experienced during the mission are due to the continuous turning on/off of the transponder antenna (used to send data to Earth) early in the mission, which followed a duty cycle of 24 hours and induced 17\,mK peak-to-peak fluctuations (over an average temperature of $\sim290\,\text{K}$) in the temperature near the back-end amplifiers.
\end{enumerate}
This is well represented by Fig.~\ref{fig:HKFourierTransforms}, which shows that the power spectrum of the thermal fluctuations measured on the focal plane and in the back-end modules has a high low-frequency part, but it has negligible power at time scales shorter than one day. (The peak caused by the transponder switching is clearly visible in the spectrum of $T_\mathrm{be}$.)

We do not expect significant variations in the gain of the radiometers on timescales shorter than these, i.e., tens of hours. To make our discussion more quantitative, we can estimate the rate of change of a temperature or output voltage $f(t)$ by means of the following parameter\footnote{If the quantity $f(t)$ is affected by statistical noise at high frequencies, as it is the case for all the quantities considered here, it is necessary to apply some kind of low pass filter to it before applying Eq.~\ref{eq:rateOfChange} in order to obtain meaningful results.}:
\begin{equation}
\label{eq:rateOfChange}
\tau_f(t) = \varepsilon\left|\frac{f(t)}{\frac{\ud f}{\ud t}(t)}\right|,
\end{equation}
which has the unit of time and quantifies the typical time required to induce a change of level $\varepsilon$ in $f$ at time $t$. Using $\varepsilon = 0.01$ (i.e., we are looking for 1\,\% changes), we find the following timescales:
\begin{enumerate}
\item Fluctuations in the temperature $T_\mathrm{fp}$ of the focal plane happen on timescales of the order of weeks.
\item The time scale for fluctuations in the temperature $T_\mathrm{be}$ of the back-end are faster during the first survey, as $\left<\tau_{T_\mathrm{be}}\right>$ is of the order of tens of hours. After the first survey\footnote{The number of sky surveys in this \Planck{} data release is two and a half.} the transponder was left on continuously, and this value increases to roughly one week.
\item Results similar to those for the back-end are found when considering the total-power voltage $V_\mathrm{ref}$, which measures the temperature of the 4\,K loads (i.e., setting $f$ equal to $V_\mathrm{ref}$ instead of $T_\mathrm{fp}$ or $T_\mathrm{be}$).
\end{enumerate}
These results motivate the need to re-calibrate each radiometer more than once per day. The most natural length of time for \Planck{} is the duration of one \emph{pointing period}, i.e., the interval between two consecutive repointings of the spacecraft, which happens roughly once per hour \citep{dupac2005} and is short enough to detect any significant change in the gains. At the same time, the interval is long enough to sample the dipole signal (our main calibration source) with good signal-to-noise ratio, since during one pointing period the telescope scans the same circle in the sky tens of times (the median value is 39 times, and 50\,\% of the pointings fall in the 36--42 range).

For this data release we chose not to explicitly consider effects due to the aging of the radiometric components and variations in the emissivity of the telescope. Such phenomena can lead to variations in the gain, noise temperatures, or an increase in the instabilities of the instrument, but we have had no clear evidence that they are significant on the relatively short time span covered by this data release (one year and a half). In September-October 2013 we have run a number of End-of-Life tests on the instrument, with the purpose of quantitatively assessing such effects: we will present the results of our analysis in a future \Planck{} data release.

%% file: P02b_2_2_beam_efficiency_window_functions.tex
As described by \citet{Bersanelli+al:1997}, \citet{Cappellini+al:2003}, and \citet{planck2011-1.4}, the calibration of LFI is referenced to the dipole signal, which is a nearly ideal calibrating source. Since \Planck{} observes the sky by spinning around the Sun-Earth axis with a speed of 1\,rpm \citep{tauber2010a}, and since the main beams are located at $\sim$85$^\circ$ from the spinning axis, the dipole induces a sinusoidal fluctuation in the time ordered data with frequency 1/60\,Hz and varying amplitude\footnote{The scanning strategy has been designed such that this amplitude never vanishes, see \citet{dupac2005}. However, as the scan axis changes, variations in the observed amplitude \emph{do} occur and they affect the accuracy of the calibration, as we discuss in Sect.~\protect\ref{sec:OSGCalibration}.} which can be used for the calibration. For this release, our reference dipole $D_\mathrm{ref}$ is the combination of the solar dipole as given by the {\it WMAP} values \citep{jarosik2010}, and the orbital dipole, derived from the known velocity of the \Planck{} spacecraft relative to the barycentre of the Solar System. We now describe the calibration procedure for LFI and, in particular, we discuss how non-ideal beams affect the process. This discussion is similar to the one carried on in \citet{planck2013-p01a}. However, here we provide a more rigorous treatment which involves a description of the relevant quantities as a function of time rather than as a function of the observing direction.

\begin{figure}
	\begin{DIFnomarkup}
	\input{three_beams_plot.tex}
	\end{DIFnomarkup}
	\caption{\label{fig:threeLFIBeams} Simulated main meam pattern for beams at the three frequencies: \texttt{LFI18M} (70\,GHz, 13\parcm41 FWHM), \texttt{LFI24M} (44\,GHz, 23\parcm23), and \texttt{LFI27M} (30\,GHz, 33\parcm06). The beams have been calculated using \texttt{GRASP}.}
\end{figure}
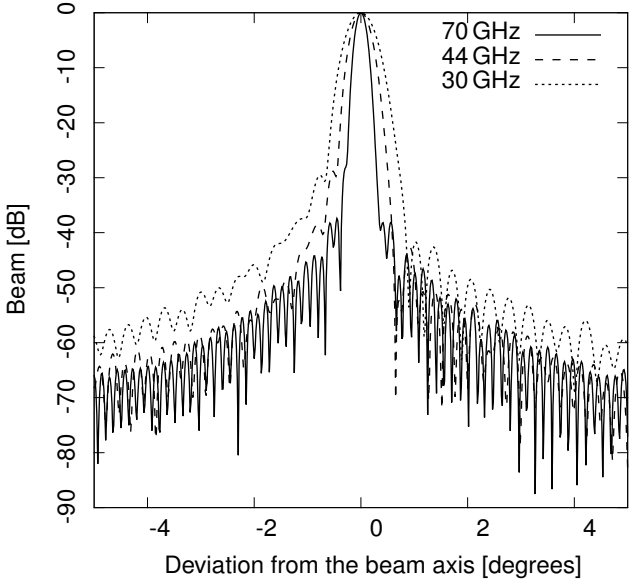

The starting point of our discussion is Eq.~\ref{eq:calibrationEquation}. In the LFI pipeline \citep{planck2013-p02} we remove\footnote{This step is based on an algorithm which starts from the approximation $T_\mathrm{sky} + D \approx D$ and then refines the solution iteratively. The real algorithm is more complex than this, as it removes $T_\mathrm{sky}$ at the same time as the least-square fit discussed later in this section: the details of the algorithm are presented in Sect.~\ref{sec:OSGCalibration}.} the $T_\mathrm{sky}$ term from the data, so that the equation reduces to the following:
\begin{equation}
\label{eq:gainFittingTrue}
V'_\mathrm{out}(t) = G (B * D) \bigl(\xversor(t), t\bigr) + M,
\end{equation}
where we indicate with $V'_\mathrm{out}$ the signal $V_\mathrm{out}$ without the $T_\mathrm{sky}$ component. This signal is compared with a model of the beam-convolved dipole, based on the beam response $B_\mathrm{model}$ and on the reference dipole $D_\mathrm{ref}$:
\begin{equation}
\label{eq:gainFittingFirstStep}
V'_\mathrm{out}(t) = \tilde G (B_\mathrm{model} * D_\mathrm{ref}) \bigl(\xversor(t), t\bigr) + \tilde M.
\end{equation}
The LFI beams are discussed in \citet{planck2013-p02d} (main beams and window functions) and \citet{planck2013-p02a} (sidelobes); Fig.~\ref{fig:threeLFIBeams} shows three typical beam profiles in the $[-5^\circ, 5^\circ]$ range. The unknown parameters $\tilde G$ and $\tilde M$ in Eq.~\ref{eq:gainFittingFirstStep} can be found by means of a least square fit between $V_\mathrm{out}$ and the right hand side. (Such fit is done once per each pointing period, i.e., the period between two consecutive repointings of the spacecraft, which happens roughly once per hour.) From the point of view of the LFI calibration, the $\tilde G$ factor is the only important parameter to estimate. To estimate it, we equate Eq.~\ref{eq:gainFittingTrue} and Eq.~\ref{eq:gainFittingFirstStep} and take the time derivative $\timeder$ of the two\footnote{Although the equation is mathematically correct, we note that it is of little use for a numerical implementation, as the denominator is a sinusoid which periodically goes to zero, thus making the quantity diverge to infinity.} sides:
\begin{equation}
\label{eq:gainReconstructionWithSidelobes}
\tilde{G} = G \frac{\timeder(B*D)}{\timeder\bigl(B_\mathrm{model} * D_\mathrm{ref}\bigr)}.
\end{equation}
A common approach to the use of this equation is to approximate the beam with a pencil\footnote{In the context of this paper, a pencil beam $B_\text{pencil}$ is a Dirac delta function centered on the beam axis $\hat{e}_b$:
\[
B_\text{pencil} (\hat{x}) = \delta\bigr(\hat{x} - \hat{e}_b\bigr).
\].} beam, $B_\mathrm{model} \approx B_\mathrm{pencil}$, as adopted by HFI \citep{planck2013-p03f} and {\it WMAP} \citep{jarosik2007}. For LFI we have computed the full $4\pi$ beams for all detectors and we developed a fast convolution routine to estimate $B_\mathrm{model} * D_\mathrm{ref}$. This was motivated by the wish to fully control dipole-coupling to the sidelobes, particularly at 30\,GHz where sidelobes are larger. The convolution of a generic beam with a dipole produces a smearing effect (due to the fact that not all the power is in the main beam) and a slight tilt in the dipole axis (resulting from asymmetries in the beam). As shown in detail in Sect.~\ref{sec:fourPiConvolver} and Appendix~\ref{sec:computingDipoleConvolutionParams}, these effects can be quantified by the length and direction, respectively, of the vector $\Svector$ defined there.

We verified \textit{a posteriori} that the net effect of the convolution does not produce a significant improvement over a pencil beam model at the present stage of the analysis (besides the tilt effect, which is $<0.01\,\%$). For this data release, therefore, in the convolution routine we rescaled the length of vector $\Svector$ to unity, which is effectively equivalent to modelling the beam as a pencil beam. The advantage of this approach is that it matches the convention on the normalization of the beam that has been assumed in the calibration of HFI data. In the following description of map calibration and associated uncertainties, therefore, we will assume $B_\mathrm{model} \approx B_\mathrm{pencil}$. Full use of the $4\pi$ convolver, including integrated frequency-dependent sidelobes within the LFI radiometric bands, will be applied for later analyses, including polarization calibration.

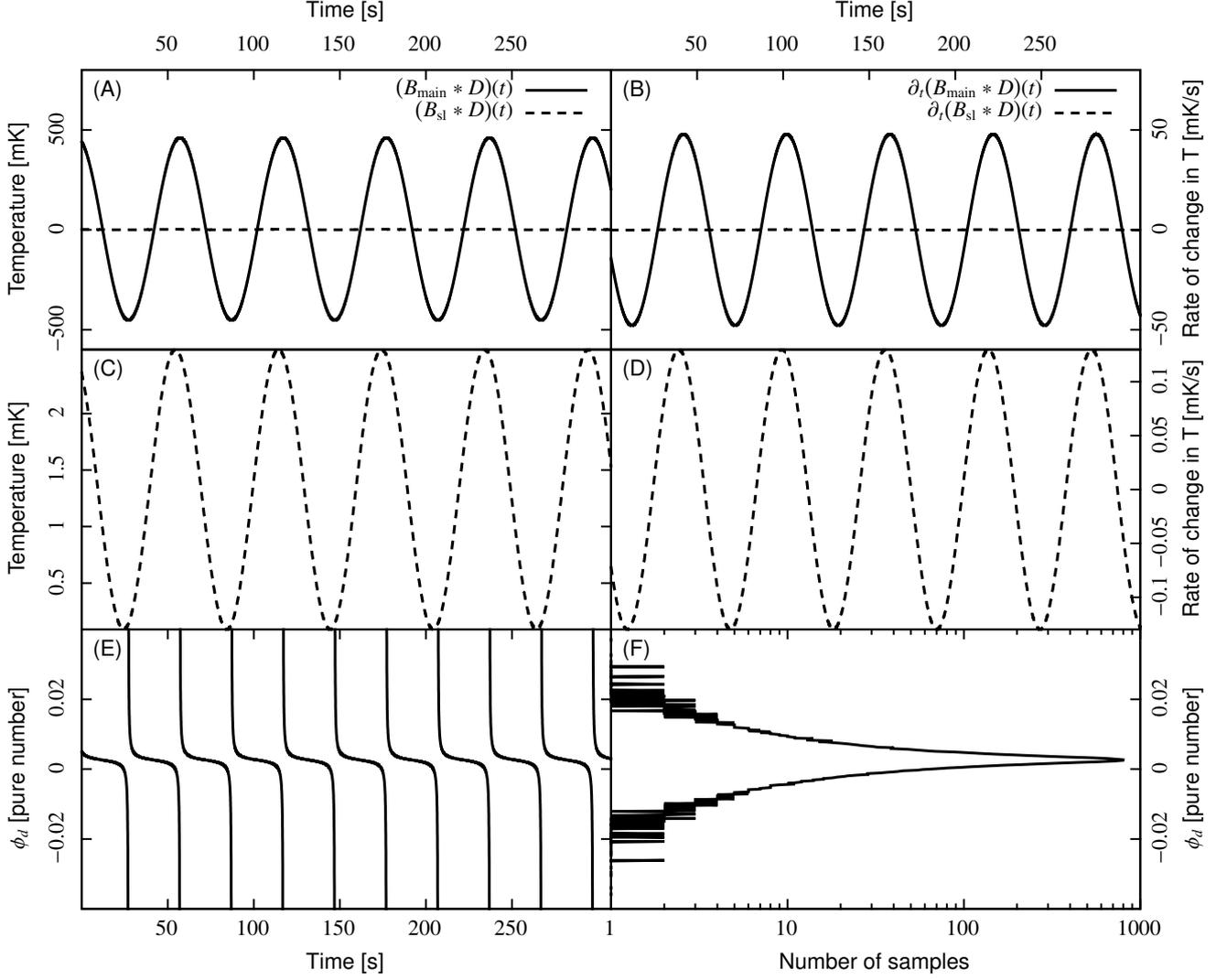
\begin{figure*}
	\begin{DIFnomarkup}
	\input{phi_d_in_time.tex}
	\end{DIFnomarkup}
	\caption{\label{fig:phiDInTime} Results of a simulation which shows how $\phi_D$ is computed. We assume to observe a $1\,\mathrm{K}$ peak-to-peak dipole in the sky for 5 minutes with a scanning strategy very similar to the one used for \Planck{}'s 30\,GHz radiometers, i.e., the sky is scanned in circles of high amplitude ($\sim 85^\circ$) with a rotation frequency $\nu = 1/60\,\mathrm{Hz}$ and a sampling frequency of $32.5\,\mathrm{Hz}$ (so that 10\,000 temperature samples are generated for each data stream). We observe the dipole using a realistic 30\,GHz beam $B = B_\mathrm{main} + B_\mathrm{side}$ with FWHM $0\pdeg5$. \textit{Panel A}: Plot of the $\bigl(B_\mathrm{main} * D\bigr)(t)$ term, which oscillates as a sinusoid with amplitude $\lesssim 0.5\,\mathrm{K}$; the term $\bigl(B_\mathrm{sl} * D\bigr)(t)$ is negligible (see panel C for a close-up). \textit{Panel B}: plot of the $\timeder\bigl(B_\mathrm{main}*D\bigr)(t)$ term, used in the definition of $\phi_D$ (Eq.~\protect\ref{eq:phiD}). \textit{Panel C}: Close-up of the $\bigl(B_\mathrm{sl} * D\bigr)(t)$ term shown in panel A. \textit{Panel D}: Close-up of the $\timeder \bigl(B_\mathrm{sl} * D\bigr)(t)$ shown in panel B. \textit{Panel E}: Value of $\phi_D$ as a function of time, calculated using the definition in Eq.~\protect\ref{eq:phiD}. \textit{Panel F}: Distribution of the 10\,000 values of $\phi_D$ plotted in panel E. Half of the values fall within the 0.19--0.34\,\% range.}
\end{figure*}

We can write the true beam $B$ as the sum of two terms, $B = B_\mathrm{main} + B_\mathrm{side}$, where $B_\mathrm{main}$ represents the contribution of what we define as the \emph{main beam} (defined as the portion of the beam within $5^\circ$ of the beam centre), and $B_\mathrm{side}$ represents the much smaller remaining part (the sidelobes). Equation~\ref{eq:gainReconstructionWithSidelobes} becomes:
\begin{equation}
\tilde{G} \approx G \frac{\timeder\bigl(B_\mathrm{main}*D + B_\mathrm{side}*D\bigr)}{\timeder\bigl(B_\mathrm{pencil} * D_\mathrm{ref}\bigr)}.
\end{equation}
The convolution of the main beam with the signal from the sky, $T=D+T_\mathrm{sky}$, is nearly identical to an ideal pencil beam except that only a fraction\footnote{The quantity $1 - f_\mathrm{side}$ is approximately equal to $\left\|\Svector\right\|$. However, as explained in Appendix~\ref{sec:computingDipoleConvolutionParams}, the vector $\Svector$ models the coupling of the full beam (including the main beam, which causes a tiny smearing effect because of its finite FWHM) with the dipole and only uses the first directional moments of the beam shape.} $1 - f_\mathrm{side}$ of the antenna gain is contained in the main beam:
\begin{equation}
B_\mathrm{main} * T \approx (1 - f_\mathrm{side}) B_\mathrm{pencil} * T \approx (1 - f_\mathrm{side}) T.
\end{equation}
So we have:
\begin{equation}
\begin{split}
\tilde{G} 
&= G \frac{(1 - f_\mathrm{side}) \timeder\bigl(B_\mathrm{pencil}*D\bigr) + \timeder\bigl(B_\mathrm{side}*D\bigr)}{\timeder \bigl(B_\mathrm{pencil} * D_\mathrm{ref}\bigr)} \\
&= G (1 - f_\mathrm{side}) (1 + \phi_D),
\end{split}
\end{equation}
where
\begin{equation}
\label{eq:phiD}
\phi_D = \frac{\timeder\bigl(B_\mathrm{side}*D\bigr)}{\timeder\bigl(B_\mathrm{main}*D\bigr)}.
\end{equation}
Fig.~\ref{fig:phiDInTime} shows the result of a simulation which illustrates how $\phi_D$ can be computed from the TODs.

Consider now a properly calibrated timeline $\tilde{T}_\mathrm{sky}$ where the dipole and the monopole terms in Eq.~\ref{eq:calibrationEquation} have been removed, so that $\tilde{T}_\mathrm{sky} = G (B * T_\mathrm{sky})$. Within each pointing period of constant $\tilde G$, we can write the relationship between the measured sky temperature $\tilde{T}_\mathrm{sky}$ and the level of the true sky temperature\footnote{Unlike $\tilde{T}_\mathrm{sky}$, $T_\mathrm{sky}$ is the sky temperature as seen through a pencil beam, and therefore it contains information at every angular scale. But obviously, in the context of the overall calibration level of LFI, we are interested only in the overall level of the sky temperature, say for angles $\gtrsim 0\pdeg5$. We thus ignore any scientific information contained in $T_\mathrm{sky}$ at smaller scale.} $T_\mathrm{sky}$ as:
\begin{equation}
\tilde{T}_\mathrm{sky} \bigl(\vec{x}(t)\bigr)
= \frac{G \bigl (B * T_\mathrm{sky}\bigr)\bigl(\vec{x}(t)\bigr)}{\tilde{G}} 
= \frac{\bigl(B * T_\mathrm{sky}\bigr)\bigl(\vec{x}(t)\bigr)}{(1 - f_\mathrm{side})(1 + \phi_D)}.
\end{equation}
Solving for the true temperature $T_\mathrm{sky}$, we find:
\begin{equation}
\label{eq:skyMapCorrectedForSL}
T_\mathrm{sky} \approx \tilde{T}_\mathrm{sky} (1 - \phi_\mathrm{sky} + \phi_D),
\end{equation}
where
\begin{equation}
\phi_\mathrm{sky} = \frac{B_\mathrm{side} * T_\mathrm{sky}}{B_\mathrm{main} * T_\mathrm{sky}} \left(\frac{T_\mathrm{sky}}{\tilde{T}_\mathrm{sky}}\right)
\end{equation}
is a time-dependent quantity which quantifies how much of the sky signal enters the beam through its sidelobes.

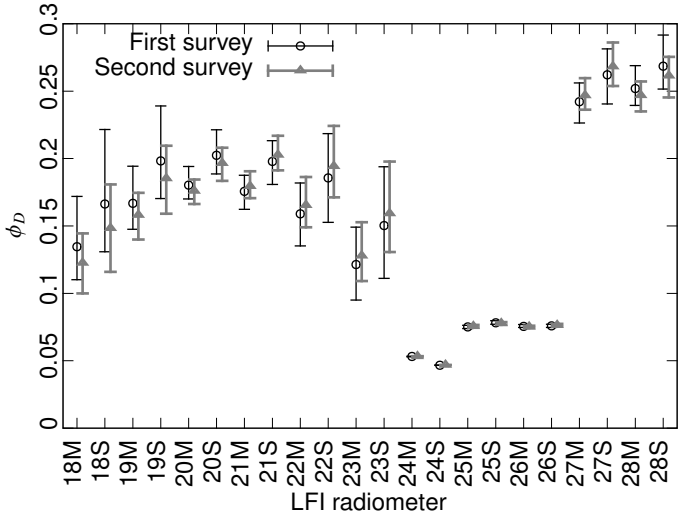
\begin{figure}
\centering
\begin{DIFnomarkup}
\input{phi_d_plot.tex}
\end{DIFnomarkup}
\caption{\label{fig:phiD} Estimated value of $\phi_D$ averaged over the first and second surveys. The central frequencies of the radiometers are 70\,GHz for \texttt{18M}\ldots\texttt{23S}, 44\,GHz for \texttt{24M}\ldots\texttt{26S}, and 30\,GHz for \texttt{27M}\ldots\texttt{28M}. In order to reduce the amount of data to consider, the time-dependent quantity $\phi_D$ was projected on a Healpix map (with \texttt{NSIDE} equal to 64) and binned. The error bars represent the first and third quartiles of the values of the 49\,152 pixels in the map, while the middle point is the median.}
\end{figure}

A precise evaluation of the correction terms $\phi_D$ and $\phi_\mathrm{sky}$ in Eq.~\ref{eq:skyMapCorrectedForSL} requires detailed simulations. We have performed such calculations for $\phi_D$  by computing the convolution of the sidelobes with the dipole from sample {\tt GRASP} full beams, leading to $\phi_D \approx 0.15\,\%$ (see Fig.~\ref{fig:phiD}). The term $\phi_\mathrm{sky}$ requires a full convolution of a sky model with the full beam, and it is frequency dependent. We performed a simulation for a 70\,GHz channel, projected the values of $\phi_\mathrm{sky}$ on a map and found values ranging between 0.05\,\% to 0.2\,\% throughout a full survey. In conclusion, the correction terms $\phi_\mathrm{sky}$ and $\phi_D$ are of the same order, within 0.2\,\%, and they tend to cancel out in Eq.~\ref{eq:skyMapCorrectedForSL}. Given that the relative uncertainties on both terms are large, we do not correct for them in the data and estimate a residual uncertainty of $0.2\,\%$ in the gain. This uncertainty is included in our overall estimated calibration uncertainty.

%% file: three_beams_plot.tex
% GNUPLOT: LaTeX picture with Postscript
\begingroup
  \fontfamily{phv}%
  \selectfont
  \makeatletter
  \providecommand\color[2][]{%
    \GenericError{(gnuplot) \space\space\space\@spaces}{%
      Package color not loaded in conjunction with
      terminal option `colourtext'%
    }{See the gnuplot documentation for explanation.%
    }{Either use 'blacktext' in gnuplot or load the package
      color.sty in LaTeX.}%
    \renewcommand\color[2][]{}%
  }%
  \providecommand\includegraphics[2][]{%
    \GenericError{(gnuplot) \space\space\space\@spaces}{%
      Package graphicx or graphics not loaded%
    }{See the gnuplot documentation for explanation.%
    }{The gnuplot epslatex terminal needs graphicx.sty or graphics.sty.}%
    \renewcommand\includegraphics[2][]{}%
  }%
  \providecommand\rotatebox[2]{#2}%
  \@ifundefined{ifGPcolor}{%
    \newif\ifGPcolor
    \GPcolorfalse
  }{}%
  \@ifundefined{ifGPblacktext}{%
    \newif\ifGPblacktext
    \GPblacktexttrue
  }{}%
  % define a \g@addto@macro without @ in the name:
  \let\gplgaddtomacro\g@addto@macro
  % define empty templates for all commands taking text:
  \gdef\gplbacktext{}%
  \gdef\gplfronttext{}%
  \makeatother
  \ifGPblacktext
    % no textcolor at all
    \def\colorrgb#1{}%
    \def\colorgray#1{}%
  \else
    % gray or color?
    \ifGPcolor
      \def\colorrgb#1{\color[rgb]{#1}}%
      \def\colorgray#1{\color[gray]{#1}}%
      \expandafter\def\csname LTw\endcsname{\color{white}}%
      \expandafter\def\csname LTb\endcsname{\color{black}}%
      \expandafter\def\csname LTa\endcsname{\color{black}}%
      \expandafter\def\csname LT0\endcsname{\color[rgb]{1,0,0}}%
      \expandafter\def\csname LT1\endcsname{\color[rgb]{0,1,0}}%
      \expandafter\def\csname LT2\endcsname{\color[rgb]{0,0,1}}%
      \expandafter\def\csname LT3\endcsname{\color[rgb]{1,0,1}}%
      \expandafter\def\csname LT4\endcsname{\color[rgb]{0,1,1}}%
      \expandafter\def\csname LT5\endcsname{\color[rgb]{1,1,0}}%
      \expandafter\def\csname LT6\endcsname{\color[rgb]{0,0,0}}%
      \expandafter\def\csname LT7\endcsname{\color[rgb]{1,0.3,0}}%
      \expandafter\def\csname LT8\endcsname{\color[rgb]{0.5,0.5,0.5}}%
    \else
      % gray
      \def\colorrgb#1{\color{black}}%
      \def\colorgray#1{\color[gray]{#1}}%
      \expandafter\def\csname LTw\endcsname{\color{white}}%
      \expandafter\def\csname LTb\endcsname{\color{black}}%
      \expandafter\def\csname LTa\endcsname{\color{black}}%
      \expandafter\def\csname LT0\endcsname{\color{black}}%
      \expandafter\def\csname LT1\endcsname{\color{black}}%
      \expandafter\def\csname LT2\endcsname{\color{black}}%
      \expandafter\def\csname LT3\endcsname{\color{black}}%
      \expandafter\def\csname LT4\endcsname{\color{black}}%
      \expandafter\def\csname LT5\endcsname{\color{black}}%
      \expandafter\def\csname LT6\endcsname{\color{black}}%
      \expandafter\def\csname LT7\endcsname{\color{black}}%
      \expandafter\def\csname LT8\endcsname{\color{black}}%
    \fi
  \fi
  \setlength{\unitlength}{0.0500bp}%
  \begin{picture}(4980.00,4520.00)%
    \gplgaddtomacro\gplbacktext{%
      \put(516,595){\rotatebox{-270}{\makebox(0,0){\strut{}-90}}}%
      \put(516,1008){\rotatebox{-270}{\makebox(0,0){\strut{}-80}}}%
      \put(516,1422){\rotatebox{-270}{\makebox(0,0){\strut{}-70}}}%
      \put(516,1835){\rotatebox{-270}{\makebox(0,0){\strut{}-60}}}%
      \put(516,2248){\rotatebox{-270}{\makebox(0,0){\strut{}-50}}}%
      \put(516,2662){\rotatebox{-270}{\makebox(0,0){\strut{}-40}}}%
      \put(516,3075){\rotatebox{-270}{\makebox(0,0){\strut{}-30}}}%
      \put(516,3488){\rotatebox{-270}{\makebox(0,0){\strut{}-20}}}%
      \put(516,3902){\rotatebox{-270}{\makebox(0,0){\strut{}-10}}}%
      \put(516,4315){\rotatebox{-270}{\makebox(0,0){\strut{} 0}}}%
      \put(1193,409){\makebox(0,0){\strut{}-4}}%
      \put(1989,409){\makebox(0,0){\strut{}-2}}%
      \put(2785,409){\makebox(0,0){\strut{} 0}}%
      \put(3581,409){\makebox(0,0){\strut{} 2}}%
      \put(4377,409){\makebox(0,0){\strut{} 4}}%
      \csname LTb\endcsname%
      \put(144,2455){\rotatebox{-270}{\makebox(0,0){\strut{}Beam [dB]}}}%
      \csname LTb\endcsname%
      \put(2785,130){\makebox(0,0){\strut{}Deviation from the beam axis [degrees]}}%
    }%
    \gplgaddtomacro\gplfronttext{%
      \csname LTb\endcsname%
      \put(3885,4148){\makebox(0,0)[r]{\strut{}70\,GHz}}%
      \csname LTb\endcsname%
      \put(3885,3962){\makebox(0,0)[r]{\strut{}44\,GHz}}%
      \csname LTb\endcsname%
      \put(3885,3776){\makebox(0,0)[r]{\strut{}30\,GHz}}%
    }%
    \gplbacktext
    \put(0,0){\includegraphics{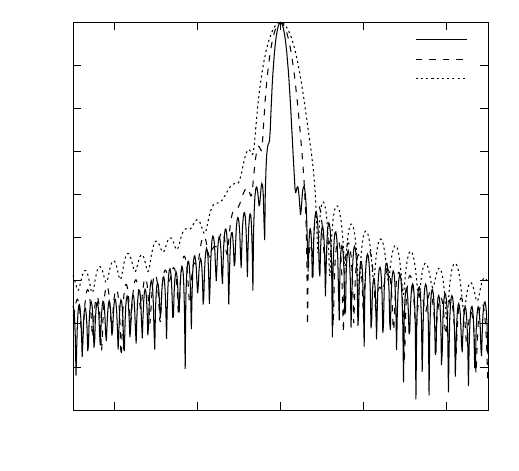}}%
    \gplfronttext
  \end{picture}%
\endgroup

%% file: phi_d_in_time.tex
% GNUPLOT: LaTeX picture with Postscript
\begingroup
  \fontfamily{phv}%
  \selectfont
  \makeatletter
  \providecommand\color[2][]{%
    \GenericError{(gnuplot) \space\space\space\@spaces}{%
      Package color not loaded in conjunction with
      terminal option `colourtext'%
    }{See the gnuplot documentation for explanation.%
    }{Either use 'blacktext' in gnuplot or load the package
      color.sty in LaTeX.}%
    \renewcommand\color[2][]{}%
  }%
  \providecommand\includegraphics[2][]{%
    \GenericError{(gnuplot) \space\space\space\@spaces}{%
      Package graphicx or graphics not loaded%
    }{See the gnuplot documentation for explanation.%
    }{The gnuplot epslatex terminal needs graphicx.sty or graphics.sty.}%
    \renewcommand\includegraphics[2][]{}%
  }%
  \providecommand\rotatebox[2]{#2}%
  \@ifundefined{ifGPcolor}{%
    \newif\ifGPcolor
    \GPcolorfalse
  }{}%
  \@ifundefined{ifGPblacktext}{%
    \newif\ifGPblacktext
    \GPblacktexttrue
  }{}%
  % define a \g@addto@macro without @ in the name:
  \let\gplgaddtomacro\g@addto@macro
  % define empty templates for all commands taking text:
  \gdef\gplbacktext{}%
  \gdef\gplfronttext{}%
  \makeatother
  \ifGPblacktext
    % no textcolor at all
    \def\colorrgb#1{}%
    \def\colorgray#1{}%
  \else
    % gray or color?
    \ifGPcolor
      \def\colorrgb#1{\color[rgb]{#1}}%
      \def\colorgray#1{\color[gray]{#1}}%
      \expandafter\def\csname LTw\endcsname{\color{white}}%
      \expandafter\def\csname LTb\endcsname{\color{black}}%
      \expandafter\def\csname LTa\endcsname{\color{black}}%
      \expandafter\def\csname LT0\endcsname{\color[rgb]{1,0,0}}%
      \expandafter\def\csname LT1\endcsname{\color[rgb]{0,1,0}}%
      \expandafter\def\csname LT2\endcsname{\color[rgb]{0,0,1}}%
      \expandafter\def\csname LT3\endcsname{\color[rgb]{1,0,1}}%
      \expandafter\def\csname LT4\endcsname{\color[rgb]{0,1,1}}%
      \expandafter\def\csname LT5\endcsname{\color[rgb]{1,1,0}}%
      \expandafter\def\csname LT6\endcsname{\color[rgb]{0,0,0}}%
      \expandafter\def\csname LT7\endcsname{\color[rgb]{1,0.3,0}}%
      \expandafter\def\csname LT8\endcsname{\color[rgb]{0.5,0.5,0.5}}%
    \else
      % gray
      \def\colorrgb#1{\color{black}}%
      \def\colorgray#1{\color[gray]{#1}}%
      \expandafter\def\csname LTw\endcsname{\color{white}}%
      \expandafter\def\csname LTb\endcsname{\color{black}}%
      \expandafter\def\csname LTa\endcsname{\color{black}}%
      \expandafter\def\csname LT0\endcsname{\color{black}}%
      \expandafter\def\csname LT1\endcsname{\color{black}}%
      \expandafter\def\csname LT2\endcsname{\color{black}}%
      \expandafter\def\csname LT3\endcsname{\color{black}}%
      \expandafter\def\csname LT4\endcsname{\color{black}}%
      \expandafter\def\csname LT5\endcsname{\color{black}}%
      \expandafter\def\csname LT6\endcsname{\color{black}}%
      \expandafter\def\csname LT7\endcsname{\color{black}}%
      \expandafter\def\csname LT8\endcsname{\color{black}}%
    \fi
  \fi
  \setlength{\unitlength}{0.0500bp}%
  \begin{picture}(10200.00,8500.00)%
    \gplgaddtomacro\gplbacktext{%
      \put(607,5562){\rotatebox{-270}{\makebox(0,0){\strut{}$-500$}}}%
      \put(607,6382){\rotatebox{-270}{\makebox(0,0){\strut{}$0$}}}%
      \put(607,7203){\rotatebox{-270}{\makebox(0,0){\strut{}$500$}}}%
      \put(1485,7881){\makebox(0,0){\strut{}$50$}}%
      \put(2187,7881){\makebox(0,0){\strut{}$100$}}%
      \put(2888,7881){\makebox(0,0){\strut{}$150$}}%
      \put(3590,7881){\makebox(0,0){\strut{}$200$}}%
      \put(4291,7881){\makebox(0,0){\strut{}$250$}}%
      \csname LTb\endcsname%
      \put(286,6546){\rotatebox{-270}{\makebox(0,0){\strut{}Temperature [mK]}}}%
      \csname LTb\endcsname%
      \put(2941,8159){\makebox(0,0){\strut{}Time [s]}}%
      \put(886,7509){\makebox(0,0)[l]{\strut{}(A)}}%
    }%
    \gplgaddtomacro\gplfronttext{%
      \csname LTb\endcsname%
      \put(4311,7528){\makebox(0,0)[r]{\strut{}$\bigl(B_\mathrm{main}*D\bigr)(t)$}}%
      \csname LTb\endcsname%
      \put(4311,7342){\makebox(0,0)[r]{\strut{}$\bigl(B_\mathrm{sl}*D\bigr)(t)$}}%
    }%
    \gplgaddtomacro\gplbacktext{%
      \csname LTb\endcsname%
      \put(9600,5562){\rotatebox{-270}{\makebox(0,0){\strut{}$-50$}}}%
      \put(9600,6382){\rotatebox{-270}{\makebox(0,0){\strut{}$0$}}}%
      \put(9600,7203){\rotatebox{-270}{\makebox(0,0){\strut{}$50$}}}%
      \put(5801,7881){\makebox(0,0){\strut{}$50$}}%
      \put(6502,7881){\makebox(0,0){\strut{}$100$}}%
      \put(7204,7881){\makebox(0,0){\strut{}$150$}}%
      \put(7906,7881){\makebox(0,0){\strut{}$200$}}%
      \put(8607,7881){\makebox(0,0){\strut{}$250$}}%
      \csname LTb\endcsname%
      \put(9860,6546){\rotatebox{-270}{\makebox(0,0){\strut{}Rate of change in T [mK/s]}}}%
      \csname LTb\endcsname%
      \put(7257,8159){\makebox(0,0){\strut{}Time [s]}}%
      \put(5202,7509){\makebox(0,0)[l]{\strut{}(B)}}%
    }%
    \gplgaddtomacro\gplfronttext{%
      \csname LTb\endcsname%
      \put(8626,7528){\makebox(0,0)[r]{\strut{}$\timeder \bigl(B_\mathrm{main}*D\bigr)(t)$}}%
      \csname LTb\endcsname%
      \put(8626,7342){\makebox(0,0)[r]{\strut{}$\timeder \bigl(B_\mathrm{sl}*D\bigr)(t)$}}%
    }%
    \gplgaddtomacro\gplbacktext{%
      \csname LTb\endcsname%
      \put(607,3480){\rotatebox{-270}{\makebox(0,0){\strut{}$0.5$}}}%
      \put(607,3945){\rotatebox{-270}{\makebox(0,0){\strut{}$1$}}}%
      \put(607,4409){\rotatebox{-270}{\makebox(0,0){\strut{}$1.5$}}}%
      \put(607,4873){\rotatebox{-270}{\makebox(0,0){\strut{}$2$}}}%
      \csname LTb\endcsname%
      \put(286,4249){\rotatebox{-270}{\makebox(0,0){\strut{}Temperature [mK]}}}%
      \put(886,5212){\makebox(0,0)[l]{\strut{}(C)}}%
    }%
    \gplgaddtomacro\gplfronttext{%
    }%
    \gplgaddtomacro\gplbacktext{%
      \csname LTb\endcsname%
      \put(9600,3364){\rotatebox{-270}{\makebox(0,0){\strut{}$-0.1$}}}%
      \put(9600,3807){\rotatebox{-270}{\makebox(0,0){\strut{}$-0.05$}}}%
      \put(9600,4249){\rotatebox{-270}{\makebox(0,0){\strut{}$0$}}}%
      \put(9600,4692){\rotatebox{-270}{\makebox(0,0){\strut{}$0.05$}}}%
      \put(9600,5135){\rotatebox{-270}{\makebox(0,0){\strut{}$0.1$}}}%
      \csname LTb\endcsname%
      \put(9860,4249){\rotatebox{-270}{\makebox(0,0){\strut{}Rate of change in T [mK/s]}}}%
      \put(5202,5212){\makebox(0,0)[l]{\strut{}(D)}}%
    }%
    \gplgaddtomacro\gplfronttext{%
    }%
    \gplgaddtomacro\gplbacktext{%
      \csname LTb\endcsname%
      \put(607,1378){\rotatebox{-270}{\makebox(0,0){\strut{}$-0.02$}}}%
      \put(607,1952){\rotatebox{-270}{\makebox(0,0){\strut{}$0$}}}%
      \put(607,2526){\rotatebox{-270}{\makebox(0,0){\strut{}$0.02$}}}%
      \put(1485,618){\makebox(0,0){\strut{}$50$}}%
      \put(2187,618){\makebox(0,0){\strut{}$100$}}%
      \put(2888,618){\makebox(0,0){\strut{}$150$}}%
      \put(3590,618){\makebox(0,0){\strut{}$200$}}%
      \put(4292,618){\makebox(0,0){\strut{}$250$}}%
      \csname LTb\endcsname%
      \put(286,1952){\rotatebox{-270}{\makebox(0,0){\strut{}$\phi_d$ [pure number]}}}%
      \csname LTb\endcsname%
      \put(2941,339){\makebox(0,0){\strut{}Time [s]}}%
      \put(886,2914){\makebox(0,0)[l]{\strut{}(E)}}%
    }%
    \gplgaddtomacro\gplfronttext{%
    }%
    \gplgaddtomacro\gplbacktext{%
      \csname LTb\endcsname%
      \put(5100,618){\makebox(0,0){\strut{}$1$}}%
      \put(6538,618){\makebox(0,0){\strut{}$10$}}%
      \put(7976,618){\makebox(0,0){\strut{}$100$}}%
      \put(9414,618){\makebox(0,0){\strut{}$1000$}}%
      \put(9600,1378){\rotatebox{-270}{\makebox(0,0){\strut{}$-0.02$}}}%
      \put(9600,1952){\rotatebox{-270}{\makebox(0,0){\strut{}$0$}}}%
      \put(9600,2526){\rotatebox{-270}{\makebox(0,0){\strut{}$0.02$}}}%
      \csname LTb\endcsname%
      \put(9860,1952){\rotatebox{-270}{\makebox(0,0){\strut{}$\phi_d$ [pure number]}}}%
      \csname LTb\endcsname%
      \put(7257,339){\makebox(0,0){\strut{}Number of samples}}%
      \put(5202,2914){\makebox(0,0)[l]{\strut{}(F)}}%
    }%
    \gplgaddtomacro\gplfronttext{%
    }%
    \gplbacktext
    \put(0,0){\includegraphics{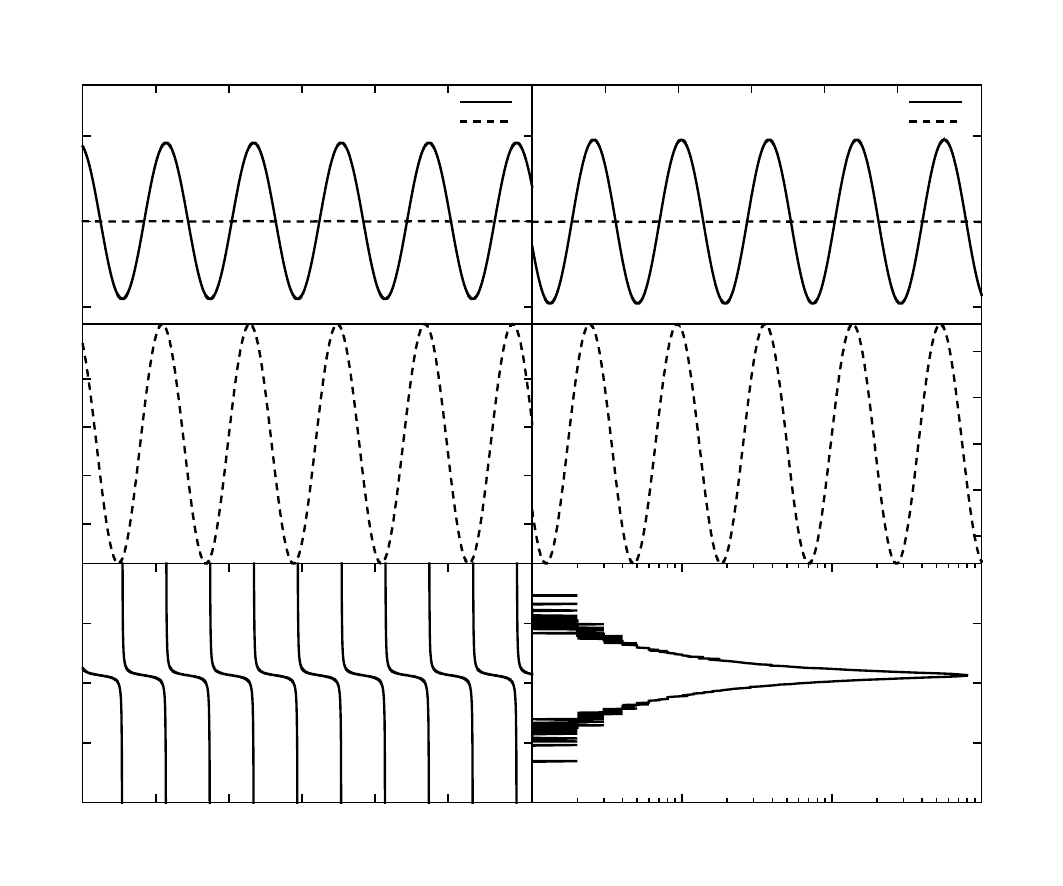}}%
    \gplfronttext
  \end{picture}%
\endgroup

%% file: phi_d_plot.tex
% GNUPLOT: LaTeX picture with Postscript
\begingroup
  \fontfamily{phv}%
  \selectfont
  \makeatletter
  \providecommand\color[2][]{%
    \GenericError{(gnuplot) \space\space\space\@spaces}{%
      Package color not loaded in conjunction with
      terminal option `colourtext'%
    }{See the gnuplot documentation for explanation.%
    }{Either use 'blacktext' in gnuplot or load the package
      color.sty in LaTeX.}%
    \renewcommand\color[2][]{}%
  }%
  \providecommand\includegraphics[2][]{%
    \GenericError{(gnuplot) \space\space\space\@spaces}{%
      Package graphicx or graphics not loaded%
    }{See the gnuplot documentation for explanation.%
    }{The gnuplot epslatex terminal needs graphicx.sty or graphics.sty.}%
    \renewcommand\includegraphics[2][]{}%
  }%
  \providecommand\rotatebox[2]{#2}%
  \@ifundefined{ifGPcolor}{%
    \newif\ifGPcolor
    \GPcolortrue
  }{}%
  \@ifundefined{ifGPblacktext}{%
    \newif\ifGPblacktext
    \GPblacktexttrue
  }{}%
  % define a \g@addto@macro without @ in the name:
  \let\gplgaddtomacro\g@addto@macro
  % define empty templates for all commands taking text:
  \gdef\gplbacktext{}%
  \gdef\gplfronttext{}%
  \makeatother
  \ifGPblacktext
    % no textcolor at all
    \def\colorrgb#1{}%
    \def\colorgray#1{}%
  \else
    % gray or color?
    \ifGPcolor
      \def\colorrgb#1{\color[rgb]{#1}}%
      \def\colorgray#1{\color[gray]{#1}}%
      \expandafter\def\csname LTw\endcsname{\color{white}}%
      \expandafter\def\csname LTb\endcsname{\color{black}}%
      \expandafter\def\csname LTa\endcsname{\color{black}}%
      \expandafter\def\csname LT0\endcsname{\color[rgb]{1,0,0}}%
      \expandafter\def\csname LT1\endcsname{\color[rgb]{0,1,0}}%
      \expandafter\def\csname LT2\endcsname{\color[rgb]{0,0,1}}%
      \expandafter\def\csname LT3\endcsname{\color[rgb]{1,0,1}}%
      \expandafter\def\csname LT4\endcsname{\color[rgb]{0,1,1}}%
      \expandafter\def\csname LT5\endcsname{\color[rgb]{1,1,0}}%
      \expandafter\def\csname LT6\endcsname{\color[rgb]{0,0,0}}%
      \expandafter\def\csname LT7\endcsname{\color[rgb]{1,0.3,0}}%
      \expandafter\def\csname LT8\endcsname{\color[rgb]{0.5,0.5,0.5}}%
    \else
      % gray
      \def\colorrgb#1{\color{black}}%
      \def\colorgray#1{\color[gray]{#1}}%
      \expandafter\def\csname LTw\endcsname{\color{white}}%
      \expandafter\def\csname LTb\endcsname{\color{black}}%
      \expandafter\def\csname LTa\endcsname{\color{black}}%
      \expandafter\def\csname LT0\endcsname{\color{black}}%
      \expandafter\def\csname LT1\endcsname{\color{black}}%
      \expandafter\def\csname LT2\endcsname{\color{black}}%
      \expandafter\def\csname LT3\endcsname{\color{black}}%
      \expandafter\def\csname LT4\endcsname{\color{black}}%
      \expandafter\def\csname LT5\endcsname{\color{black}}%
      \expandafter\def\csname LT6\endcsname{\color{black}}%
      \expandafter\def\csname LT7\endcsname{\color{black}}%
      \expandafter\def\csname LT8\endcsname{\color{black}}%
    \fi
  \fi
  \setlength{\unitlength}{0.0500bp}%
  \begin{picture}(4980.00,3960.00)%
    \gplgaddtomacro\gplbacktext{%
      \csname LTb\endcsname%
      \put(298,715){\rotatebox{-270}{\makebox(0,0){\strut{} 0}}}%
      \csname LTb\endcsname%
      \put(298,1222){\rotatebox{-270}{\makebox(0,0){\strut{} 0.05}}}%
      \csname LTb\endcsname%
      \put(298,1728){\rotatebox{-270}{\makebox(0,0){\strut{} 0.1}}}%
      \csname LTb\endcsname%
      \put(298,2235){\rotatebox{-270}{\makebox(0,0){\strut{} 0.15}}}%
      \csname LTb\endcsname%
      \put(298,2742){\rotatebox{-270}{\makebox(0,0){\strut{} 0.2}}}%
      \csname LTb\endcsname%
      \put(298,3248){\rotatebox{-270}{\makebox(0,0){\strut{} 0.25}}}%
      \csname LTb\endcsname%
      \put(298,3755){\rotatebox{-270}{\makebox(0,0){\strut{} 0.3}}}%
      \csname LTb\endcsname%
      \put(477,613){\rotatebox{-270}{\makebox(0,0)[r]{\strut{}18M}}}%
      \csname LTb\endcsname%
      \put(685,613){\rotatebox{-270}{\makebox(0,0)[r]{\strut{}18S}}}%
      \csname LTb\endcsname%
      \put(894,613){\rotatebox{-270}{\makebox(0,0)[r]{\strut{}19M}}}%
      \csname LTb\endcsname%
      \put(1102,613){\rotatebox{-270}{\makebox(0,0)[r]{\strut{}19S}}}%
      \csname LTb\endcsname%
      \put(1310,613){\rotatebox{-270}{\makebox(0,0)[r]{\strut{}20M}}}%
      \csname LTb\endcsname%
      \put(1518,613){\rotatebox{-270}{\makebox(0,0)[r]{\strut{}20S}}}%
      \csname LTb\endcsname%
      \put(1726,613){\rotatebox{-270}{\makebox(0,0)[r]{\strut{}21M}}}%
      \csname LTb\endcsname%
      \put(1935,613){\rotatebox{-270}{\makebox(0,0)[r]{\strut{}21S}}}%
      \csname LTb\endcsname%
      \put(2143,613){\rotatebox{-270}{\makebox(0,0)[r]{\strut{}22M}}}%
      \csname LTb\endcsname%
      \put(2351,613){\rotatebox{-270}{\makebox(0,0)[r]{\strut{}22S}}}%
      \csname LTb\endcsname%
      \put(2559,613){\rotatebox{-270}{\makebox(0,0)[r]{\strut{}23M}}}%
      \csname LTb\endcsname%
      \put(2768,613){\rotatebox{-270}{\makebox(0,0)[r]{\strut{}23S}}}%
      \csname LTb\endcsname%
      \put(2976,613){\rotatebox{-270}{\makebox(0,0)[r]{\strut{}24M}}}%
      \csname LTb\endcsname%
      \put(3184,613){\rotatebox{-270}{\makebox(0,0)[r]{\strut{}24S}}}%
      \csname LTb\endcsname%
      \put(3392,613){\rotatebox{-270}{\makebox(0,0)[r]{\strut{}25M}}}%
      \csname LTb\endcsname%
      \put(3601,613){\rotatebox{-270}{\makebox(0,0)[r]{\strut{}25S}}}%
      \csname LTb\endcsname%
      \put(3809,613){\rotatebox{-270}{\makebox(0,0)[r]{\strut{}26M}}}%
      \csname LTb\endcsname%
      \put(4017,613){\rotatebox{-270}{\makebox(0,0)[r]{\strut{}26S}}}%
      \csname LTb\endcsname%
      \put(4225,613){\rotatebox{-270}{\makebox(0,0)[r]{\strut{}27M}}}%
      \csname LTb\endcsname%
      \put(4433,613){\rotatebox{-270}{\makebox(0,0)[r]{\strut{}27S}}}%
      \csname LTb\endcsname%
      \put(4642,613){\rotatebox{-270}{\makebox(0,0)[r]{\strut{}28M}}}%
      \csname LTb\endcsname%
      \put(4850,613){\rotatebox{-270}{\makebox(0,0)[r]{\strut{}28S}}}%
      \csname LTb\endcsname%
      \put(28,2235){\rotatebox{-270}{\makebox(0,0){\strut{}$\phi_D$}}}%
      \csname LTb\endcsname%
      \put(2663,130){\makebox(0,0){\strut{}LFI radiometer}}%
    }%
    \gplgaddtomacro\gplfronttext{%
      \csname LTb\endcsname%
      \put(1801,3588){\makebox(0,0)[r]{\strut{}First survey}}%
      \csname LTb\endcsname%
      \put(1801,3402){\makebox(0,0)[r]{\strut{}Second survey}}%
    }%
    \gplbacktext
    \put(0,0){\includegraphics{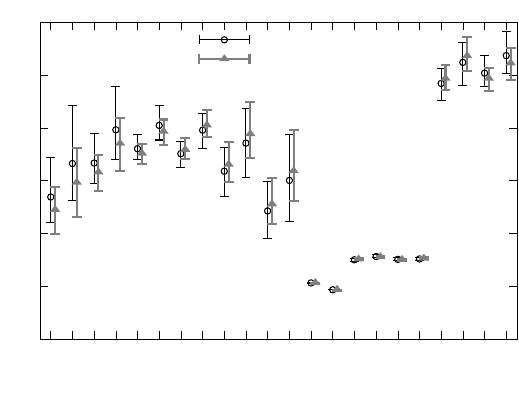}}%
    \gplfronttext
  \end{picture}%
\endgroup

%% file: P02b_2_3_colour_corrections.tex
\begin{table*}[tmb]
\begingroup
\newdimen\tblskip \tblskip=5pt
\caption{Multiplicative colour corrections $\cc(\alpha)$ for individual LFI Radiometer Chain Assemblies and for the band average maps.}
\label{tbl:colourCorrections}
\nointerlineskip
\vskip -3mm
\footnotesize
\setbox\tablebox=\vbox{
   \newdimen\digitwidth 
   \setbox0=\hbox{\rm 0} 
   \digitwidth=\wd0 
   \catcode`*=\active 
   \def*{\kern\digitwidth}
   \newdimen\signwidth 
   \setbox0=\hbox{+} 
   \signwidth=\wd0 
   \catcode`!=\active 
   \def!{\kern\signwidth}
\halign{
% Template
\hbox to 2.5cm{#\leaderfil}\tabskip 0.5em & % Detector
\hfil# &   % alpha = -2.0
\hfil# &   % alpha = -1.5beamShape
\hfil# &   % alpha = -1.0
\hfil# &   % alpha = -0.5
\hfil# &   % alpha =  0.0
\hfil# &   % alpha =  0.5
\hfil# &   % alpha =  1.0
\hfil# &   % alpha =  1.5
\hfil# &   % alpha =  2.0
\hfil# &   % alpha =  2.5
\hfil# &   % alpha =  3.0
\hfil# &   % alpha =  3.5
\hfil#\tabskip=0pt\cr % alpha =  4.0
\noalign{\doubleline}
\omit\hfil RCA \hfil &
\multispan{13}\hfil Spectral index $\alpha$ \hfil \cr
\omit                   & % Empty
\omit\hfil $-2.0$ \hfil &
\omit\hfil $-1.5$ \hfil &
\omit\hfil $-1.0$ \hfil &
\omit\hfil $-0.5$ \hfil &
\omit\hfil $ 0.0$ \hfil &
\omit\hfil $ 0.5$ \hfil &
\omit\hfil $ 1.0$ \hfil &
\omit\hfil $ 1.5$ \hfil &
\omit\hfil $ 2.0$ \hfil &
\omit\hfil $ 2.5$ \hfil &
\omit\hfil $ 3.0$ \hfil &
\omit\hfil $ 3.5$ \hfil &
\omit\hfil $ 4.0$ \hfil \cr
\noalign{\vskip 3pt\hrule\vskip 5pt}

LFI-18& $0.948$& $0.961$& $0.972$& $0.981$& $0.988$& $0.994$& $0.997$& $0.998$& $0.997$& $0.995$& $0.990$& $0.983$& $0.975$ \cr
LFI-19& $0.856$& $0.878$& $0.899$& $0.919$& $0.939$& $0.957$& $0.975$& $0.991$& $1.006$& $1.020$& $1.032$& $1.043$& $1.053$ \cr
LFI-20& $0.889$& $0.908$& $0.925$& $0.941$& $0.956$& $0.970$& $0.983$& $0.994$& $1.003$& $1.011$& $1.018$& $1.023$& $1.027$ \cr
LFI-21& $0.917$& $0.933$& $0.947$& $0.960$& $0.971$& $0.981$& $0.989$& $0.996$& $1.001$& $1.004$& $1.006$& $1.006$& $1.004$ \cr
LFI-22& $1.024$& $1.026$& $1.027$& $1.026$& $1.023$& $1.018$& $1.011$& $1.003$& $0.993$& $0.982$& $0.969$& $0.955$& $0.940$ \cr
LFI-23& $0.985$& $0.991$& $0.996$& $0.999$& $1.001$& $1.002$& $1.002$& $1.000$& $0.997$& $0.993$& $0.988$& $0.982$& $0.975$ \cr
\noalign{\vskip 2pt}
{\bf 70 GHz}& $0.938$& $0.951$& $0.963$& $0.973$& $0.982$& $0.988$& $0.994$& $0.997$& $0.999$& $0.999$& $0.998$& $0.995$& $0.991$ \cr
\noalign{\vskip 6pt}
LFI-24& $0.978$& $0.984$& $0.988$& $0.993$& $0.996$& $0.998$& $0.999$& $1.000$& $0.999$& $0.998$& $0.996$& $0.993$& $0.989$ \cr
LFI-25& $0.967$& $0.974$& $0.980$& $0.985$& $0.990$& $0.994$& $0.996$& $0.999$& $1.000$& $1.000$& $1.000$& $0.999$& $0.997$ \cr
LFI-26& $0.957$& $0.966$& $0.973$& $0.980$& $0.985$& $0.990$& $0.995$& $0.998$& $1.000$& $1.001$& $1.002$& $1.002$& $1.000$ \cr
\noalign{\vskip 2pt}
{\bf 44 GHz}& $0.968$& $0.975$& $0.981$& $0.986$& $0.990$& $0.994$& $0.997$& $0.999$& $1.000$& $1.000$& $0.999$& $0.998$& $0.995$ \cr
\noalign{\vskip 6pt}
LFI-27& $0.948$& $0.959$& $0.969$& $0.978$& $0.985$& $0.991$& $0.995$& $0.998$& $1.000$& $1.000$& $0.998$& $0.995$& $0.991$ \cr
LFI-28& $0.946$& $0.958$& $0.968$& $0.977$& $0.985$& $0.991$& $0.996$& $0.998$& $1.000$& $0.999$& $0.997$& $0.993$& $0.988$ \cr
\noalign{\vskip 2pt}
{\bf 30 GHz}& $0.947$& $0.959$& $0.969$& $0.977$& $0.985$& $0.991$& $0.995$& $0.998$& $1.000$& $1.000$& $0.998$& $0.994$& $0.989$ \cr 
\noalign{\vskip 5pt\hrule\vskip 3pt}}}
\endPlancktablewide
%\tablenote a Footnote a.\par
%\tablenote b Footnote b.\par
\endgroup
\end{table*}

The raw differential signal $V = V_\mathrm{sky} - r V_\mathrm{ref}$ measured by a \Planck\ radiometer can be written as:
\begin{equation}
\label{eq:raw_voltage}
\begin{split}
V &= G \int g(\nu) T_{\rm RJ}(\nu)\, \ud\nu = \frac{G}{2k_B} \int g(\nu) I(\nu) \lambda^2 \,\ud\nu \\
&= G' \int \tau(\nu) I(\nu) \,\ud\nu,
\end{split}
\end{equation}
where $G$ is the overall gain, $g(\nu)$ is the bandpass, $T_{\rm RJ}(\nu)$ is the differential Rayleigh-Jeans brightness temperature averaged over the beam, and $G'$ is defined such that $2 k_B G' \tau(\nu) = G g(\nu) \lambda^2$. The constant $G'$ would be the gain if bandpasses were defined via the transmission coefficient $\tau(\nu) \propto g(\nu)\lambda^2$ instead of $g(\nu)$, as HFI does \citep{planck2013-p03f}.

Since we calibrate using the CMB dipole, the calibration signal measured
in any given pointing period is:
\begin{equation}
\label{eq:colourCorrectionDoverK}
\frac{D}{K} = G \int g(\nu) D\, \eta_{\Delta T}(\nu) \,\ud\nu,
\end{equation}
where D is the dipole amplitude in thermodynamic units and $\eta_{\Delta T}(\nu)$
is the conversion factor from CMB temperature to RJ temperature. Hence the
calibrated sky map temperature is:
\begin{equation}
\tilde{T} = K V = 
\frac{\int g(\nu) T_{\rm RJ}(\nu)\, \ud\nu}{\int g(\nu) \eta_{\Delta T}(\nu)\, \ud\nu}.
\label{eq:T_calibrated}
\end{equation}
If we are observing pure CMB fluctuations, then $T_{\rm RJ}(\nu) = \Delta T\, \eta_{\Delta T}(\nu)$, and hence we have $\tilde{T} = \Delta T$ as expected.
If we are observing foreground emission with intensity power-law index $\alpha$,
then $T_{\rm RJ}(\nu) = T_0 \,(\nu/\nu_0)^{\alpha-2}$, and 
\begin{equation}
\tilde{T} = T_0 
\frac{\int g(\nu) (\nu/\nu_0)^{\alpha-2} \, \ud\nu}{\int g(\nu) \eta_{\Delta T}(\nu)\, \ud\nu}.
\end{equation}
The LFI colour correction is defined as:
\begin{equation}
\cc(\alpha) = 
\frac{\int g(\nu) \eta_{\Delta T}(\nu)\, \ud\nu}{\eta_{\Delta T}(\nu_0) \int g(\nu) (\nu/\nu_0)^{\alpha-2} \, \ud\nu}.
\label{eq:cc}
\end{equation}
Hence the colour-corrected temperature is:
\begin{equation}
\cc(\alpha)\tilde{T} = T_0 / \eta_{\Delta T}(\nu_0).
\label{eq:correctedKCMB}
\end{equation}
This gives the {\em thermodynamic} 
brightness differential temperature (units ``K$_{\rm CMB}$'')
at the reference frequency $\nu_0$. To get to the standard (Rayleigh-Jeans)
brightness temperature at the same frequency we use $\eta_{\Delta T}(\nu_0)$ (see Eq.~\ref{eq:colourCorrectionDoverK}):
\begin{equation}
T_0 [{\rm K_{\rm RJ}}] = \tilde{T} [{\rm K_{\rm CMB}}] \, \eta_{\Delta T}(\nu_0) \, \cc(\alpha).
\end{equation}
For the present series of \Planck\ papers, the reference frequencies $\nu_0$
are defined as precisely 28.4, 44.1, and 70.4\,GHz for the three LFI bands;
in the \Planck{} Early Release papers, slightly different values were used.

Values for $\cc(\alpha)$ for the expected range of foreground spectral indices are listed in Table~\ref{tbl:colourCorrections}. Note that this definition of $\cc$ is inverted relative to the colour correction $\cc_\text{ER}(\alpha)$ used in the \Planck{} Early Release. We note additionally that $\cc(\alpha)$ quoted here are not exactly equal to $1/\cc_\text{ER}(\alpha)$ from \citet{planck2011-1.4} and \citet{planck2011-1.6}, due both to the change in reference frequency and also to a small error in the earlier estimates.

Colour corrections at intermediate spectral indices may be derived accurately
from a quadratic fit to the values in Table~\ref{tbl:colourCorrections}. In addition, the data release includes the Unit conversion and Colour Correction ({\tt UcCC}) IDL package \citep{planck2013-p03d} which calculates colour corrections and unit conversions using the band-averaged bandpass information stored in the Reduced Instrument MOdel \citep[RIMO, see][]{planck2013-p28} file also included in the data release.

From Eq.~\ref{eq:cc} it appears that the absolute scaling of the bandpass
is irrelevant for the colour correction, although 
Eq.~\ref{eq:raw_voltage}--\ref{eq:T_calibrated} require that:
\begin{equation}
\int g(\nu)\,\eta_{\Delta T}(\nu)\,\ud\nu = 1,
\label{eq:bp_norm}
\end{equation}
for consistency with the definition of gain, $G$, and calibration
factor, $K$, elsewhere in this paper. In practice, all the colour corrections
listed in Table~\ref{tbl:colourCorrections} are derived from averages across
two or more bandpasses: values for individual Radiometric Chain Assemblies (RCA) require averaging the
main and side radiometers in each RCA, and the 
response of each radiometer is the average of the two independent detectors 
\citep{zonca2009}. For consistency, the weighting for these averages must duplicate the procedure used to average the data, as described in \citet{planck2013-p02}. We recall here the procedure: (i) calibrate the individual data streams to give equal response to the CMB; (ii) combine the raw data for the detectors for each radiometer using fixed weights based on the  inverse variance measured early in the mission, incorporating fixed calibration values; (iii) recalibrate the data using
time-dependent factors as described in the present paper; (iv) combine the
two radiometers in each RCA with equal weights to minimize polarization
leakage; and  (v) combine the data from individual RCAs with inverse-variance
weights (fixed for the whole mission). When combining the bandpasses,
the initial calibration step is equivalent to normalizing each component 
bandpass according to Eq.~\ref{eq:bp_norm}. The band-averaged bandpass
stored in the RIMO, however, normalized the bandpasses using 
$\int g(\nu)\,\ud\nu = 1$. The difference is minor, since 
$\eta_{\Delta T}$ varies by only a very small amount within any of the 
LFI bands, but it accounts for a small
($<0.1\,\%$) difference between the $\cc(\alpha)$
values listed here and those derivable from the RIMO bandpasses.

Our best estimate of the uncertainty in the values of $\cc(\alpha)$, dominated by bandpass uncertainty \citep{zonca2009}, comes from an indirect method, as follows.
The two radiometers in each RCA, known as
the main- and side-arms, are sensitive to orthogonal polarizations. 
The bandpasses for the two arms differ, leading
to different colour corrections. The polarization signal is derived from
the difference of the calibrated signals from the two arms \citep{leahy2010};
unpolarized foreground emission does not precisely cancel due to the diffential
colour factors. This is the ``bandpass leakage'' effect. This leakage can be
estimated from the flight data, as described by \citet{planck2013-p02}, 
but it can also be estimated from the pre-launch bandpass models that we use
here to calculate $\cc(\alpha)$. 
To a good approximation, we can write the leakage factor as 
\begin{equation}
\left(\cc_\mathrm{S}(\alpha) - \cc_\mathrm{M}(\alpha)\right)/2 = 
(\beta-\beta_\mathrm{CMB})\,a,
\end{equation}
where the $a$-factors depend solely on the
bandpass profile, $\beta = \alpha-2$ is the temperature spectral index,
and $\beta_\mathrm{CMB}$ is the in-band spectral index of the CMB, which we
can take as zero for present purposes.  
The flight measurements of the $a$-factors are demonstrably 
more accurate than the pre-launch (``QUCS'') estimates, and so 
\begin{equation}
(\beta- \beta_\mathrm{CMB})(a_\mathrm{QUCS} - a_\mathrm{flight}) 
\approx (\delta \cc_\mathrm{S}(\alpha) - \delta \cc_\mathrm{M}(\alpha))/2,
\end{equation}
where $\delta \cc$ is the colour correction error. Statistically
\begin{equation}
\left<\left(\delta \cc_\mathrm{S}(\alpha) - 
\delta \cc_\mathrm{M}(\alpha)\right)^2\right> =
\left<\left(\delta \cc_\mathrm{S}(\alpha) +
\delta \cc_\mathrm{M}(\alpha)\right)^2\right> 
= \sigma_\cc^2,
\end{equation}
where $\sigma_\cc$ is the colour correction error for one radiometer 
pair, i.e. one RCA. Hence for individual RCAs, $\sigma_\cc \approx |\beta|\,\sigma_a$. There are too few pairs for this approach to give an accurate value for $\sigma_\cc$, especially as the error sources for the bandpasses in each 
band are different; therefore results from the three bands cannot be combined. Our rough estimates for $\sigma_a$ are 0.14\,\%, 0.46\,\%, and 0.51\,\% in the 30, 44, and 70\,GHz bands, respectively. For the band-averaged maps, the errors in the colour corrections are reduced by $\sqrt{N_\mathrm{RCA}}$, giving overall rough uncertainties of $0.1|\beta|$,\,\%, $0.3|\beta|$\,\%, and $0.2|\beta|$\,\%, respectively. Given typical values of $\beta$ of around $-2$ to $-3$, this gives uncertainties in the colour corrections of a few tenths of a percent of the tabulated value at 30 GHz, and 0.5--1\,\% at 44 and 70 GHz.

Progress towards in-flight calibration of the colour corrections is discussed further in Sect.~\ref{sec:sourceFluxes}.

%% file: P02b_3_0_introduction.tex
\begin{figure*}[tbf]
\centering
\begin{DIFnomarkup}
\input{summary_gain_plot.tex}
\end{DIFnomarkup}
\caption{\label{fig:gainCurve} Variation in time of a few quantities relevant for calibration, for radiometer LFI21M (70\,GHz, left) and LFI27M (30\,GHz, right). Vertical dashed lines mark boundaries between complete sky surveys. All temperatures are thermodynamic. \textit{Panel A:} Calibration constant $K$ estimated using the expected amplitude of the CMB dipole. Note that the uncertainty associated with the estimate changes with time, according to the amplitude of the dipole as seen in each ring. \textit{Panel B:} Expected peak-to-peak difference of the dipole signal. The shape of the curve depends on the scanning strategy of \Planck{} (see also Fig.~\protect\ref{fig:dipoleRingsInTheSky}), and it is strongly correlated with the uncertainty in the gain constant (see panel A). The two dots indicate at which time the data used to produce the plots in Fig.~\protect\ref{fig:dipoleFits} were taken. \textit{Panel C:} 4\,K total-power output voltages $V$. Note that the noise level for $V$ is much more stable than it is for $K$, as it does not depend on the dipole amplitude. (All the large jumps seen in this figure occurred because of some change in the operational state of the spacecraft; e.g., 456 days after launch we switched off the primary 20\,K sorption cooler and turned on the redundant unit.) \textit{Panel D:} The calibration constants $K$ used to actually calibrate the data for this \Planck{} data release are either derived by applying a smoothing filter to the raw gains in panel A (70 and 44\,GHz, e.g., LFI21M, left), as described in Sect.~\protect\ref{sec:OSGCalibration}, or through Eq.~\protect\ref{eq:gainAsFunctionOfV} and the 4\,K total-power voltages in panel C (30\,GHz, e.g., LFI27M, right)}
\end{figure*}
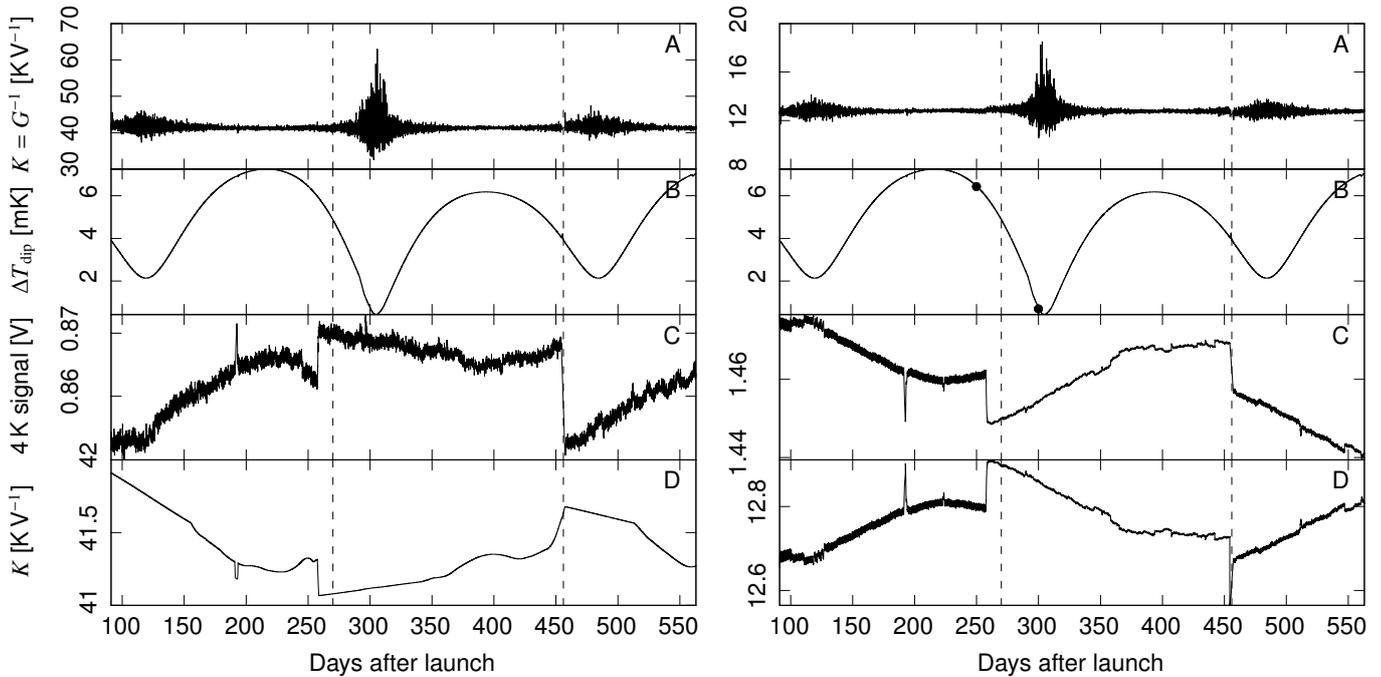

There is not a unique way to produce an estimate of the calibration constant $K = G^{-1}$ (see Eq.~\ref{eq:calibrationEquation}). In picking the method to use, one has to consider a number of elements:
\begin{enumerate}
\item the calibration should be as accurate as possible;
\item calibrated data should also be self-consistent; for example, with the same beam and the same pointing direction and orientation in the sky, the value of $T_\mathrm{sky}$ at any frequency should not depend on when the measurement was done, nor on the detector (under the hypothesis that everything else, e.g., the bandpass, can be assumed to be the same);
\item the estimated shape of $K(t)$ should be motivated by a physically meaningful model of the radiometer.
\end{enumerate}
To help in the classification of the calibration methods discussed in this work, we write:
\begin{equation}
\label{eq:KKzeroAndXi}
K (t) = K_0 \bigl( 1 + \xi(t) \bigr),
\end{equation}
thus decomposing $K$ into a constant term $K_0$ and a unitless time-varying quantity $\xi(t)$. Typically $\xi(t)$ varies by a few percent per year. This allows us to categorize calibration methods into the following families: (1) \emph{absolute methods} produce an estimate for $K_0$; (2) \emph{relative methods} estimate how $\xi$ changes with time; and (3) some methods are able to estimate $K(t)$ directly and can therefore be considered both absolute and relative.

\begin{figure}
	\includegraphics{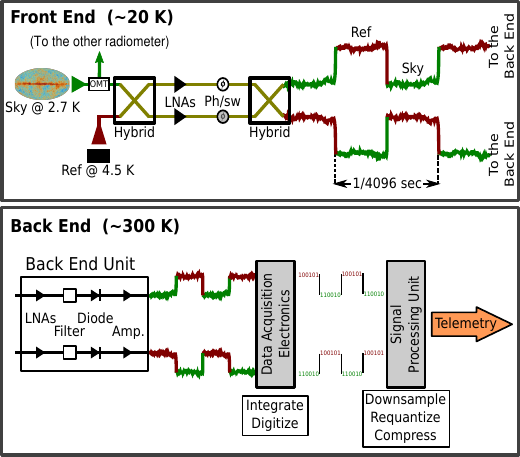}
	\caption{\label{fig:LFIradiometer} Schematics of a LFI radiometer. The radiation entering the feed horns in the 20\,K front end is split by an OrthoMode transducer (OMT) into its two linearly polarized components. The two signals feed two twin radiometers. (Only one of them is shown in this figure.) The 2.7\,K signal is mixed with the signal emitted by a reference blackbody at 4.5\,K by an hybrid coupler before being amplified by a Low Noise Amplifier (LNA). A 4096\,Hz phase switch induces an alternating 0$^\circ$/180$^\circ$ phase shift, so that the outputs of the second hybrid are two sequences of sky/reference signals, which both propagate to the warm back-end. Here they are further amplified and measured, before being compressed into packets and sent to Earth. Refer to \citet{mennella2010} for more details.}
\end{figure}

We implemented two calibration schemes for the LFI radiometers:
\begin{itemize}
    \item The \emph{OSG dipole calibration} is an improved version of the pipeline used to calibrate the data used in the \Planck{} Early and Intermediate data releases \citep{planck2011-1.4}. It is based on the OSG algorithm (``Optimal Search for Gains'') and relies on the signal of the solar and orbital dipoles as observed by the spacecraft. This method is only weakly affected by non-idealities in the radiometers (e.g., non-linearities in the ADCs, see Sect.~\ref{sec:ADCNonIdealities}), but optical effects (e.g., sidelobes, see Sect.~\ref{sec:optics}) can induce systematic errors in the reconstruction of the gain.
    \item In the \emph{4\,K calibration} we use the dipole only to fix the absolute level of calibration ($K_0$ in Eq.~\ref{eq:KKzeroAndXi}), but we estimate gain changes (the $\xi$ term) using the 4\,K total-power output of the radiometers. (See Fig.~\ref{fig:LFIradiometer} for a schematic of a LFI radiometer.) Unlike the OSG dipole calibration, this scheme is sensitive to ADC non-linearities, but it provides an estimate for the gain, $K$, which is independent\footnote{Note however that, even if the calibration constants $K_i$ are not affected by uncertainties in beam shapes, this is not necessarily true for the calibrated data. After having applied the calibration factors, we must remove the dipole signal convolved with the beam from the data, and therefore any uncertainty in the beam shape will lead to a systematic error in the calibrated data, even if the 4\,K calibration is used.} of optical effects, assuming that the optical properties remain constant. (As Fig.~4 in \citet{planck2013-p02d} shows, measurements of the main beam characteristics from planet transits during the whole \Planck{} nominal mission show no trace of systematic variations with time.)
\end{itemize}
The chief reason why we decided to use two different calibration schemes was to improve the self-consistency of the maps. We were not able to derive a calibration method that was robust against both radiometric non-idealities and optical effects in time for this data release. Since optical effects are most significant in the 30\,GHz channels, we decided to apply the 4\,K calibration to these radiometers instead of the other one (which was the baseline for the Early and Intermediate \Planck/LFI papers). We provide a summary of the calibration methods used for each LFI frequency in Table~\ref{tbl:calibrationMethodsOverview}. In Sect.~\ref{sec:optics} we will discuss the level of consistency between the two calibration methods in the context of the treatment of optical systematics.

The outline of this section is as follows. The OSG dipole calibration and the 4\,K calibration are explained in Sect.~\ref{sec:OSGCalibration} and \ref{sec:dVV}. In Sect.~\ref{sec:orbitalDipole} we present the current status of our efforts to use the orbital dipole signal for calibrating data. Finally, Sect.~\ref{sec:zeroLevel} deals with the task of setting the zero-level of \Planck's LFI maps.

\begin{table*}[tmb]
\begingroup
\newdimen\tblskip \tblskip=5pt
\caption{\label{tbl:calibrationMethodsOverview} Methods used to calibrate the radiometers.}
\nointerlineskip
\vskip -3mm
\footnotesize
\setbox\tablebox=\vbox{
   \newdimen\digitwidth 
   \setbox0=\hbox{\rm 0} 
   \digitwidth=\wd0 
   \catcode`*=\active 
   \def*{\kern\digitwidth}
   \newdimen\signwidth 
   \setbox0=\hbox{+} 
   \signwidth=\wd0 
   \catcode`!=\active 
   \def!{\kern\signwidth}
{
\halign{
\hfil #\hfil\tabskip=1em&
\hfil #\hfil &
\hfil #\hfil &
\vtop{\hsize 4.0cm\noindent\hangafter=1\hangindent=1em\strut #\strut\par} &
# \hfill\tabskip=0em\cr
\noalign{\doubleline}
\omit\hfil Frequency\hfil&
\multispan2\hfil Calibration\hfil&
\omit\hfil Primary systematics\hfil&
\omit\hfil Relevant paper sections\hfil\cr
&
\omit\hfil Absolute\hfil&
\omit\hfil Relative\hfil&
&
\cr
\noalign{\vskip 3pt\hrule\vskip 5pt}
30\,GHz& CMB dipole& 4\,K & ADC non-linearities, variations of $T_\mathrm{noise}$ with time & Sect.~\protect\ref{sec:OSGCalibration}, \protect\ref{sec:dVV}, \protect\ref{sec:ADCNonIdealities}\cr
44\,GHz& CMB dipole& CMB dipole& Optical systematics, bandpass response of the radiometers& Sect.~\protect\ref{sec:OSGCalibration}, \protect\ref{sec:optics}\cr
70\,GHz& CMB dipole& CMB dipole& Same as for 44\,GHz & Sect.~\protect\ref{sec:OSGCalibration}, \protect\ref{sec:optics}\cr
\noalign{\vskip 5pt\hrule\vskip 3pt}}}}
\endPlancktablewide                 % ends two-column \halign
\endgroup
\end{table*}

%% file: summary_gain_plot.tex
% GNUPLOT: LaTeX picture with Postscript
\begingroup
  \fontfamily{phv}%
  \selectfont
  \makeatletter
  \providecommand\color[2][]{%
    \GenericError{(gnuplot) \space\space\space\@spaces}{%
      Package color not loaded in conjunction with
      terminal option `colourtext'%
    }{See the gnuplot documentation for explanation.%
    }{Either use 'blacktext' in gnuplot or load the package
      color.sty in LaTeX.}%
    \renewcommand\color[2][]{}%
  }%
  \providecommand\includegraphics[2][]{%
    \GenericError{(gnuplot) \space\space\space\@spaces}{%
      Package graphicx or graphics not loaded%
    }{See the gnuplot documentation for explanation.%
    }{The gnuplot epslatex terminal needs graphicx.sty or graphics.sty.}%
    \renewcommand\includegraphics[2][]{}%
  }%
  \providecommand\rotatebox[2]{#2}%
  \@ifundefined{ifGPcolor}{%
    \newif\ifGPcolor
    \GPcolorfalse
  }{}%
  \@ifundefined{ifGPblacktext}{%
    \newif\ifGPblacktext
    \GPblacktexttrue
  }{}%
  % define a \g@addto@macro without @ in the name:
  \let\gplgaddtomacro\g@addto@macro
  % define empty templates for all commands taking text:
  \gdef\gplbacktext{}%
  \gdef\gplfronttext{}%
  \makeatother
  \ifGPblacktext
    % no textcolor at all
    \def\colorrgb#1{}%
    \def\colorgray#1{}%
  \else
    % gray or color?
    \ifGPcolor
      \def\colorrgb#1{\color[rgb]{#1}}%
      \def\colorgray#1{\color[gray]{#1}}%
      \expandafter\def\csname LTw\endcsname{\color{white}}%
      \expandafter\def\csname LTb\endcsname{\color{black}}%
      \expandafter\def\csname LTa\endcsname{\color{black}}%
      \expandafter\def\csname LT0\endcsname{\color[rgb]{1,0,0}}%
      \expandafter\def\csname LT1\endcsname{\color[rgb]{0,1,0}}%
      \expandafter\def\csname LT2\endcsname{\color[rgb]{0,0,1}}%
      \expandafter\def\csname LT3\endcsname{\color[rgb]{1,0,1}}%
      \expandafter\def\csname LT4\endcsname{\color[rgb]{0,1,1}}%
      \expandafter\def\csname LT5\endcsname{\color[rgb]{1,1,0}}%
      \expandafter\def\csname LT6\endcsname{\color[rgb]{0,0,0}}%
      \expandafter\def\csname LT7\endcsname{\color[rgb]{1,0.3,0}}%
      \expandafter\def\csname LT8\endcsname{\color[rgb]{0.5,0.5,0.5}}%
    \else
      % gray
      \def\colorrgb#1{\color{black}}%
      \def\colorgray#1{\color[gray]{#1}}%
      \expandafter\def\csname LTw\endcsname{\color{white}}%
      \expandafter\def\csname LTb\endcsname{\color{black}}%
      \expandafter\def\csname LTa\endcsname{\color{black}}%
      \expandafter\def\csname LT0\endcsname{\color{black}}%
      \expandafter\def\csname LT1\endcsname{\color{black}}%
      \expandafter\def\csname LT2\endcsname{\color{black}}%
      \expandafter\def\csname LT3\endcsname{\color{black}}%
      \expandafter\def\csname LT4\endcsname{\color{black}}%
      \expandafter\def\csname LT5\endcsname{\color{black}}%
      \expandafter\def\csname LT6\endcsname{\color{black}}%
      \expandafter\def\csname LT7\endcsname{\color{black}}%
      \expandafter\def\csname LT8\endcsname{\color{black}}%
    \fi
  \fi
  \setlength{\unitlength}{0.0500bp}%
  \begin{picture}(10200.00,4920.00)%
    \gplgaddtomacro\gplbacktext{%
      \put(514,3826){\rotatebox{-270}{\makebox(0,0){\strut{} 30}}}%
      \put(514,4099){\rotatebox{-270}{\makebox(0,0){\strut{} 40}}}%
      \put(514,4373){\rotatebox{-270}{\makebox(0,0){\strut{} 50}}}%
      \put(514,4646){\rotatebox{-270}{\makebox(0,0){\strut{} 60}}}%
      \put(514,4919){\rotatebox{-270}{\makebox(0,0){\strut{} 70}}}%
      \put(876,3640){\makebox(0,0){\strut{}}}%
      \put(1338,3640){\makebox(0,0){\strut{}}}%
      \put(1801,3640){\makebox(0,0){\strut{}}}%
      \put(2263,3640){\makebox(0,0){\strut{}}}%
      \put(2725,3640){\makebox(0,0){\strut{}}}%
      \put(3187,3640){\makebox(0,0){\strut{}}}%
      \put(3649,3640){\makebox(0,0){\strut{}}}%
      \put(4111,3640){\makebox(0,0){\strut{}}}%
      \put(4574,3640){\makebox(0,0){\strut{}}}%
      \put(5036,3640){\makebox(0,0){\strut{}}}%
      \csname LTb\endcsname%
      \put(142,4372){\rotatebox{-270}{\makebox(0,0){\strut{}$K = G^{-1}$ [K\,V$^{-1}$]}}}%
      \put(5054,4733){\makebox(0,0)[r]{\strut{}A}}%
    }%
    \gplgaddtomacro\gplfronttext{%
    }%
    \gplgaddtomacro\gplbacktext{%
      \put(656,2984){\rotatebox{-270}{\makebox(0,0){\strut{} 2}}}%
      \put(656,3304){\rotatebox{-270}{\makebox(0,0){\strut{} 4}}}%
      \put(656,3625){\rotatebox{-270}{\makebox(0,0){\strut{} 6}}}%
      \put(876,2547){\makebox(0,0){\strut{}}}%
      \put(1338,2547){\makebox(0,0){\strut{}}}%
      \put(1801,2547){\makebox(0,0){\strut{}}}%
      \put(2263,2547){\makebox(0,0){\strut{}}}%
      \put(2725,2547){\makebox(0,0){\strut{}}}%
      \put(3187,2547){\makebox(0,0){\strut{}}}%
      \put(3649,2547){\makebox(0,0){\strut{}}}%
      \put(4111,2547){\makebox(0,0){\strut{}}}%
      \put(4574,2547){\makebox(0,0){\strut{}}}%
      \put(5036,2547){\makebox(0,0){\strut{}}}%
      \put(142,3279){\rotatebox{-270}{\makebox(0,0){\strut{}$\Delta T_\mathrm{dip}$ [mK]}}}%
      \put(5054,3639){\makebox(0,0)[r]{\strut{}B}}%
    }%
    \gplgaddtomacro\gplfronttext{%
    }%
    \gplgaddtomacro\gplbacktext{%
      \put(514,2118){\rotatebox{-270}{\makebox(0,0){\strut{} 0.86}}}%
      \put(514,2592){\rotatebox{-270}{\makebox(0,0){\strut{} 0.87}}}%
      \put(876,1454){\makebox(0,0){\strut{}}}%
      \put(1338,1454){\makebox(0,0){\strut{}}}%
      \put(1801,1454){\makebox(0,0){\strut{}}}%
      \put(2263,1454){\makebox(0,0){\strut{}}}%
      \put(2725,1454){\makebox(0,0){\strut{}}}%
      \put(3187,1454){\makebox(0,0){\strut{}}}%
      \put(3649,1454){\makebox(0,0){\strut{}}}%
      \put(4111,1454){\makebox(0,0){\strut{}}}%
      \put(4574,1454){\makebox(0,0){\strut{}}}%
      \put(5036,1454){\makebox(0,0){\strut{}}}%
      \put(142,2186){\rotatebox{-270}{\makebox(0,0){\strut{}4\,K signal [V]}}}%
      \put(5054,2546){\makebox(0,0)[r]{\strut{}C}}%
    }%
    \gplgaddtomacro\gplfronttext{%
    }%
    \gplgaddtomacro\gplbacktext{%
      \put(656,546){\rotatebox{-270}{\makebox(0,0){\strut{} 41}}}%
      \put(656,1093){\rotatebox{-270}{\makebox(0,0){\strut{} 41.5}}}%
      \put(656,1639){\rotatebox{-270}{\makebox(0,0){\strut{} 42}}}%
      \put(876,360){\makebox(0,0){\strut{}100}}%
      \put(1338,360){\makebox(0,0){\strut{}150}}%
      \put(1801,360){\makebox(0,0){\strut{}200}}%
      \put(2263,360){\makebox(0,0){\strut{}250}}%
      \put(2725,360){\makebox(0,0){\strut{}300}}%
      \put(3187,360){\makebox(0,0){\strut{}350}}%
      \put(3649,360){\makebox(0,0){\strut{}400}}%
      \put(4111,360){\makebox(0,0){\strut{}450}}%
      \put(4574,360){\makebox(0,0){\strut{}500}}%
      \put(5036,360){\makebox(0,0){\strut{}550}}%
      \put(142,1092){\rotatebox{-270}{\makebox(0,0){\strut{}$K$ [K\,V$^{-1}$]}}}%
      \put(2974,81){\makebox(0,0){\strut{}Days after launch}}%
      \put(5054,1453){\makebox(0,0)[r]{\strut{}D}}%
    }%
    \gplgaddtomacro\gplfronttext{%
    }%
    \gplgaddtomacro\gplbacktext{%
      \put(5501,3826){\rotatebox{-270}{\makebox(0,0){\strut{} 8}}}%
      \put(5501,4190){\rotatebox{-270}{\makebox(0,0){\strut{} 12}}}%
      \put(5501,4555){\rotatebox{-270}{\makebox(0,0){\strut{} 16}}}%
      \put(5501,4919){\rotatebox{-270}{\makebox(0,0){\strut{} 20}}}%
      \put(5863,3640){\makebox(0,0){\strut{}}}%
      \put(6325,3640){\makebox(0,0){\strut{}}}%
      \put(6787,3640){\makebox(0,0){\strut{}}}%
      \put(7249,3640){\makebox(0,0){\strut{}}}%
      \put(7711,3640){\makebox(0,0){\strut{}}}%
      \put(8174,3640){\makebox(0,0){\strut{}}}%
      \put(8636,3640){\makebox(0,0){\strut{}}}%
      \put(9098,3640){\makebox(0,0){\strut{}}}%
      \put(9560,3640){\makebox(0,0){\strut{}}}%
      \put(10022,3640){\makebox(0,0){\strut{}}}%
      \put(10040,4733){\makebox(0,0)[r]{\strut{}A}}%
    }%
    \gplgaddtomacro\gplfronttext{%
    }%
    \gplgaddtomacro\gplbacktext{%
      \put(5643,2984){\rotatebox{-270}{\makebox(0,0){\strut{} 2}}}%
      \put(5643,3304){\rotatebox{-270}{\makebox(0,0){\strut{} 4}}}%
      \put(5643,3625){\rotatebox{-270}{\makebox(0,0){\strut{} 6}}}%
      \put(5863,2547){\makebox(0,0){\strut{}}}%
      \put(6325,2547){\makebox(0,0){\strut{}}}%
      \put(6787,2547){\makebox(0,0){\strut{}}}%
      \put(7249,2547){\makebox(0,0){\strut{}}}%
      \put(7711,2547){\makebox(0,0){\strut{}}}%
      \put(8174,2547){\makebox(0,0){\strut{}}}%
      \put(8636,2547){\makebox(0,0){\strut{}}}%
      \put(9098,2547){\makebox(0,0){\strut{}}}%
      \put(9560,2547){\makebox(0,0){\strut{}}}%
      \put(10022,2547){\makebox(0,0){\strut{}}}%
      \colorrgb{0.00,0.00,0.00}%
      \put(10040,3639){\makebox(0,0)[r]{\strut{}B}}%
    }%
    \gplgaddtomacro\gplfronttext{%
    }%
    \gplgaddtomacro\gplbacktext{%
      \put(5501,1657){\rotatebox{-270}{\makebox(0,0){\strut{} 1.44}}}%
      \put(5501,2245){\rotatebox{-270}{\makebox(0,0){\strut{} 1.46}}}%
      \put(5863,1454){\makebox(0,0){\strut{}}}%
      \put(6325,1454){\makebox(0,0){\strut{}}}%
      \put(6787,1454){\makebox(0,0){\strut{}}}%
      \put(7249,1454){\makebox(0,0){\strut{}}}%
      \put(7711,1454){\makebox(0,0){\strut{}}}%
      \put(8174,1454){\makebox(0,0){\strut{}}}%
      \put(8636,1454){\makebox(0,0){\strut{}}}%
      \put(9098,1454){\makebox(0,0){\strut{}}}%
      \put(9560,1454){\makebox(0,0){\strut{}}}%
      \put(10022,1454){\makebox(0,0){\strut{}}}%
      \put(10040,2546){\makebox(0,0)[r]{\strut{}C}}%
    }%
    \gplgaddtomacro\gplfronttext{%
    }%
    \gplgaddtomacro\gplbacktext{%
      \put(5643,658){\rotatebox{-270}{\makebox(0,0){\strut{} 12.6}}}%
      \put(5643,1289){\rotatebox{-270}{\makebox(0,0){\strut{} 12.8}}}%
      \put(5863,360){\makebox(0,0){\strut{}100}}%
      \put(6325,360){\makebox(0,0){\strut{}150}}%
      \put(6787,360){\makebox(0,0){\strut{}200}}%
      \put(7249,360){\makebox(0,0){\strut{}250}}%
      \put(7711,360){\makebox(0,0){\strut{}300}}%
      \put(8174,360){\makebox(0,0){\strut{}350}}%
      \put(8636,360){\makebox(0,0){\strut{}400}}%
      \put(9098,360){\makebox(0,0){\strut{}450}}%
      \put(9560,360){\makebox(0,0){\strut{}500}}%
      \put(10022,360){\makebox(0,0){\strut{}550}}%
      \csname LTb\endcsname%
      \put(7961,81){\makebox(0,0){\strut{}Days after launch}}%
      \put(10040,1453){\makebox(0,0)[r]{\strut{}D}}%
    }%
    \gplgaddtomacro\gplfronttext{%
    }%
    \gplbacktext
    \put(0,0){\includegraphics{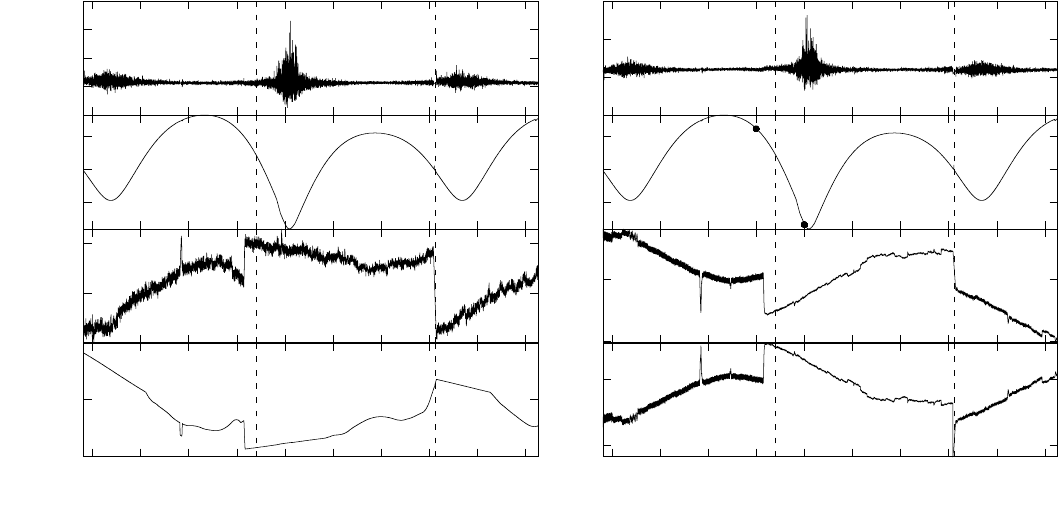}}%
    \gplfronttext
  \end{picture}%
\endgroup

%% file: P02b_3_1_kinematic_dipole.tex
The combination of the motions of the spacecraft, the Earth, and the Solar System with respect to the CMB produce a dipole signal with an amplitude of $(3.355 \pm 0.008)\,\mathrm{mK}$ \citep{hinshaw2009} on the full sky. Like {\it COBE} \citep{1996ApJ.Kogut} and {\it WMAP} \citep{jarosik2010}, we use this signal to estimate the value of $K = G^{-1}$ in Eq.~\ref{eq:calibrationEquation}. We indicate the gain estimates produced using this method with $K_\mathrm{dip}$. To avoid the contamination of the dipole signal with residual Galactic emission, we employ a iterative technique that removes this sky signal from the calibration process.

\begin{table}[tmb]
\begingroup
\newdimen\tblskip \tblskip=5pt
\caption{\label{tab:monopoleAndDipoleParameters} Parameters used in the model of the dipole signal.}
\nointerlineskip
\vskip -3mm
\footnotesize
\setbox\tablebox=\vbox{
   \newdimen\digitwidth 
   \setbox0=\hbox{\rm 0} 
   \digitwidth=\wd0 
   \catcode`*=\active 
   \def*{\kern\digitwidth}
   \newdimen\signwidth 
   \setbox0=\hbox{+} 
   \signwidth=\wd0 
   \catcode`!=\active 
   \def!{\kern\signwidth}
\halign{
\hbox to 2.0cm{#\leaderfil}\tabskip 1.0em&
\hfil #\hfil&
\hfil #\hfil\tabskip=0pt\cr
\noalign{\doubleline}
\omit\hfil Parameter\hfil & Value& Source\cr
\noalign{\vskip 3pt\hrule\vskip 5pt}
\multispan3\hfil CMB monopole\hfil \cr
\noalign{\vskip 3pt}
$T_\mathrm{CMB}$& $2.725 \pm 0.002\,\mathrm{K}$& \citet{mather1999}\cr \noalign{\vskip 8pt}
\multispan3\hfil Solar speed$^a$\hfil \cr
\noalign{\vskip 3pt}
$l$& $263\pdeg99 \pm 0\pdeg 14$& \citet{jarosik2010}\cr
$b$& $48\pdeg26 \pm 0\pdeg03$& \citet{jarosik2010}\cr
$v_\mathrm{Sun}$& $369.0 \pm 0.9\,\mathrm{km\,s^{-1}}$& \citet{jarosik2010}\cr
\noalign{\vskip 8pt}
\multispan3\hfil Velocity of the spacecraft$^b$\hfil\cr
\noalign{\vskip 3pt}
$\left<v_\mathrm{Planck}\right>$& $30.0\pm 0.4\,\mathrm{km\,s^{-1}}$&\cr
$\min v_\mathrm{Planck}$& $29.39\,\mathrm{km\,s^{-1}}$&\cr
$\max v_\mathrm{Planck}$& $30.60\,\mathrm{km\,s^{-1}}$&\cr
\noalign{\vskip 5pt\hrule\vskip 3pt}}}
\endPlancktable
\tablenote a Relative to the CMB rest frame.\par
\tablenote b The values reported here are representative of the overall speed of the spacecraft. The calibration code uses the full timestream of velocity components $(v_x, v_y, v_z)$, sampled once every minute.\par
\endgroup
\end{table}

The basic steps of this calibration procedure are as follows:
\begin{enumerate}
\item We combine the velocity of the spacecraft with respect to the Sun $\vec{v}_\mathrm{Planck}$ and the velocity of the Sun with respect to the CMB $\vec{v}_\mathrm{Sun}$ to estimate the amplitude and alignment of the dipole in the sky in a given direction $\xversor$:
  \begin{equation}
  \label{eq:dipole:fundamental}
    D(\xversor, t) = T_\mathrm{CMB}\left(\frac1{\gamma(t)\,\bigl(1 - \vec{\beta}(t)\cdot \xversor\bigr)} - 1\right),
  \end{equation}
where $T_\mathrm{CMB}$ is the temperature of the CMB monopole, $\boldsymbol\beta = \bigl(\vec{v}_\mathrm{Sun} + \vec{v}_\mathrm{Planck}\bigr)/c$, $\gamma = (1 - \beta^2)^{-1/2}$, $\vec{v}_\mathrm{Sun}$ is the velocity of the Solar System with respect to the CMB rest frame, and $\vec{v}_\mathrm{Planck}$ is the spacecraft's velocity with respect to the Solar System's barycentre. Table~\ref{tab:monopoleAndDipoleParameters} reports the numerical values used in the pipeline.

\item We produce discrete time-ordered data (TOD) of the expected overall dipole signal (thermodynamic temperature) at time $t_i$:
  \begin{equation}
    D_i = D\bigl(\xversor (t_i), t_i\bigr),
  \end{equation}
where $i$ ranges from 1 to $N \sim 10^5$ (the number of samples in a pointing period). As we explained in Sect.~\ref{sec:beamEfficiencyAndWindowFunctions}, the computation of the expected dipole signal takes into account the shape of the beams, following a method described in Sect.~\ref{sec:fourPiConvolver}.

\item Using pointing information, we project both $V_i$ (the voltage $V$ at time $i$, as used in Eq.~\ref{eq:calibrationEquation}) and $D_i$ on a {\tt HEALPix} map \citep{Gorski+2005} with $N_\text{side} = 256$. Multiple hits on the same pixels are averaged in both cases. The result is a pair of maps, $V^\text{map}_k$ and $D^\text{map}_k$, with $k$ being the pixel index\footnote{Most of the pixels in the maps contain no data, since during one pointing period the beam paints a thin circle in the sky. We assume hereafter that the index $k$ runs only through the pixels which have been hit at least once.}.

\begin{figure}
\centering
\includegraphics[width=88mm]{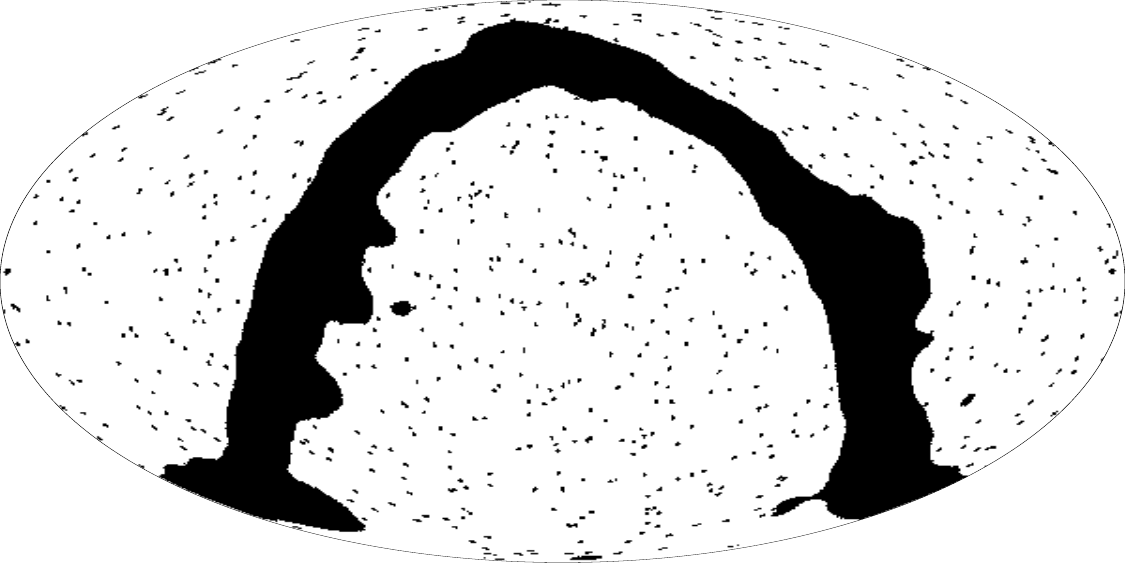}
\caption{\label{fig:calibrationMask} Mask used in the application of Eq.~\protect\ref{eq:dipoleFit} to 30\,GHz data (Ecliptic coordinates). The mask hides point sources and the strong emission of the Galactic plane. What is left is that part of the sky where the only significant emissions are the CMB dipole and the CMB itself. Similar masks have been used for 44 and 70\,GHz data.}
\end{figure}

\item\label{enum:weightedLeastSquares} We use weighted least squares to estimate $K$ in Eq.~\ref{eq:calibrationEquation} from the correlation between the signal in volt $V^\text{map}_k$ with $D^\text{map}_k$:
	\begin{equation}
	\label{eq:dipoleFit}
	V^\text{map}_k = K_\text{dip}\,D^\text{map}_k + \epsilon,
	\end{equation}
where $K_\text{dip}$ and $\epsilon$ are the parameters used in the fit. Each sample $k$ is weighted according to the number of hits per pixel. In computing the fit, we use a frequency-dependent mask to avoid those pixels where a strong non-Gaussian signal other than the dipole is expected. Such masks are the union of a Galactic mask \citep[CG80, see][]{planck2013-p06} and a point source mask (conceptually similar to the point source masks used in component separation, but the masking radius is kept fixed at 32\,\arcmin). See Fig.~\ref{fig:calibrationMask} for an example (30\,GHz). In Sect.~\ref{sec:interchannelCalConsistency} we explore the effect of changing the mask on the calibration.

\item\label{enum:mademoiselle} The fit described in the previous step suffers from the presence of noise in the sky signal, $V^\text{map}$ (mainly due to the CMB and to Galactic emission that was not properly masked). Therefore we apply an iterative algorithm, named \texttt{Mademoiselle}, that uses destriping techniques to iteratively improve the calibration. The algorithm subtracts the convolution between the dipole signal and the $4\pi$ beam from $V^\text{map}$, thus obtaining a set of maps (one per each pointing period $m$) which estimate the sky signal alone, $T^\text{sky}$:
\begin{equation}
T^\text{sky}_k = \bigl( V^\text{map}_k - K_\text{dip} D^\text{map}_k - \epsilon\bigr) K_{\text{dip}},
\end{equation}
with $k$ being the pixel index. (Here again each map covers only a tiny fraction of the whole sky, i.e., the circle covered during each pointing period.) We apply a destriping algorithm to the set of $m$ maps of $T^\text{sky}$ and use the result to apply a correction to $K_\text{dip}$ and $\epsilon$. We iterate this process until a convergence criterion is satisfied. The result of Mademoiselle is a new set of gains, $K_i$ (again, one per pointing period,) and offsets, $\epsilon_i$. See also the section about TOD processing in the \Planck{} explanatory supplement \citep{planck2013-p28}.

\item\label{enum:waveletSmoothingFilter} The gains produced by this procedure need to be further processed in order to reduce the statistical noise. We applied an adaptive smoothing filter based on wavelets, which smooths more around dipole minima and less around maxima \citep[see][]{planck2013-p02}. In those cases where the noise is too high for the filter to produce meaningful results (i.e., near dipole minima), we substituted the gains with a straight line. This can be seen in the left panel (D) of Fig.~\ref{fig:gainCurve}, near 100, 300, and 500 days after launch.

\item\label{enum:dipoleCleaning} Once a set of $K_i$ gains (one per pointing period) is produced, the pipeline calibrates the timelines and subtracts the dipole signal, so that in the calibrated timestreams the Galactic signal and the CMB fluctuations are the only relevant astrophysical components.
\end{enumerate}
Details of the code implementation are given in \citet{planck2013-p02}.

\begin{figure}
	\includegraphics{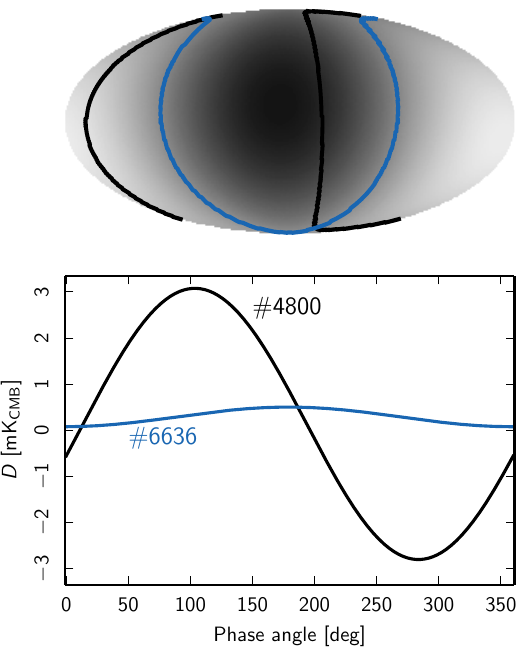}
	\caption{\label{fig:dipoleRingsInTheSky} {\it Top}: dipole signal due to the motion of the Solar System only (in Ecliptic coordinates). The points observed by LFI27M during pointing periods 4800 (250 days after launch) and 6636 (300 days) are shown with black and blue lines, respectively. (In Fig.~\protect\ref{fig:gainCurve}, panel B on the right, we indicate the times when these two periods occurred with black dots.) {\it Bottom}: dipole signal along the same two pointing periods as a function of the North phase angle. Note the difference in the amplitude of the two sinusoidal waves. Such differences are due to the \Planck's scanning strategy and determine the statistical error in the estimation of the calibration constant $K=G^{-1}$ (Eq.~\ref{eq:calibrationEquation}). Refer also to Fig.~\protect\ref{fig:dipoleFits} for details about how $K$ is computed for these two pointing periods.}
\end{figure}

\begin{figure}
	\includegraphics{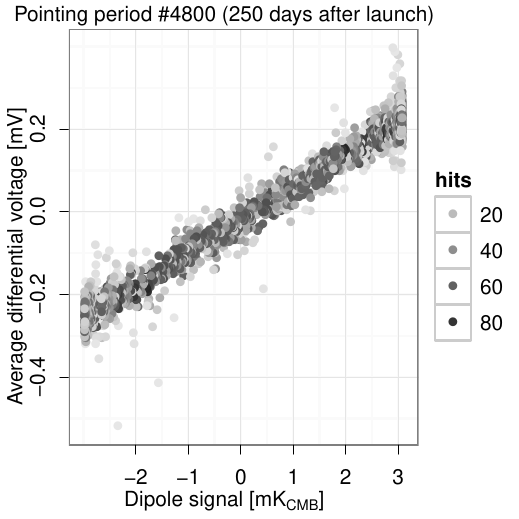}

	\includegraphics{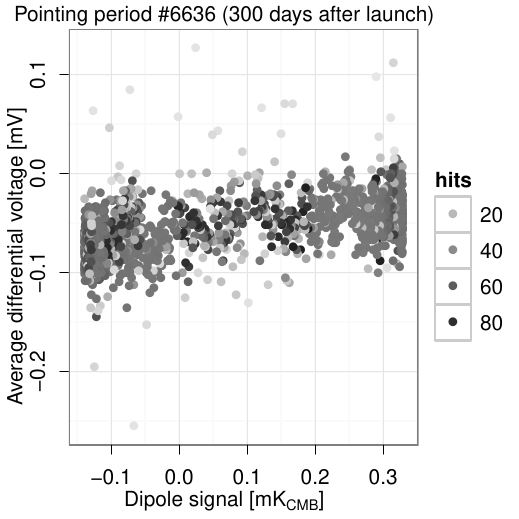}
	
	\caption{\label{fig:dipoleFits} The iterative calibration procedure fits the differential voltages with the expected dipolar signal (K$_\mathrm{CMB}$), both projected on a map, and it calculates the gain according to Eq.~\protect\ref{eq:dipoleFit}. As Fig.~\protect\ref{fig:dipoleRingsInTheSky} shows, depending on the position of the circle with respect to the axis of the dipole signal, the range spanned by the signal can be either large (dipole maximum, \emph{top}) or small (dipole minimum, \emph{bottom}). Of course, in the latter case the estimation of the gain suffers from a larger statistical error. }
\end{figure}

The accuracy in gain reconstruction depends critically on the orientation of the spacecraft with respect to the dipole axis, i.e.\ the value of the product $(\vec{v}_\mathrm{Sun} + \vec{v}_\mathrm{Planck}) \cdot\xversor$. We speak of a \emph{dipole minimum} when the spacecraft's orientation is such that the value of the scalar product reaches a minimum, and a \emph{dipole maximum} when reaches a maximum. Figure~\ref{fig:gainCurve} shows this idea for two LFI radiometers, LFI21M (left) and LFI27M (right): in panel A we show the values of $K_\mathrm{dip}$ as a function of time, while in panel B we show the expected amplitude of the dipole $\Delta T_\mathrm{dip}$ as seen in the circle in the sky (\emph{ring}) which is scanned during each pointing period. It is clear that the noise in $K_\mathrm{dip}$ is mainly due to the variation of $\Delta T_\mathrm{dip}$. The reason why the peak-to-peak difference of the measured dipole signal varies is represented in Fig.~\ref{fig:dipoleRingsInTheSky}, which shows which parts of the sky are observed by LFI27M during two pointing periods, one (4800) near a dipole maximum, and another (6636) near a dipole minimum. Finally, Fig.~\ref{fig:dipoleFits} shows the data used to compute the weighted linear regression presented in item~\ref{enum:weightedLeastSquares} above for the same pointing periods.

%% file: P02b_3_2_4K_calibration.tex
Gain changes in a LFI radiometer can be modelled using the emission of the internal 4\,K reference load as measured by the radiometer itself \citep{valenziano2009}.
 
We indicate with $V_\mathrm{ref}$ the output voltage which measures the temperature of the 4\,K reference load. Eq.~\ref{eq:calibrationEquation} changes via the transformation $B * \bigl(T_\mathrm{sky} + D\bigr) \rightarrow T_\mathrm{ref}$:
\begin{equation}
\begin{split}
V_\mathrm{ref} &= G\Bigl(T_\mathrm{ref} + T_\mathrm{noise} + \epsilon_\mathrm{iso} \bigl(B * T_\mathrm{sky} + D + T^\mathrm{sky}_\mathrm{noise})\Bigr) \\
&= G\Bigl(T_\mathrm{ref} + T_\mathrm{noise} + \epsilon_\mathrm{iso} \Sigma_\mathrm{sky}\bigr),
\end{split}
\end{equation}
where the term $T_\mathrm{noise}$ is analogous to the monopole term $M$ in Eq.~\ref{eq:calibrationEquation}, and we include a possible leakage from the sky signal $\Sigma_\mathrm{sky} = B * T_\mathrm{sky} + D + T^\mathrm{sky}_\mathrm{noise}$ (the term $T^\mathrm{sky}_\mathrm{noise}$ represents the noise temperature associated with the measurement of the sky signal). Solving for $K = G^{-1}$ yields a formula that can be used to estimate $K$ from the ratio between the sum of temperatures and the 4\,K reference voltage:
\begin{equation}
\label{eq:kFordVV}
K_{4\,\mathrm{K}}(t) = \frac{T_\mathrm{ref} + \epsilon_\mathrm{iso} \Sigma_\mathrm{sky} + T_\mathrm{noise}}{V_\mathrm{ref}} \equiv \frac{T_\mathrm{tot}}{V_\mathrm{ref}},
\end{equation}
where $\epsilon_\mathrm{iso}$ is the isolation \citep[for its definition see][]{villa2010} and the term $\epsilon_\mathrm{iso} \Sigma_\mathrm{sky}$ represents the amount of signal coming from the sky and leaking into the reference load signal. We express the term $V_\mathrm{ref}^{-1}$ as:
\begin{equation}
\label{eq:voltageFirstOrderExpansion}
\begin{split}
V_\mathrm{ref}^{-1}&=V_\mathrm{ref,0}^{-1}\left(1+\frac{V_\mathrm{ref}-V_\mathrm{ref,0}}{V_\mathrm{ref,0}}\right)^{-1} \\
&\approx V_\mathrm{ref,0}^{-1}\left(1 - \frac{V_\mathrm{ref}-V_\mathrm{ref,0}}{V_\mathrm{ref,0}}\right) \\
&= V_\mathrm{ref,0}^{-1}\left(2 - \frac{V_\mathrm{ref}}{V_\mathrm{ref,0}}\right),
\end{split}
\end{equation}
where we applied a first-order Taylor expansion\footnote{The error from the first-order approximation in Eq.~\ref{eq:voltageFirstOrderExpansion} is of the order of 0.05\,\%.} for $V_\mathrm{ref}$ around the value $V_\mathrm{ref,0}$. We substitute it into Eq.~\ref{eq:kFordVV} to derive the following equation:
\begin{equation}
\label{eq:gainAsFunctionOfV}
K_{\mathrm{4\,K}}(t) \approx \frac{T_\mathrm{ref,0} + \epsilon_\mathrm{iso} \Sigma_\mathrm{sky,0} + T_\mathrm{noise,0}}{V_\mathrm{ref,0}} \left( 2 - \frac{V_\mathrm{ref}(t)}{V_\mathrm{ref,0}}\right).
\end{equation}
(Ideally, variations in $T_\mathrm{ref}$, in $\epsilon_\mathrm{iso} T_\mathrm{ref}$, and in $T_\mathrm{noise}$ should be negligible; we will quantify their variation later.) In principle, Eq.~\ref{eq:gainAsFunctionOfV} would be enough to estimate $K$ at any time, if one knew all the terms in the right side of the equation with the desired accuracy (e.g., $0.1\,\%$). Since this is not our case, to find the unknown factor in the equation we use the estimate of $K(t)$ from the dipole (see Sect.~\ref{sec:OSGCalibration}) to perform a weighted linear least squares from which we estimate $K_0$:
\begin{equation}
K_\mathrm{dip}(t) = K_0 \times \left( 2 - \frac{V_\mathrm{ref}(t)}{V_\mathrm{ref,0}}\right).
\end{equation}
(Note that here we use the ``raw'' values for $K_\mathrm{dip}(t)$, i.e., before applying the smoothing filter described in Sect.~\ref{sec:OSGCalibration}, point~\ref{enum:waveletSmoothingFilter}.) The average value\footnote{In principle any value for $V_\mathrm{ref,0}$ might be used. However, the fact that we derived Eq.~\ref{eq:gainAsFunctionOfV} using a Taylor expansion of $K(t)$ around $V_\mathrm{ref,0}$ suggest that we use a value of the voltage that is the closest to all the values of $V(t)$ we expect during the survey, in order to improve the accuracy of the first-order approximation. Depending on what we mean by ``closest'', we might either choose the median or the mean of $V(t)$; in this context we found that the difference between these quantities is probably negligible, as it is always less than 0.5\,\%.} of $V_\mathrm{ref}(t)$ over the whole set of data to be calibrated is used for $V_\mathrm{ref,0}$, and the weights $w_i$ in the linear squares fit are proportional to the amplitude of the dipole signal $\Delta T_\mathrm{dip}$ seen by the radiometer during the $i$-th pointing, that is, $w_i \propto \Delta T_\mathrm{dip}$.  According to our classification based on Eq.~\ref{eq:KKzeroAndXi}, this calibration scheme is relative (with $\xi(t) = \bigl(V_\mathrm{ref,0} - V_\mathrm{ref}(t)\bigr)/V_\mathrm{ref,0}$) but not absolute, as we need $K_\mathrm{dip}(t)$ in order to estimate $K_0$.

As an example, see Fig.~\ref{fig:gainCurve}, panel D (right), which shows the values of $K_\mathrm{4\,\mathrm{K}}$ calculated for LFI27M (a 30\,GHz radiometer). Unlike the case for LFI21M (panel D, left), which required a smoothing filter to be applied to the raw gains in panel A, here we use the total-power voltages without any filtering, as their error does not depend on the amplitude of the observed dipole in the sky, but only on the stability of the 4\,K reference load, which is better by orders of magnitude.

\begin{figure}[tbf]
	\centering
	\begin{DIFnomarkup}
	\input{temperature_change_events.tex}
	\end{DIFnomarkup}
	\caption{\label{fig:temperatureChangeEvents} Variation of the temperature of the back-end with time, as measured by sensor \texttt{LM207332} (placed near the back-end amplifiers of LFI27M and 27S). The two major thermal events that happened during the period of the nominal mission are clearly visible here: (1) 258 days after launch the transponder antenna switched from a on/off duty cycle of 24 h to being always kept on, and (2) 456 days after launch we switched to the other sorption cooler used to cool the focal plane.}
\end{figure}
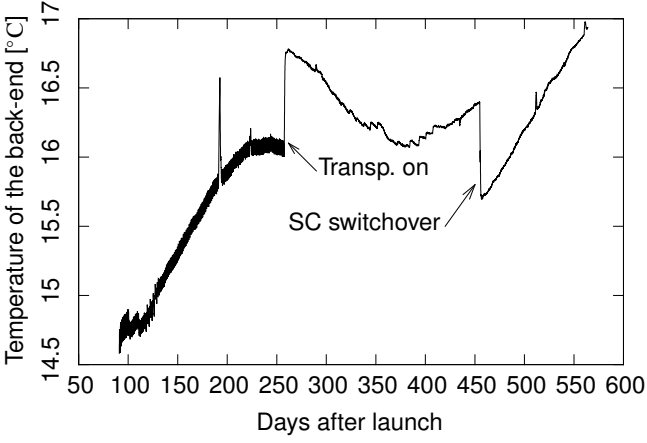

When we presented Eq.~\ref{eq:gainAsFunctionOfV}, we began with the assumption that variations in time of the three terms $T_\mathrm{ref}$, $\epsilon_\mathrm{iso} \Sigma_\mathrm{sky}$, and $T_\mathrm{noise}$ were negligible, in order not to induce significant systematic effects in the determination of the gain. We will now quantify how much they are expected to change during the mission. (Refer to Sect.~\ref{sec:timeScale} for a general discussion of time scales and LFI calibration.) If we write $V_\mathrm{ref} (t) - V_\mathrm{ref,0}$ as $\delta V_\mathrm{ref}$, and similarly for $\delta T_\mathrm{ref}$, $\delta \Sigma_\mathrm{sky}$, $\delta T_\mathrm{noise}$, and $\delta K$, from Eq.~\ref{eq:gainAsFunctionOfV} we derive the following expression:
\begin{equation}
\frac{\delta K(t)}{K_0} = -\frac{\delta V_\mathrm{ref}}{V_\mathrm{ref,0}} + \frac{\delta T_\mathrm{ref} + \epsilon_\mathrm{iso}\,\delta \Sigma_\mathrm{sky} + \delta T_\mathrm{noise}}{T_\mathrm{tot,0}},
\end{equation}
with $T_\mathrm{tot,0} = T_\mathrm{ref,0} + \epsilon_\mathrm{iso}\,T_\mathrm{sky,0} + T_\mathrm{noise,0} \approx 20\,\mathrm{K}$ for the LFI 30\,GHz radiometers. This equation expresses fluctuations in $K$ in terms of fluctuations in $V_\mathrm{ref}$ and the three terms in the numerator of the second fraction on the right side, whose sum should be much smaller than $T_\mathrm{tot,0}$. In the case of LFI's 30\,GHz radiometers, the following considerations apply:
\begin{enumerate}
\item The variation of $T_\mathrm{ref}$ during \Planck{}'s nominal mission is of the order of a few mK, therefore making the term $\delta T_\mathrm{ref} / T_\mathrm{tot,0}$ negligible (hourly variations are of the order of 0.1\,mK).
\item The value of $\epsilon_\mathrm{iso}$ is generally less than 0.1 \citep{villa2010}. The biggest contribution to $\delta \Sigma_\mathrm{sky}$ comes from the solar dipole $D$ ($\max \delta D \approx 3.5\,\mathrm{mK}$). Such variations have an impact of only about $0.01\,\%$ on $K_{4\,\mathrm{K}}$.
\item For $\delta T_\mathrm{noise}/T_\mathrm{tot,0}$ the approximation is not so good, because of the dependence of $T_\mathrm{noise}$ on the temperature of the focal plane \citep{terenzi2009b}. Two events triggered significant changes in the satellite thermal environment. Firstly, starting 258 days after launch, the transponder of the communication system was always kept on, instead of being switched on just for transmission, in order to increase thermal stability. Secondly, 456 days after launch, the first sorption cooler reached its end-of-life and the second sorption cooler was switched on (see Fig.~\ref{fig:temperatureChangeEvents}). These major changes of the thermal environment are expected to have an impact on $T_\mathrm{noise}$ of 0.5--1\,\%. This cannot be considered negligible and has to be corrected for.
\end{enumerate}

The simplest correction strategy for this last effect would be to apply the 4\,K calibration model separately for each section of the data. However, this technique is more subject to the systematic effects known to affect the dipole calibration.  Therefore a more robust approach is to create a map of each section independently, removing both the cosmological and orbital dipoles, then fit for a residual dipole while masking the Galactic emission, and finally estimate the correction factor $\xi_\mathrm{4\,K}$ as:
\begin{equation}
    \xi_\mathrm{4\,K} (i) = 1 - \frac{\Delta T_\mathrm{fit}(i)}{\Delta T_\mathrm{sky}},
\end{equation}
where $\Delta T_\mathrm{fit}(i)$ is the dipole fitted on each section $i$ and $\Delta T_\mathrm{sky}$ is the cosmological dipole. For example, if the residual dipole is $0.1\,\%$ of the cosmological dipole, it means we are over-calibrating by $0.1\,\%$, so we need to correct the calibration by that factor. Table~\ref{tbl:4kcalfixfac} lists the residual dipole values and the corrections applied.

\begin{table*}[tmb]                 % table* is a two-column table.  Drop the * for one column.
\begingroup
\newdimen\tblskip \tblskip=5pt
\caption{\label{tbl:4kcalfixfac}Dipole residuals and correction factors applied to the 4\,K calibration at 30\,GHz in response to major thermal events.}
\nointerlineskip
\vskip -3mm
\footnotesize
\setbox\tablebox=\vbox{
   \newdimen\digitwidth 
   \setbox0=\hbox{\rm 0} 
   \digitwidth=\wd0 
   \catcode`*=\active 
   \def*{\kern\digitwidth}
   \newdimen\signwidth 
   \setbox0=\hbox{+} 
   \signwidth=\wd0 
   \catcode`!=\active 
   \def!{\kern\signwidth}
\halign{
\hfill #\hfill\tabskip 2em&
\hfill #\hfill&
\hfill #\hfill&
\hfill #\hfill&
\hfill #\hfill\/\tabskip=0pt\cr
\noalign{\doubleline}
Time range& Dipole LFI27& Dipole LFI28& Correction LFI27& Correction LFI28\cr
[days]&[\microK]&[\microK]&[\%]&[\%]\cr
\noalign{\vskip 3pt\hrule\vskip 5pt}
*91-257& $-17.03$& *$-8.91$& *$0.58$& 0.30\cr
258-454& $-17.62$& $-13.06$& *$0.60$& 0.44\cr
455-563& *$20.66$& *$-6.71$& $-0.70$& 0.23\cr
\noalign{\vskip 2pt\hrule\vskip 3pt}}}
\endPlancktablewide
\endgroup
\end{table*}

%% file: temperature_change_events.tex
% GNUPLOT: LaTeX picture with Postscript
\begingroup
  \fontfamily{phv}%
  \selectfont
  \makeatletter
  \providecommand\color[2][]{%
    \GenericError{(gnuplot) \space\space\space\@spaces}{%
      Package color not loaded in conjunction with
      terminal option `colourtext'%
    }{See the gnuplot documentation for explanation.%
    }{Either use 'blacktext' in gnuplot or load the package
      color.sty in LaTeX.}%
    \renewcommand\color[2][]{}%
  }%
  \providecommand\includegraphics[2][]{%
    \GenericError{(gnuplot) \space\space\space\@spaces}{%
      Package graphicx or graphics not loaded%
    }{See the gnuplot documentation for explanation.%
    }{The gnuplot epslatex terminal needs graphicx.sty or graphics.sty.}%
    \renewcommand\includegraphics[2][]{}%
  }%
  \providecommand\rotatebox[2]{#2}%
  \@ifundefined{ifGPcolor}{%
    \newif\ifGPcolor
    \GPcolorfalse
  }{}%
  \@ifundefined{ifGPblacktext}{%
    \newif\ifGPblacktext
    \GPblacktexttrue
  }{}%
  % define a \g@addto@macro without @ in the name:
  \let\gplgaddtomacro\g@addto@macro
  % define empty templates for all commands taking text:
  \gdef\gplbacktext{}%
  \gdef\gplfronttext{}%
  \makeatother
  \ifGPblacktext
    % no textcolor at all
    \def\colorrgb#1{}%
    \def\colorgray#1{}%
  \else
    % gray or color?
    \ifGPcolor
      \def\colorrgb#1{\color[rgb]{#1}}%
      \def\colorgray#1{\color[gray]{#1}}%
      \expandafter\def\csname LTw\endcsname{\color{white}}%
      \expandafter\def\csname LTb\endcsname{\color{black}}%
      \expandafter\def\csname LTa\endcsname{\color{black}}%
      \expandafter\def\csname LT0\endcsname{\color[rgb]{1,0,0}}%
      \expandafter\def\csname LT1\endcsname{\color[rgb]{0,1,0}}%
      \expandafter\def\csname LT2\endcsname{\color[rgb]{0,0,1}}%
      \expandafter\def\csname LT3\endcsname{\color[rgb]{1,0,1}}%
      \expandafter\def\csname LT4\endcsname{\color[rgb]{0,1,1}}%
      \expandafter\def\csname LT5\endcsname{\color[rgb]{1,1,0}}%
      \expandafter\def\csname LT6\endcsname{\color[rgb]{0,0,0}}%
      \expandafter\def\csname LT7\endcsname{\color[rgb]{1,0.3,0}}%
      \expandafter\def\csname LT8\endcsname{\color[rgb]{0.5,0.5,0.5}}%
    \else
      % gray
      \def\colorrgb#1{\color{black}}%
      \def\colorgray#1{\color[gray]{#1}}%
      \expandafter\def\csname LTw\endcsname{\color{white}}%
      \expandafter\def\csname LTb\endcsname{\color{black}}%
      \expandafter\def\csname LTa\endcsname{\color{black}}%
      \expandafter\def\csname LT0\endcsname{\color{black}}%
      \expandafter\def\csname LT1\endcsname{\color{black}}%
      \expandafter\def\csname LT2\endcsname{\color{black}}%
      \expandafter\def\csname LT3\endcsname{\color{black}}%
      \expandafter\def\csname LT4\endcsname{\color{black}}%
      \expandafter\def\csname LT5\endcsname{\color{black}}%
      \expandafter\def\csname LT6\endcsname{\color{black}}%
      \expandafter\def\csname LT7\endcsname{\color{black}}%
      \expandafter\def\csname LT8\endcsname{\color{black}}%
    \fi
  \fi
  \setlength{\unitlength}{0.0500bp}%
  \begin{picture}(4980.00,3400.00)%
    \gplgaddtomacro\gplbacktext{%
      \put(432,595){\rotatebox{-270}{\makebox(0,0){\strut{} 14.5}}}%
      \put(432,1115){\rotatebox{-270}{\makebox(0,0){\strut{} 15}}}%
      \put(432,1635){\rotatebox{-270}{\makebox(0,0){\strut{} 15.5}}}%
      \put(432,2155){\rotatebox{-270}{\makebox(0,0){\strut{} 16}}}%
      \put(432,2675){\rotatebox{-270}{\makebox(0,0){\strut{} 16.5}}}%
      \put(432,3195){\rotatebox{-270}{\makebox(0,0){\strut{} 17}}}%
      \put(609,409){\makebox(0,0){\strut{} 50}}%
      \put(978,409){\makebox(0,0){\strut{} 100}}%
      \put(1348,409){\makebox(0,0){\strut{} 150}}%
      \put(1717,409){\makebox(0,0){\strut{} 200}}%
      \put(2087,409){\makebox(0,0){\strut{} 250}}%
      \put(2456,409){\makebox(0,0){\strut{} 300}}%
      \put(2826,409){\makebox(0,0){\strut{} 350}}%
      \put(3195,409){\makebox(0,0){\strut{} 400}}%
      \put(3565,409){\makebox(0,0){\strut{} 450}}%
      \put(3934,409){\makebox(0,0){\strut{} 500}}%
      \put(4304,409){\makebox(0,0){\strut{} 550}}%
      \put(4673,409){\makebox(0,0){\strut{} 600}}%
      \csname LTb\endcsname%
      \put(162,1895){\rotatebox{-270}{\makebox(0,0){\strut{}Temperature of the back-end [${}^\circ\mathrm{C}$]}}}%
      \csname LTb\endcsname%
      \put(2641,130){\makebox(0,0){\strut{}Days after launch}}%
      \put(2407,2051){\makebox(0,0)[l]{\strut{}Transp. on}}%
      \put(3318,1635){\makebox(0,0)[r]{\strut{}SC switchover}}%
    }%
    \gplgaddtomacro\gplfronttext{%
    }%
    \gplbacktext
    \put(0,0){\includegraphics{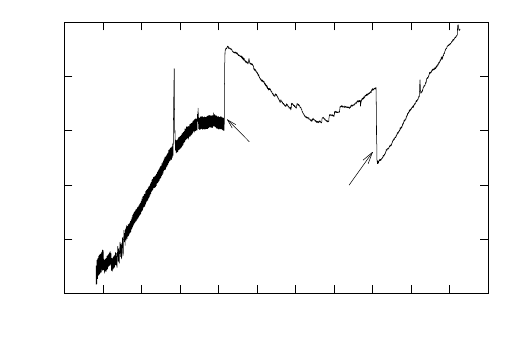}}%
    \gplfronttext
  \end{picture}%
\endgroup

%% file: P02b_3_3_orbital_dipole.tex
The CMB dipole induced by the peculiar velocity of the solar system relative to the CMB last scattering surface is a superb tool for LFI calibration, thanks to its high signal to noise source for all the frequency channels of LFI. Unfortunately it is not a fundamental cosmological parameter and has no predicted amplitude and direction: therefore, it must be calibrated against something else which is known absolutely. The relative velocity of the \Planck{} satellite with respect to the solar system barycentre is known very precisely and adds a further dipole signal with a known amplitude and direction to the data. As the satellite is located at $L_2$, its velocity relative to the CMB rest frame is dominated by the Earth's motion around the Sun and ranges between 29.6\,\kms and 30.6\,\kms, and the amplitude of the induced dipole signal is around $250\,\muK$. This is an order of magnitude smaller than the cosmological dipole, and it is too weak to use on a pointing-by-pointing basis. Therefore, some form of relative calibration must be used to tie together data from each pointing period.

For the {\it WMAP} radiometers, passively-cooled to 100K, in the 5-year and 7-year analyses \citep{hinshaw2009,jarosik2010} an adequate solution for relative calibration was obtained using a simple parametrized fit based on detector voltages and hot and cold stage temperatures. For the LFI radiometers, which are cryogenically cooled to 20\,K and use blackbody reference loads cooled to 4\,K, a more complex thermal model is needed for a high precision parametric solution based on housekeeping information. In this analysis, the approximate method described in Sect.~\ref{sec:dVV} is implemented, together with the OSG approach reported in Sect.~\ref{sec:OSGCalibration}. 

Our starting point for using the orbital dipole is the time-ordered-data, since the direction of the induced dipole is in the direction the satellite is traveling, and it is therefore time-dependent. Furthermore, one must use TODs which have not been processed with the calibration module, since that module removes both dipoles from the data. As most of the pixels
on the sky are observed with two observations separated by
approximately six months, where the velocity is in opposite directions,
the orbital dipole signal cancels out to first order on the full
mission sky maps. The time-ordered data are binned into rings for each
pointing period, based on the rotation phase angle of the satellite. In
this way all the relevant information is stored in a more compact and
accessible form, with the advantage that the orbital dipole appears as
a near sinusoid variation with the same phase in each ring. 

A good relative calibration is required to remove the gain differences
the between rings. For this we use the gains used in the final
analysis as these are the best available. For the 70 and 44\,GHz
channels this may appear to be potentially circular as they are
already calibrated on the solar dipole, but tests introducing a
scaling factor into the relative gains shows that the same factor is
recovered in the absolute gain. In this way any tension between the
{\it WMAP} dipole used in the calibration and the orbital dipole would be
revealed as a non-unity correction factor.

Both solar and orbital dipoles can then be fitted in the time line,
with the direction and amplitude of the solar dipole left free and the
amplitude of the orbital dipole providing the final calibration. Currently, uncertainties in the ring-to-ring relative calibration
and far-sidelobes limit the accuracy of this calibration to
$0.26\,\%,0.16\,\%$ and $0.64\,\%$ at 70, 44 and 30\,GHz, respectively,
based on variations in the results. As this does not offer a significant
improvement over the {\it WMAP} dipole, this analysis has therefore
been postponed to a subsequent release of the LFI data when
the characterization of far sidelobes and gain variations is more
mature. Table~\ref{tbl:prelim44GHzdipoles} reports the results of a preliminary analysis of correction factors and solar dipole
parameters for the 44\,GHz channels (where the effect of far-sidelobes
is least).

\begin{table}[tmb]
\begingroup
\newdimen\tblskip \tblskip=5pt
\caption{\label{tbl:prelim44GHzdipoles} Preliminary estimates of the CMB dipole in the 44\,GHz channels.}
\nointerlineskip
\vskip -3mm
\footnotesize
\setbox\tablebox=\vbox{
   \newdimen\digitwidth \setbox0=\hbox{\rm
   0} \digitwidth=\wd0 \catcode`*=\active \def*{\kern\digitwidth}
   \newdimen\signwidth 
   \setbox0=\hbox{+} 
   \signwidth=\wd0 
   \catcode`!=\active 
   \def!{\kern\signwidth}

{
\halign{
\hfil #\tabskip=1em &
\hfil #\hfil &
\hfil #\hfil &
\hfil #\hfil &
\hfil #\hfil &
\hfil #\hfil\tabskip=0em\cr
\noalign{\doubleline}
\omit\hfil Horn&
\omit\hfil Arm\hfil&
\omit\hfil Correction$^a$\hfil&
\omit\hfil $v_\mathrm{Sun}\,[\mathrm{km\,s^{-1}}]$\hfil&
\omit\hfil $l$\hfil&
\omit\hfil $b$\hfil\cr
\noalign{\vskip 3pt\hrule\vskip 5pt}
24&M& 1.0005 & 367.76& $263\pdeg96$& $48\pdeg54$\cr
  &S& 1.0000 & 368.03& $263\pdeg96$& $48\pdeg50$\cr
25&M& 1.0006 & 367.97& $263\pdeg93$& $48\pdeg45$\cr
  &S& 0.9967 & 369.35& $264\pdeg01$& $48\pdeg28$\cr
26&M& 1.0025 & 367.44& $263\pdeg87$& $48\pdeg42$\cr
  &S& 1.0004 & 368.33& $263\pdeg92$& $48\pdeg41$\cr
\noalign{\vskip 5pt\hrule\vskip 3pt}}}}
\endPlancktable
\tablenote a The ratio of expected orbital dipole to measured dipole. \par
\endgroup
\end{table}

%% file: P02b_3_4_zero_level.tex
The monopole of \Planck{} maps is unconstrained: adding a constant to the time-ordered data leaves the likelihood of the map unchanged. It is therefore conventional to adjust the monopole of the maps in order to attain plausible absolute values of the high Galactic latitude diffuse Galactic foreground signal, and plausible values of the frequency scaling of this Galactic signal from channel to channel.

For LFI we have implemented two different approaches to evaluate the zero level. The first considers a circular region of sky centred on $(l, b) = (344\pdeg47, -77\pdeg08)$ with radius of one degree\footnote{We chose this region since it is nearly free of bright point sources and is well representative of the Galactic diffuse emission.}, on which the monopole matching is performed. We subtracted the estimated contribution of the CMB (smoothed to the angular resolution of each channel) from each LFI frequency map prior to matching diffuse foreground amplitudes.  The average value of the CMB-subtracted pixels within the above mask is adjusted to match following values: 35\,\muK{} at 30\,GHz, 15\,\muK{} at 44\,GHz, and 18\,\muK{} at 70\,GHz.

The other approach we implemented is similar to the one adopted by {\it WMAP} and described by \citet{bennett2003b,bennett2012}. The procedure is the following: (1) we smooth the maps at $1^\circ$ angular resolution, (2) we subtract the estimated CMB signal, again smoothed at the common resolution, and (3) we fit the observed variation with Galactic latitude assuming a plane-parallel model for the Galactic emission. We consider only the behaviour on the southern hemisphere (in the range $-90^\circ < b < -15^\circ$)  in order to avoid possible contamination from the high-galactic latitude structures present in the Galactic emission such as the North Polar Spur. We divide the map into stripes of constant latitude $2.5^\circ$ wide and we evaluate the mean level of CMB subtracted signal. Then, this is fitted with a cosecant model of the form $T = A\csc b + B$. The offset is the value by which the intercept $B$ is zero.

The results yielded by the two methods are in agreement. We set the zero levels of the maps released by the \Planck{} collaboration to the estimates obtained using the second method. The numerical values are reported in Table~\ref{tab:accuracyBudgetResult}.

%% file: P02b_4_1_impact_of_optics_on_calibration.tex
The optical response of each radiometer, including the effect caused by the presence of the telescope and the baffles, impacts the estimation of the calibration constants. Apart from the effects due to the coupling of the CMB dipole with the full beam response described in Sect.~\ref{sec:beamEfficiencyAndWindowFunctions}, an important effect (particularly significant at 30\,GHz) is the fact that the brightness of the Galactic plane produces a non-trivial spurious signal which biases the fit with the dipole.

\begin{figure}
\centering
\includegraphics[width=88mm]{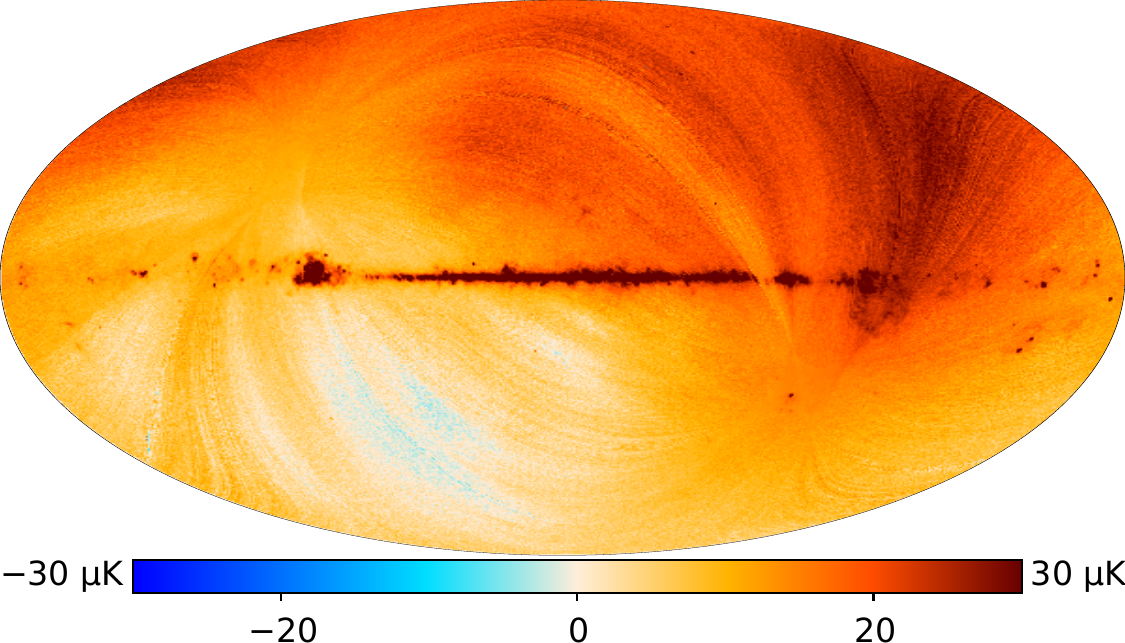}
\caption{\label{fig:OSGdVVDifference} Difference between the 30\,GHz LFI map calibrated using the OSG calibration (Sect.~\protect\ref{sec:OSGCalibration}) and the nominal 30\,GHz map (calibrated using the 4\,K method, see Sect.~\protect\ref{sec:dVV}). The residual dipole visible in the map is due only to the latter calibration, as the nominal method removes any residual dipole. A few stripes are visible as well: they are caused by Galactic straylight affecting the fit with the dipole (OSG calibration), as hinted by the fact that such stripes are positive.}
\end{figure}

Consequently, for the current release, we have decided to calibrate the 30\,GHz channels using the 4\,K calibration, as described in Sect.~\ref{sec:dVV}. This procedure avoids the spurious signals induced by the sidelobes\footnote{Note however that the two methods produce maps that show a remarkable agreement: their peak-to-peak difference, if we neglect a few outlying pixels, is of the order of a few tens of \muK, thus of the same order of magnitude as the rms\ of the survey difference maps discussed in Sect.~\ref{sec:nullTests}.} (see Fig.~\ref{fig:OSGdVVDifference}). In a future release, we plan to use the fully modelled sidelobe patterns to correct the timelines in the iterative calibration sequence, thereby using the same calibration scheme for all three frequency bands. We will also treat polarization calibration.

The optical modelling of LFI, including extensive discussion of the sidelobes, is covered in \citet{planck2013-p02d}.

%% file: P02b_4_2_beam_convolution.tex
\def\DeltaTDipoleDirac{\Delta T_{\delta}}
\def\DeltaTDipole{\Delta T}
\def\Pointing{\vec{x}_{\mathrm{b}}}
\def\PointingE{\vec{x}}
\def\DipoleE{\vec{D}_{\mathrm{E}}}
\def\DipoleEv{\vec{D}_{\mathrm{E}}}
\def\RotBeamObs{\tens{U}}
\def\BeamRd{B}
\def\betarad{\beta_{\mathrm{rad}}}
\newcommand{\RadiometerM}[1]{\tens{M}_{#1}}
\def\RadiometerA{\tens{A}}
\def\vsun{\vec{v}_{\mathrm{Sun}}}
\def\vplanck{\vec{v}_{\mathrm{Planck}}}
\def\PointingBx{x_\mathrm{b}}
\def\PointingBy{y_\mathrm{b}}
\def\PointingBz{z_\mathrm{b}}
\def\Tcmb{T_\mathrm{CMB}}
\def\deg{^{\circ}}
\def\HEALpix{{\tt HEALPix}}
\def\Nside{N_\mathrm{side}}

In Sect.~\ref{sec:beamEfficiencyAndWindowFunctions} we mentioned that the procedure is based on a convolution of a reference dipole with the beam shape. This procedure can be time-consuming, as it must be done once for each sample measured by each radiometer. Fortunately, the spatial symmetry of the dipole signal allows us to derive a fast and elegant method to properly take into account the shape of the full beams when using the dipole to calibrate the data. We describe the mathematical details of the model in Appendix~\ref{sec:computingDipoleConvolutionParams}; here we provide a physical interpretation.

The model applies a convolution with the full beam to Eq.~\ref{eq:dipole:fundamental}, and it takes advantage of the high level of symmetry in the dipole to greatly simplify the calculation. The result of the convolution is still a dipole signal, but its direction and amplitude are slightly different from the dipole before the convolution, for the following reasons:
\begin{enumerate}
\item the amplitude changes because of all the power entering the beam from directions other than the beam axis;
\item the direction changes because of asymmetries in the beam shape.
\end{enumerate}
Such changes are characterized by a constant vector $\Svector$, which only depends on the beam shape. The maximum difference between the amplitudes of the convolved dipole and of the dipole in the sky, under the simplifying hypothesis that $\vsun + \vec{v}_{\mathrm{Planck}} \approx \vsun$, is:
\begin{equation}
\label{eq:dipoleAmplitudeCorrection}
\Delta T_{\mathrm{sc}}=\Tcmb\left|\frac{\vsun}{c}\right|\bigl(1-\left|\Svector\right|\bigr).
\end{equation}
The direction of $\Svector$ in the beam reference frame represents the deviation of the effective axis of the dipole signal with respect to the axis of the dipole in the sky. A convenient measure of the overall deflection is represented by the deflection angle
\begin{equation}
\label{eq:DipoleAxisDeflection}
\delta_{\mathrm{z}} = \arccos(\vec{e}_z\cdot\Svector/|\Svector|),
\end{equation}
where $\vec{e}_z$ is the nominal direction of the beam.

\begin{table}[tmb]
\begingroup
\newdimen\tblskip \tblskip=5pt
\caption{\label{tbl:dipoleDeflection} Dipole deflection and correction due to sidelobes.}
\nointerlineskip
\vskip -3mm
\footnotesize
\setbox\tablebox=\vbox{
   \newdimen\digitwidth 
   \setbox0=\hbox{\rm 0} 
   \digitwidth=\wd0 
   \catcode`*=\active 
   \def*{\kern\digitwidth}
   \newdimen\signwidth 
   \setbox0=\hbox{+} 
   \signwidth=\wd0 
   \catcode`!=\active 
   \def!{\kern\signwidth}
{
\halign{
\hbox to 3.0cm{#\leaderfil}\tabskip 0.5em&
\hfil #\hfil&
\hfil #\hfil\tabskip=0em\cr
\noalign{\doubleline}
\omit\hfil Radiometer\hfil&
\omit\hfil $\delta_z$\tablefootmark{a}\hfil&
\omit\hfil $\Delta T_\mathrm{sc}$\tablefootmark{b}\hfil\cr
\noalign{\vskip 3pt\hrule\vskip 5pt}
\noalign{\vskip 2pt}
LFI18M& *5.94& *7.19\cr
LFI18S& *7.35& *9.10\cr
LFI19M& *6.83& *8.49\cr
LFI19S& *7.92& 10.09\cr
LFI20M& *9.43& 11.56\cr
LFI20S& 10.10& 12.65\cr
LFI21M& *9.46& 11.59\cr
LFI21S& *9.78& 12.31\cr
LFI22M& *6.96& *8.62\cr
LFI22S& *8.20& 10.38\cr
LFI23M& *5.68& *6.90\cr
LFI23S& *7.23& *8.98\cr
\noalign{\vskip 2pt}
{\bf 70\,GHz (mean)}& *7.91& *9.82\cr
\noalign{\vskip 6pt}
LFI24M& *2.37& *2.95\cr
LFI24S& *1.43& *1.98\cr
LFI25M& *1.31& *1.95\cr
LFI25S& *0.88& *1.49\cr
LFI26M& *1.31& *1.91\cr
LFI26S& *0.90& *1.44\cr
\noalign{\vskip 2pt}
{\bf 44\,GHz (mean)}& *1.37& *1.95\cr
\noalign{\vskip 6pt}
LFI27M& *8.54& 11.11\cr
LFI27S& *7.92& 10.78\cr
LFI28M& *8.66& 11.25\cr
LFI28S& *7.92& 10.77\cr
\noalign{\vskip 2pt}
{\bf 30\,GHz (mean)}& *8.26& 10.98\cr
\noalign{\vskip 5pt\hrule\vskip 3pt}}}}
\endPlancktable
\tablenote a Deflection in the polar axis of the dipole, in arcmin (Eq.~\ref{eq:dipoleAmplitudeCorrection}).\par
\tablenote b Correction to the amplitude of the dipole, in $\muK_\mathrm{CMB}$ (Eq.~\ref{eq:DipoleAxisDeflection}).\par
\endgroup
\end{table}

\begin{figure}
\centering
\includegraphics{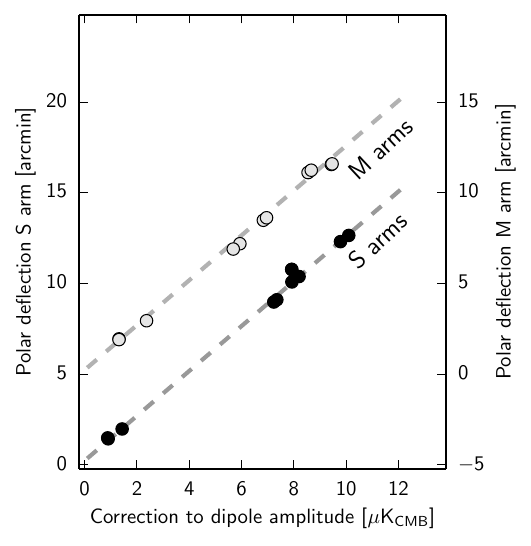}
\caption{\label{fig:dipoleSFactor} 
Regression between dipole deflection (arcmin) and dipole amplitude correction (\muK) for each radiometer. Numbers are taken from Table~\protect\ref{tbl:dipoleDeflection}. Side arms are in black, main arms are in white and have been shifted by $5\arcm$ to improve readability.}
\end{figure}

Figure~\ref{fig:dipoleSFactor} plots the corresponding $\delta_{\mathrm{z}}$ versus $\Delta T_{\mathrm{sc}}$ for each detector, and the numerical values are tabulated in Table~\ref{tbl:dipoleDeflection}. There is an evident linear correlation between the two quantities, with the best fit being
\begin{equation}
\delta_\mathrm{z} = 1.2368\,\Delta T_\mathrm{sc} + 0.2329.
\end{equation}
The deflection angles are between $1\parcm3$ and $12\parcm6$, so that
the largest correction is 13\,\muK. It is interesting to note the slight level of asymmetry between the side and main arms of each feed-horn. This might lead to differences in corrections nearly as large as 1\,\muK. An example is given by the two arms of LFI28. In contrast, the two arms of LFI21 show no asymmetries.

As we already stated in Sect.~\ref{sec:beamEfficiencyAndWindowFunctions}, for this data release we have neglected $\Delta T_\mathrm{sc}$ (thus setting $\left\|\Svector\right\| = 1$), but we have considered the tilt effect given by $\delta_\mathrm{z}$. The tilt was considered not only when using Eq.~\ref{eq:dipole:fundamental} (i.e., fitting timestreams with the expected dipole signal), but also when we cleaned the calibrated data of the dipole itself (see point \ref{enum:dipoleCleaning} in Sect.~\ref{sec:OSGCalibration}).

%% file: P02b_4_3_ADC_systematic_effects.tex
As stated in Sect.~\ref{sec:calibrationMethods}, the calibration process also includes the conversion of the analogue detector voltages by the ADCs. This should be absorbed into the final radiometer calibration, but the ADCs can suffer a differential linearity problem, where the voltage step between binary levels varies slightly. The ADC employed use a digital to analogue circuit (DAC) and a successive approximation algorithm to match the voltages, so an imperfect step associated with a particular binary bit will lead to a recurring non-linear glitch when that bit changes. This can lead to a small but sharp deviation in the response curve as shown in Fig.~10 of \citet{planck2013-p02a}. Since the slope of the response curve determines the effective calibration, a large gradient change can happen over a small range of voltage, potentially becoming the dominant source of gain uncertainty.

\begin{figure}[tbf]
\centering
\includegraphics{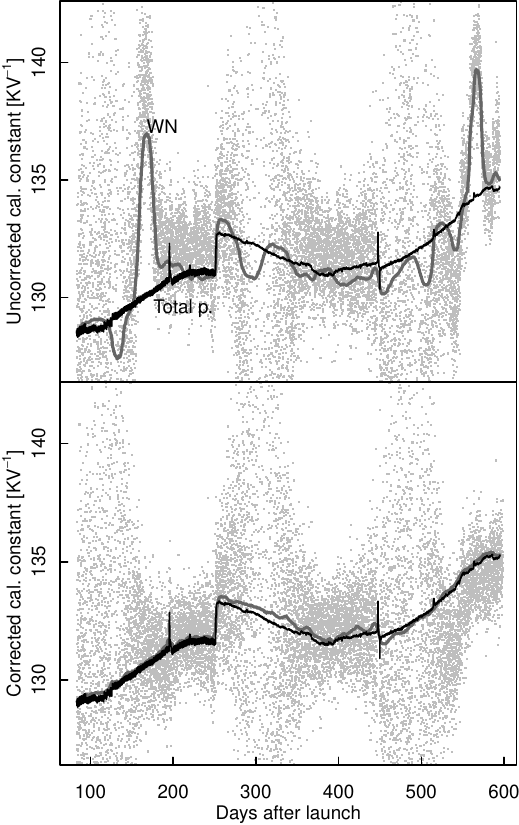}
\caption{Time evolution of the calibration constant used to convert detector voltages into a thermodynamic temperature in Kelvin. The top panel shows the comparison for ADC uncorrected LFI25S-11 diode data between the raw dipole gains (grey dots) and the reciprocal of total-power sky detector voltage (black line), the latter scaled to have the same mean level as the gains. Two highly significant departures can be seen around day 160 and 560 together with lower level variations over most of the time range. The thick dark grey line  shows inverse white noise estimates, again scaled to the same gain level, but filtered with a 100 ring moving median filter to reduce the scatter. As expected, this follows the dipole gains. The lower panel shows the same thing, but after the ADC correction has been applied to the TOD data.}
\label{fig:gains_2511_before_after_ADC}
\end{figure}

Such an effect is seen in the LFI radiometers for some channels, especially those with low back-end gain, where the quantization levels will be more evident. For this reason the radiometers of 44\,GHz are the most affected, as shown in the upper panel of Fig.~\ref{fig:gains_2511_before_after_ADC} for the worst affected LFI25S-11 channel \citep[for more information, refer to][]{planck2013-p02a}.

The correction for this effect requires the reconstruction of the response curves, which can be inferred from the related difference in behaviour between white noise and detector voltages. The reason is that the white noise, like the dipole and sky temperature fluctuations, is sensitive to the gradient of the response curve, whereas the detector voltages are just determined by the point on the curve. This is explained in Appendix~A of \citet{planck2013-p02a} and an example for LFI25S-11 is shown in that paper together with estimates of ADC residual errors in terms of systematic effects at the map level. The implementation in the pipeline uses splines\footnote{We use the spline functions provided by the GNU Scientific Library \citep{GSLRefMan}.}, fitted to pre-calculated template curves of the non-linearities. The result of correcting the TOD data is shown in the lower panel of Fig.~\ref{fig:gains_2511_before_after_ADC}, where the dipole gains and inverse sky voltage and white noise are much more consistent.

\begin{figure}[tbf]
\includegraphics[width=88mm]{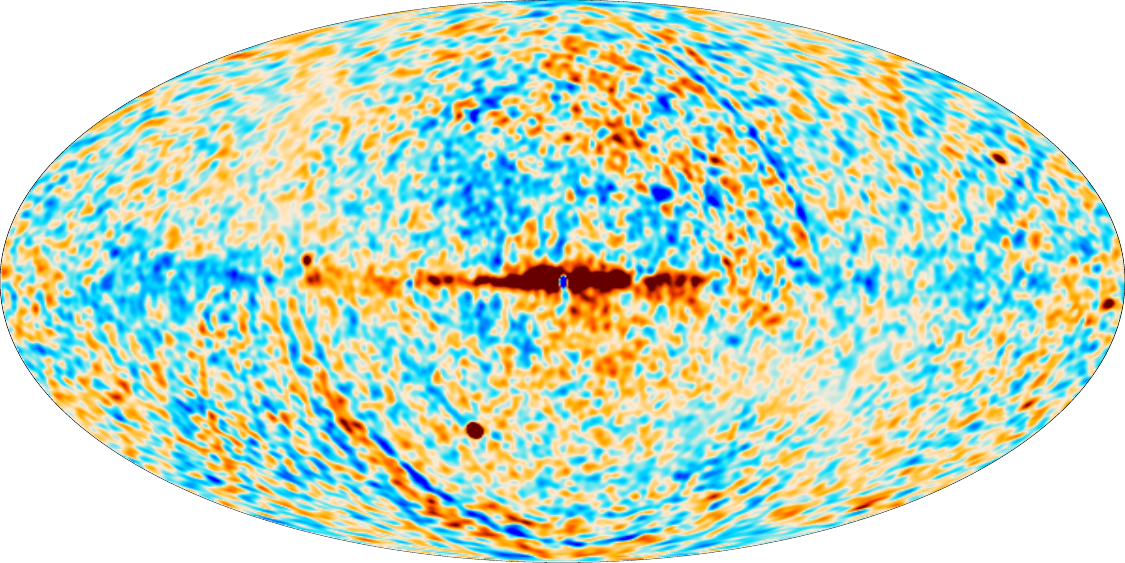}
\includegraphics[width=88mm]{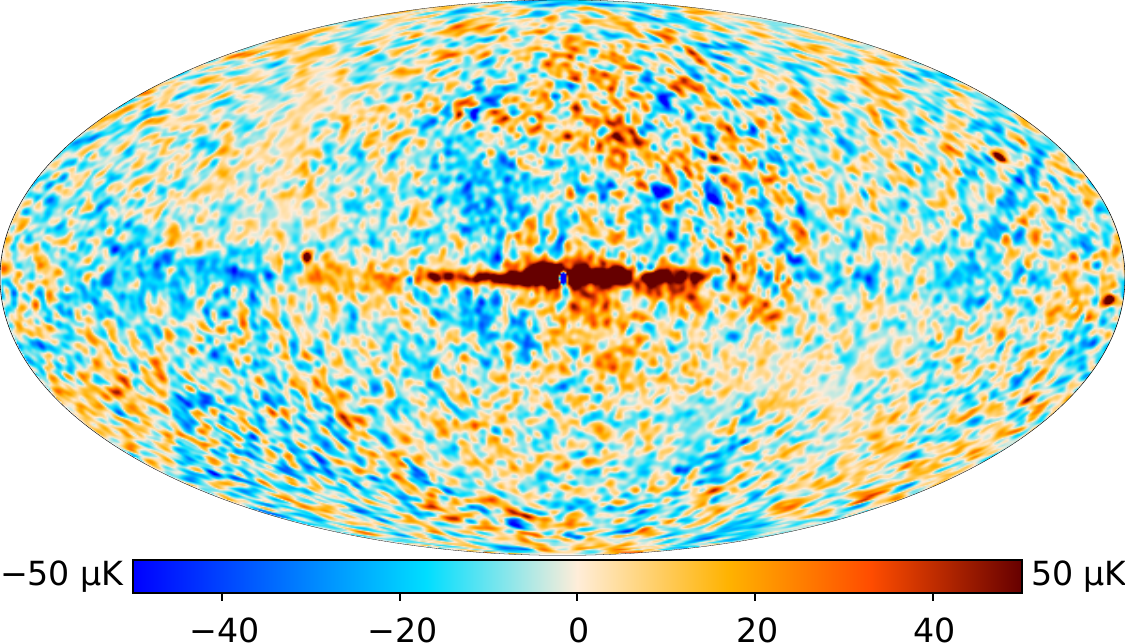}
\caption{Three degree smoothed map of the difference between the main and side arm of LFI25 before (top) and after (bottom) ADC non-linearity correction. Stripes can seen which are stronger in the region where the solar dipole peaks, as a result of residuals from calibration uncertainties. This relative uncertainty between radiometer arms directly affects polarization as shown here, since the main and side arm difference represents a pseudo-polarization.}
\label{fig:ADC_25M_25S_stripes}
\end{figure}

In the case where calibration used the level of the 4\,K load, the uncorrected data would produce uncertainties directly in the calibration of the order of 5\,\% leading to stripes of about 150\,\muK, due to the solar dipole being incorrectly removed. Using a dipole based calibration would correct the ADC non-linearity, but the need to low-pass filter it for an acceptable signal-to-noise ratio would mean that gain variations shorter than the filter timescale were not corrected.  This in turn would result in fine striping at the level of 20--30\,\muK. Further, since the ADC non-linearity can differ between the two arms of a radiometer, there is also a direct impact on polarization through differential calibration errors.  This is illustrated in Fig.~\ref{fig:ADC_25M_25S_stripes}, where we show the difference between maps made from the main and side arms of radiometer LFI25.

It should be pointed out that the LFI and HFI ADC non-linearity problems are completely different, due to the different implementations of fast switching schemes. LFI use phase switching to swap sky and reference load signals, whereas HFI use a square wave modulation and AC coupling to avoid $1/f$ contributions from the back-end electronics. More details about the ADCs used by HFI are found in \citet{planck2013-p03f}.

%% file: P02b_5_accuracy_introduction.tex
\begin{table*}[tmb]
\begingroup
\newdimen\tblskip \tblskip=5pt
\caption{\label{tab:accuracyBudgetOverview} Elements considered in deriving the accuracy of the calibration.}
\nointerlineskip
\vskip -3mm
\footnotesize
\setbox\tablebox=\vbox{
   \newdimen\digitwidth 
   \setbox0=\hbox{\rm 0} 
   \digitwidth=\wd0 
   \catcode`*=\active 
   \def*{\kern\digitwidth}
   \newdimen\signwidth 
   \setbox0=\hbox{+} 
   \signwidth=\wd0 
   \catcode`!=\active 
   \def!{\kern\signwidth}
\halign{
% Template
\hbox to 4.5cm{#\leaderfil}\tabskip 2.0em &
\hfil #\hfil &
#\hfil&
#\hfil\tabskip=0pt\cr
\noalign{\doubleline}
\omit\hfil Type of uncertainty\hfil &
\omit\hfil Applies to\hfil &
\omit\hfil Method used to assess the accuracy \hfil&
\omit\hfil Reference\hfill\cr
\noalign{\vskip 3pt\hrule\vskip 5pt}
\omit\hfil Absolute\hfil \cr
\noalign{\vskip 3pt}
Standard& All sky& Propagation of {\it WMAP} errors&Sect.~\ref{sec:WMAPerrorPropagation}\cr
Zero level& All sky& Comparison with {\it WMAP} values&Sect.~\ref{sec:zeroLevel}\cr
Beam uncertainty& All sky& {\tt GRASP} model of the beams&Sect.~\ref{sec:beamUncertainties}\cr
Sidelobe convolution effect& All sky& Simulations& Sect.~\ref{sec:beamEfficiencyAndWindowFunctions}\cr
Colour corrections& Galactic areas& Comparison of ground/flight bandpass leakages& Sect.~\ref{sec:colourCorrections}\cr
\noalign{\vskip 8pt}
\omit\hfil Relative\hfil \cr
\noalign{\vskip 3pt}
Statistical/algorithmical errors& All sky& Simulations&Sect.~\ref{sec:noiseAndGain}\cr
Known systematics& All sky& Simulations&Sect.~\ref{sec:impactOfKnownSystematics}\cr
Unknown systematics& All sky& Null tests&Sect.~\ref{sec:impactOfUnknownSystematics}\cr
\noalign{\vskip 5pt\hrule\vskip 3pt}}}
\endPlancktablewide                 % ends two-column \halign
\endgroup
\end{table*}

\begin{table*}[tmb]
\begingroup
\newdimen\tblskip \tblskip=5pt
\caption{\label{tab:accuracyBudgetResult} Accuracy in the calibration of LFI data.}
\nointerlineskip
\vskip -3mm
\footnotesize
\setbox\tablebox=\vbox{
   \newdimen\digitwidth 
   \setbox0=\hbox{\rm 0} 
   \digitwidth=\wd0 
   \catcode`*=\active 
   \def*{\kern\digitwidth}
   \newdimen\signwidth 
   \setbox0=\hbox{+} 
   \signwidth=\wd0 
   \catcode`!=\active 
   \def!{\kern\signwidth}
\halign{
% Template
\hbox to 6.5cm{#\leaderfil}\tabskip 2.0em &
\hfil #\hfil &
\hfil #\hfil &
\hfil #\hfil &
\hfil #\hfil\tabskip=0pt\cr
\noalign{\doubleline}
\omit\hfil Type of uncertainty\hfil &
\omit\hfil Applies to\hfil &
\omit\hfil 30\,GHz\hfil &
\omit\hfil 44\,GHz\hfil &
\omit\hfil 70\,GHz\hfil\cr
\noalign{\vskip 3pt\hrule\vskip 5pt}
\omit\hfil Absolute\hfil \cr
\noalign{\vskip 3pt}
Standard\tablefootmark{a}& All sky& 0.25\,\%& 0.25\,\%& 0.25\,\% \cr
Zero level [$\muK_\mathrm{CMB}$]& All sky& $-300.84 \pm 2.23$& $-22.83 \pm 0.78$& $-28.09 \pm 0.64$\cr
Beam uncertainty& All sky& 0.5\,\%& 0.1\,\%& 0.3\,\%\cr
Sidelobe convolution effect& All sky& 0.2\,\%& 0.2\,\%& 0.2\,\%\cr
Colour corrections& Galactic areas& $\left|\alpha - 2\right|\,0.1$\,\%& $\left|\alpha - 2\right|\,0.3$\,\%& $\left|\alpha - 2\right|\,0.2$\,\%\cr
\noalign{\vskip 8pt}
\omit\hfil Relative\hfil \cr
\noalign{\vskip 3pt}
Statistical/algorithmical errors\tablefootmark{b} [$\muK_\mathrm{CMB}$\,pixel$^{-1}$]& All sky& 4.3& 4.7& 6.5\cr
Known systematics\tablefootmark{c}& All sky & 0.1\,\% & 0.1\,\% & 0.1\,\%\cr
Unknown systematics\tablefootmark{d} [$\muK_\mathrm{CMB}$\,pixel$^{-1}$] & CMB areas & $<*8.8$ & $<5.2$ & $<*9.5$\cr
\omit & Galactic region& $<17.0$& $<9.8$& $<13.1$\cr
Unknown systematics\tablefootmark{e} ($50<\ell<250$)& All sky& 0.2\,\%& 0.2\,\%& 0.1\,\%\cr
\noalign{\vskip 8pt}
\omit\hfil Total\hfil \cr
\noalign{\vskip 3pt}
CMB areas\tablefootmark{f} [$\muK_\mathrm{CMB}$\,pixel$^{-1}$]& & $<*8.5$ & $<*7.1$ & $<*8.2$\cr
Galactic region\tablefootmark{f} [$\muK_\mathrm{CMB}$\,pixel$^{-1}$]& & $<38.5$& $<13.7$& $<16.8$\cr
Sum of absolute and relative errors\tablefootmark{g}& All sky& 0.82\,\%& 0.55\,\%& 0.62\,\%\cr 
\noalign{\vskip 5pt\hrule\vskip 3pt}}}
\endPlancktablewide
\tablenote a Error on the estimation of the calibration constant $K_\mathrm{dip}$ (Eq.~\ref{eq:dipoleFit}).\par
\tablenote b Peak-to-peak differences, as reported in \citet{planck2013-p02a}.\par
\tablenote c Error in the estimation of the calibration constant $K$.\par
\tablenote d Scaled rms\ of the value of pixels in odd-odd survey half-difference maps minus the rms\ due to statistical noise. We used here $N_\mathrm{side}=128$ for 30 and 44\,GHz, and $N_\mathrm{side}=256$ for 70\,GHz.\par
\tablenote e Estimated from inter-channel comparisons on the cross-spectra.\par
\tablenote f Scaled rms\ of the value of pixels in odd-even survey half-difference maps minus the rms\ due to statistical noise. As above, we used here $N_\mathrm{side}=128$ for 30 and 44\,GHz, and $N_\mathrm{side}=256$ for 70\,GHz.\par
\tablenote g Sum of the error from the standard uncertainty and the square root of the squared sum of the following errors: (1) beam uncertainty; (2) sidelobe convolution effect; and (3) unknown systematics (as measured from the power spectrum at $50 < \ell < 250$).\par
\endgroup
\end{table*}

In this section we present results of consistency checks sensitive to calibration inaccuracies, as well as simulations which quantify the estimated level of calibration systematics in the data.
Table~\ref{tab:accuracyBudgetOverview} shows an overview of the elements that have been taken into account in estimating the calibration accuracy for LFI maps. Table~\ref{tab:accuracyBudgetResult} quantifies the impact of each element in the overall accuracy budget for the calibration.

%% file: P02b_5_1_noise_and_gain.tex
\subsubsection{Propagation of statistical errors from {\it WMAP}'s dipole estimate}
\label{sec:WMAPerrorPropagation}

As stated in Sect.~\ref{sec:beamEfficiencyAndWindowFunctions}, we have calibrated LFI data using the CMB dipole as measured by {\it WMAP} \citep{jarosik2010}. In our calibration we used both the speed of the Solar System in the CMB rest frame ($v_\mathrm{Sun}$ in Table~\ref{tab:monopoleAndDipoleParameters}) as well as the direction of motion ($l$, $b$). In this section we estimate how the quoted errors on these parameters propagate through the calculation of the calibration constant $K_\mathrm{dip}$ in Eq.~\ref{eq:dipoleFit}.

\begin{figure}
	\centering
	\includegraphics{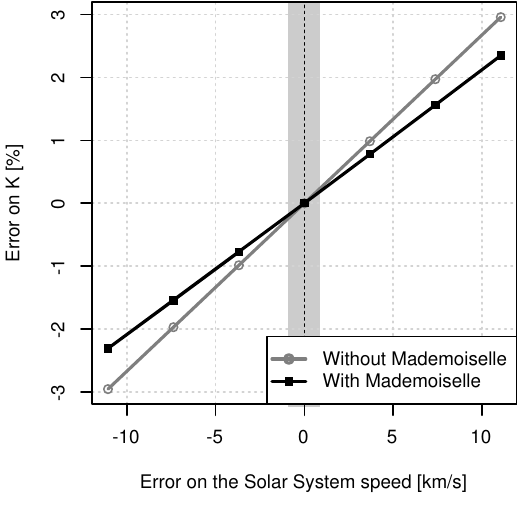}
	\includegraphics{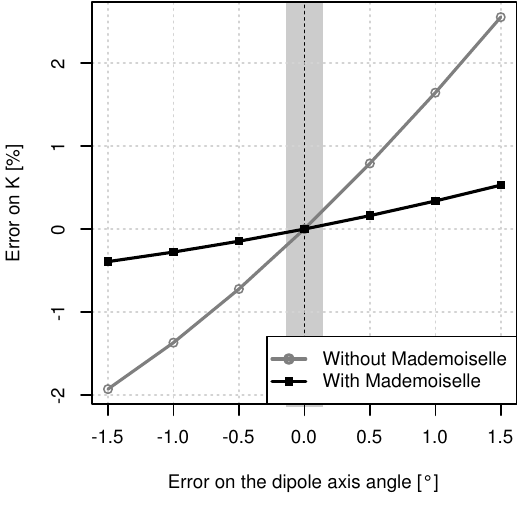}
	\caption{\label{fig:mademoiselleAndGains} Variation in the average level of the calibration constants of LFI27M, with (black line) or without (grey line) applying the \texttt{Mademoiselle} iterative code, as a function of the error in the Solar System speed (top) and the displacement of the dipole axis (bottom). The two plots are normalized so that all the lines go through the point $(0, 0)$. The width of the vertical grey band corresponds to the error quoted by \citet{jarosik2010} for $v_\mathrm{Sun}$ (top) and $l$ (bottom).}
\end{figure}

We ran the LFI calibration pipeline a number of times for a subset\footnote{Specifically, we considered LFI18M, LFI18S (70\,GHz), LFI24M, LFI24S (44\,GHz), LFI27M, and LFI28M (30\,GHz).} of radiometers; for each run we changed either $v_\mathrm{Sun}$ or $l$ from the nominal value listed in Table~\ref{tab:monopoleAndDipoleParameters}, and we compared the new average level $\left<K_\mathrm{dip}'\right>$ of the calibration constant with respect to the level $\left<K_\mathrm{dip}\right>$ of the nominal calibration. We also ran these tests without applying the iterative calibration code that cures the bias induced by the CMB and Galactic signal in dipole fitting (\texttt{Mademoiselle}; see point~\ref{enum:mademoiselle} in Sect.~\ref{sec:OSGCalibration}), in order to quantify the improvement in the stability of the calibration achieved by the iterative code.

Figure~\ref{fig:mademoiselleAndGains} shows the results of our simulations in the case of LFI27M (results for the other radiometers are in nearly perfect agreement with this). \texttt{Mademoiselle}'s ability to reduce the discrepancy $\left<K_\mathrm{dip}'\right> - \left<K_\mathrm{dip}\right>$ is clear if one compares the error on $K_\mathrm{dip}$ with and without using \texttt{Mademoiselle}. Considering {\it WMAP}'s uncertainties on $v_\mathrm{Sun}$ and $l$ and the results of our simulations, we can estimate that the relative error $\sigma_{K_\mathrm{dip}} / K_\mathrm{dip}$ due to the uncertainties in $v_\mathrm{Sun}$, $l$, and $b$ is
\begin{equation}
\begin{split}
\frac{\sigma_{K_\mathrm{dip}}}{K_\mathrm{dip}} &= 0.19\,\%\textrm{ ($v_\mathrm{Sun}$)} + 0.05\,\%\textrm{ ($l$)} + 0.01\,\%\textrm{ ($b$)} \\
&= 0.25\,\%,
\end{split}
\end{equation}
where the error for $b$ has been estimated by simple proportionality using the error on $l$. (These numbers are the same for all the radiometers we analysed.)

\subsubsection{Simulation of noise in dipole fitting}

We have run a set of simulations to assess how much the statistical noise in the radiometric signal affects gain reconstruction using the CMB dipole signal. We first decalibrate the input map or dipole with a fiducial gain, the 4\,K calibration, and then we simulate how the dipole calibration pipeline reconstructs those gains, i.e., we add the statistical noise, cut the low-dipole regions and then apply a smoothing filter.

We simulated separately the impact of the statistical noise acting on the modelled dipole and the Galaxy/CMB contribution; we model the input dipole using the {\it WMAP} direction and amplitude for the cosmological dipole (see Table~\ref{tab:monopoleAndDipoleParameters}) and NASA JPL Horizons \Planck{} velocity for the orbital dipole. Moreover, we used \Planck{} full frequency maps as the input Galactic and dipole-removed CMB signal.

We use a fiducial gain in order to make our simulation as realistic as possible, including the impact of the calibration pipeline on the underlying ``true'' gain. The 4\,K calibration (see Sect.~\ref{sec:dVV}) offers a good choice, because it also includes realistic fluctuations on shorter timescales, which are expected to be smoothed out by the dipole calibration algorithm.

After decalibrating the input signal with the fiducial gain, we add a realization of the expected statistical gain error coming from the dipole fit due to white noise and to Galaxy masking.
We model the statistical gain error according to the following equation:
\begin{equation}
\label{eq:gainErrorModel}
\frac{\sigma_{G,i}}{G_i} = \frac{\sqrt2 \sigma_0}{D_i \sqrt{\tau_i \eta_{\mathrm{mask},i}}} \sqrt{1 + \left(\frac{f_0}{f_\mathrm{knee}}\right)^\alpha},
\end{equation}
where $i$ is an index identifying a pointing period, $\sigma_{G,i}/{G_i}$ is the relative error on the gain $G_i$, $\sigma_0$ is the noise-equivalent temperature of the radiometer, $D_i$ is the amplitude of the dipole signal during pointing period $i$ (in Rayleigh-Jeans units), $\tau_i$ is the integration time of pointing period $i$, $0 \leq \eta_{\mathrm{mask},i} \leq 1$ is a dimensionless coefficient that takes into account the loss of integration time due to Galaxy masking within pointing period $i$, $f_0$ is the frequency of the dipole sinusoid seen by \Planck{}'s scanning strategy ($f_0 = 16.7\,\mathrm{mHz}$), and $f_\mathrm{knee}$ and $\alpha$ are the constants that characterize the $1/f$ noise for a given radiometer. The output of this step is equivalent to the ``raw'' dipole calibration as computed by the \texttt{Mademoiselle} iterative calibration software \citep[see Sect.~\ref{sec:OSGCalibration}, point~\ref{enum:mademoiselle}, and the][]{planck2013-p02}.

\begin{figure}
\centering
\includegraphics[width=88mm]{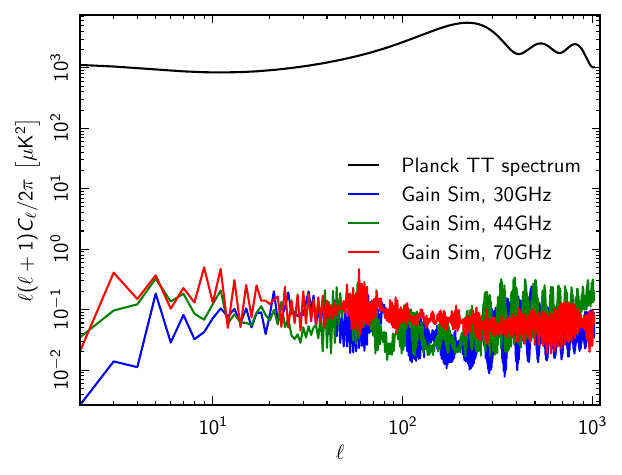}
\caption{\label{fig:gainSimulationSpectra} Power spectrum of the simulated artefacts caused by errors in gain reconstruction (see Sect.~\protect\ref{sec:noiseAndGain}), compared with \Planck's best-fit $TT$ power spectrum model.}
\end{figure}

The last step is to cut the low-dipole regions, replace them with a straight line, and then heavily smooth the data with a moving average filter of about 200 pointing periods. This is a simplified version of the wavelet-based filtering in the pipeline (Sect.~\ref{sec:OSGCalibration}, point~\ref{enum:waveletSmoothingFilter}). We run destriping on the output timelines and produce surveys and nominal mission maps of the noise for each frequency. Figure~\ref{fig:gainSimulationSpectra} shows the power spectrum of the artefacts produced by this simulation for the same timespan covered by this \Planck{} data release.

%% file: P02b_5_2_adc_effect.tex
\subsubsection{ADC non-linearities}
\label{sec:ADCnonLinearitySystematics}
We have run a set of simulations on the non-linearity of the ADCs (see Sect.~\ref{sec:ADCNonIdealities}) in order to quantify its impact on the calibration of LFI. The simulations are described by \citet{planck2013-p02a}, but we briefly recall the procedure here.

After we characterized the non-linearity of the ADCs to estimate the response curve $R(V'_i)$ of the ADC for the 44 converters, we produced time-ordered data by summing Galactic, dipole, and CMB signals, as well as $1/f$ noise. To do this we considered \Planck's scanning strategy and the measured characteristics of each LFI detector. We de-calibrated the TODs using the same gains used to calibrate flight data. We then produced two sets of maps:
\begin{enumerate}
\item the first set uses these TODs and produces maps, which are immune from the ADC effects;
\item the second set filters the TODs using the inverse of the response function $R$ in order to simulate the ADC non-linearity, and then uses the same procedure used for flight data to estimate a new response function, $R'$ (obviously, $R \approx R'$); finally the ADC correction code is applied to the data, and new maps are produced.
\end{enumerate}
We considered the difference between the two sets of maps to be a reasonable estimate of the level of errors due to the uncertainties in the procedure for estimating ADC non-linearities. These differences reveal that such uncertainties introduce an error in the estimation of the calibration constant $K$ (Eq.~\ref{eq:calibrationEquation}) which is of the order of $0.1\,\%$.

\subsubsection{Beam uncertainties}
\label{sec:beamUncertainties}

In Table~\ref{tab:accuracyBudgetResult} we quantify the error due to the fact that we considered \emph{monochromatic} beams when estimating the vector $\Svector$ in Eq.~\ref{eq:radiometerS}. We have estimated the beam response of LFI's radiometers (see Sect.~\ref{sec:fourPiConvolver}) over their bandwidth using the {\tt GRASP} software \citep[see][for more information about this]{planck2013-p02a} and we have computed a new set of vectors $\Svector'(\nu)$, which of course depend on the frequency $\nu$. The errors we report under ``Beam uncertainties'' are equal to the average of the value $\left\|\Svector\right\| - \bigl\|\left<\Svector'(\nu)\right>\bigr\|$ for the 22 radiometers, grouped according to their centre frequency.

%% file: P02b_5_3_1_interchannel_calibration.tex
A sensitive way to test the relative calibration among the different LFI radiometers  is to make independent maps and compare their angular power spectra in the vicinity of the CMB acoustic peak.  This has the advantage of testing the calibration on a source with the relevant amplitude and spectrum for our cosmology goals.  To reduce the effects of noise bias on the power spectrum estimation, we intercompare the radiometers using cross-spectra; since the noise is quite uncorrelated among the radiometers, we obtain clean spectra this way. In order to compare any pair of radiometers, we first mask the Galaxy, point sources, and all unobserved pixels. We then calculate the cross-spectrum of maps made by each radiometer with maps of a third radiometer, and take the ratio. By comparing all possible combinations we average over any small impact from the choice of the third map. For example:
\begin{equation}
\label{eq:xspectraEquation}
A_{\mathrm{LFI18S},\mathrm{LFI19M}} = \dfrac{C_l^{\mathrm{LFI18S} \times \mathrm{other}}}{C_l^{\mathrm{LFI19M} \times \mathrm{other}}}.
\end{equation}
In this case, ``other'' runs over LFI18M, LFI18S, and both radiometers of horns 20, 21, 22, and 23.

\begin{figure}[tbf]
\centering
\includegraphics[width=88mm]{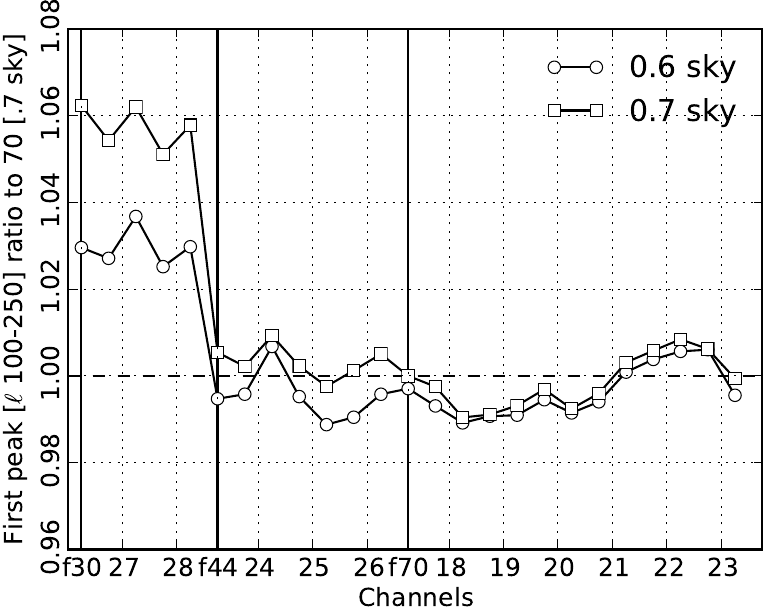}
\caption{Intercomparison of the relative calibration of the LFI radiometers, as measured on the CMB acoustic peak by averaging the power spectrum from $100 < \ell < 250$. Note that we display both main arm  (slightly to the left of each dotted, vertical line) and side arm (to the right) radiometers for each horn, as they are all calibrated independently. We also include the full frequency maps, labelled as $f30$, $f44$, and $f70$. LFI 70\,GHz full frequency at 70\,\% sky coverage is taken as reference.}
\label{fig:70ghz_xspectrum_gain_comparison}
\end{figure}

We show the results of this in Fig.~\ref{fig:70ghz_xspectrum_gain_comparison}, which captures the relative gain deviation among all the radiometers, based on the first acoustic peak. The horizontal axis is the horn number, with points representing the main and side arm radiometers placed just left and just right of the corresponding horn number.  Figure~\ref{fig:70ghz_xspectrum_gain_comparison} also includes the gain deviations averaged over the full frequency, labelled as $f30$, $f44$ and $f70$.
Interchannel consistency is everywhere within 1\,\% of the mean at 44 and 70\,GHz, while 30\,GHz is severely affected by residual Galactic signal, as highlighted by the strong reduction between the 70\,\% and 60\,\% sky coverage masks\footnote{These masks are the same used for component separation \citep{planck2013-p06}.}. This  analysis includes corrections for the beams, the scanning strategy and mask deconvolution. In this figure, the deviations from the mean calibration appear to be correlated among radiometers. Most likely this is due to the fact that the main source of systematic errors in our calibration comes from our knowledge of the beam window functions. Since the radiometer labeling is related to the focal plane layout, there may well be a correlation between mis-estimated beam shape and radiometer number. Such systematic errors in beam estimation can come from small errors in the optical element positions or mirror figures. Improved
beam models are a primary goal for future releases of the Planck data.

\begin{figure}[tbf]
\centering
\includegraphics[width=88mm]{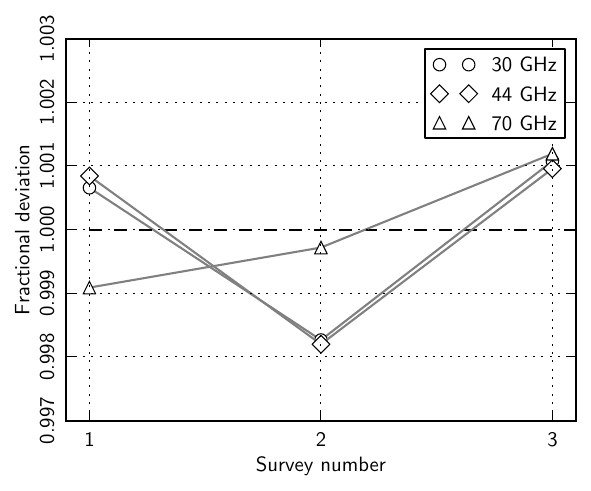}
\caption{LFI cross-spectra between surveys. As in Fig.~\protect\ref{fig:70ghz_xspectrum_gain_comparison}, ratios of the power spectra are computed in the range $50 < \ell < 250$.} 
\label{fig:xspectrum_gain_comparison_survey}
\end{figure}

Figure~\ref{fig:xspectrum_gain_comparison_survey} shows the results of a similar comparison between surveys, in this case comparing the full frequency estimates of the acoustic peak for surveys 1, 2, and 3.  The full frequency comparison is even more striking, with each band remaining within $\pm 0.15\,\%$  over the three independent surveys. Of course, many systematics which plague calibration may be common to all channels, particularly within a frequency cohort; however, $1/f$ noise, receiver degradation, and certain other sources of systematic uncertainties in the temperature calibration seem to be controlled to below 0.15\,\% .

%% file: P02b_5_3_2_null_tests.tex
We have developed a number of tests to highlight possible issues related to instrumental systematic effects which are not properly corrected within the pipeline, or are related to known changes in the operating conditions of the instrument or to intrinsic instrument properties coupled with the sky, like stray-light from sidelobes. Such tests are also discussed by \citet{planck2013-p02a}; here we concentrate on how those tests can be used to estimate the impact of calibration errors in LFI maps.

The idea behind the many tests we ran for this \Planck{} data release is to compare pairs of distinct estimates of the sky signal, that should \emph{ideally} be the same, such as:
\begin{itemize}
\item comparing the sky signal measured during two surveys (not necessarily consecutive), we call this kind of test \emph{survey-difference null tests};
\item comparing the sky signal measured by two radiometers, we call this a \emph{radiometer-difference null test};
\item comparing the map produced using only data taken in the first half  of each pointing period with data taken in the other half. This is a \emph{half-ring null test}.
\end{itemize}
Clearly, it is not trivial to determine the cause of each discrepancy in each comparison. The following criteria are useful in interpreting the results of these null tests:
\begin{enumerate}
\item The results of radiometer-difference null tests can reveal artefacts due to imbalances in the frequency response, as well as to their different optical systematics. Such optical effects can be caused by an error in pointing reconstruction (which is therefore unrelated to calibration), but also by two different ways of picking up the astrophysical signal (including the dipole) through the sidelobes. The latter effect can be critical for calibration, as the subtraction of an improper dipole signal leads to unwanted residuals in the map.

\item Survey-difference null tests are not affected by frequency response mismatches. However, they can be affected by the optical effects described above. Because of the scanning strategy employed by the \Planck{} spacecraft, such systematics will be present only if an odd number survey is compared with an even survey, or \textit{vice versa}, since odd surveys share the same scanning strategy, and the same applies for even surveys. (This is true for the first four surveys, including those considered here, but not for subsequent surveys, where the precession phase angle was changed.)

\item Half-ring null tests are able to spot systematics at high frequency (more than $1/60\,\mathrm{Hz}$, as this is the spinning frequency of the \Planck{} spacecraft). Thus, calibration errors are unlikely to affect this kind of test, since we expect the gain of the radiometers to change on much longer time scales (days or weeks). In fact, we expect half-ring null test to be sensitive only to the level of unavoidable white noise in the data.
\end{enumerate}

\begin{figure}
\centering
\includegraphics[width=88mm]{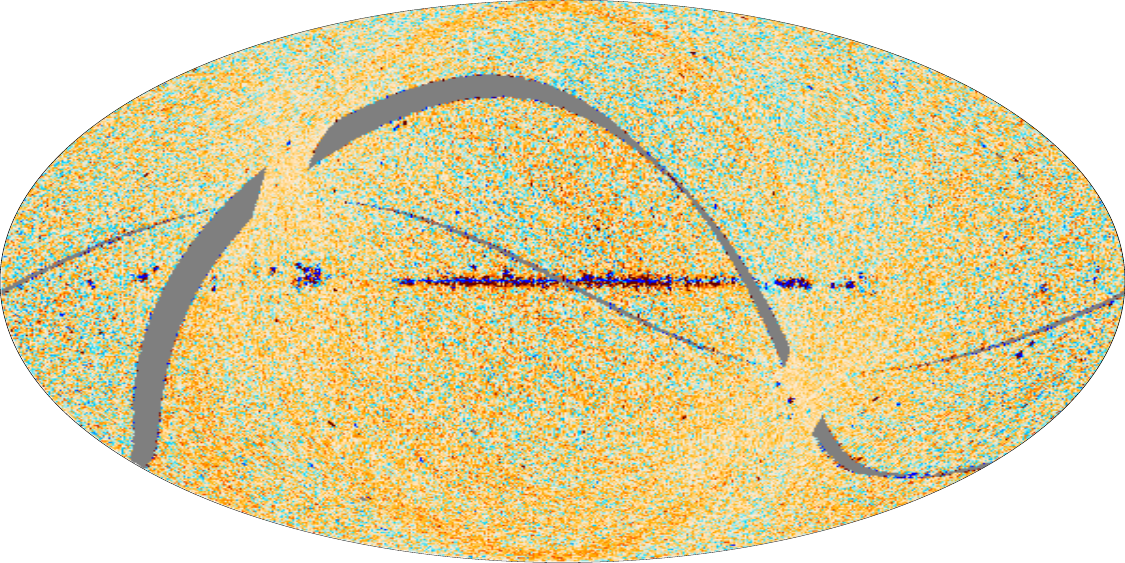}
{
\setlength{\fboxrule}{0pt}
\fbox{\includegraphics[width=88mm]
{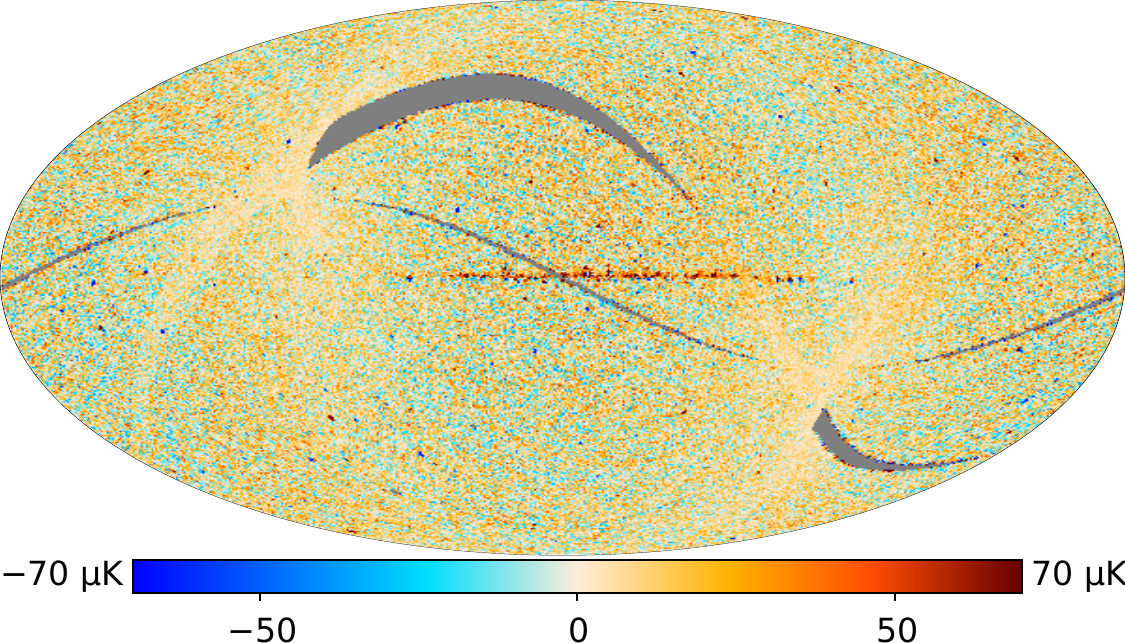}}
\fbox{\includegraphics[width=88mm]{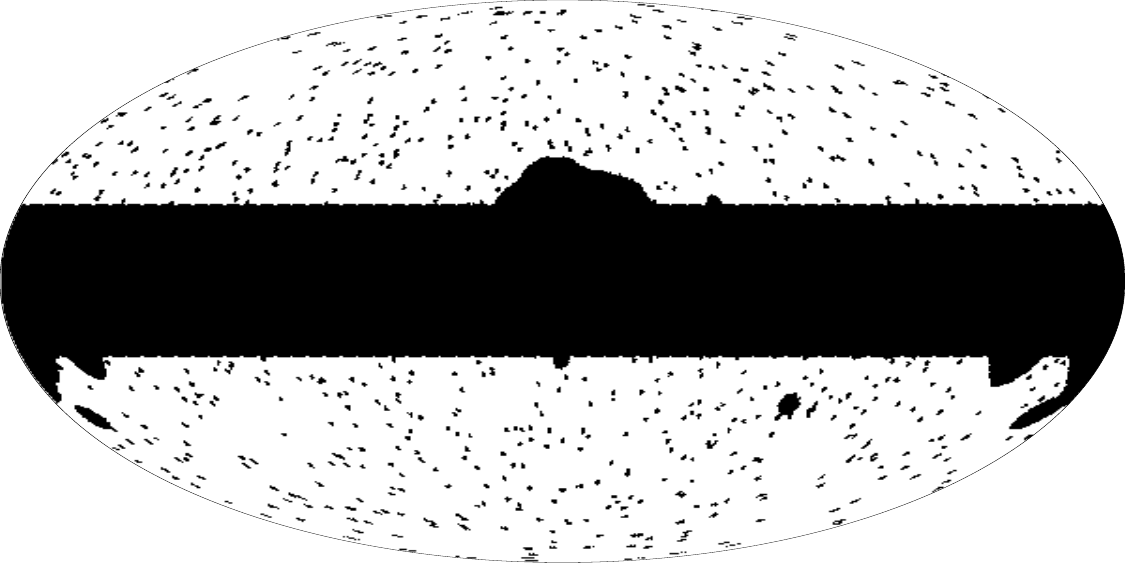}}
}
\caption{\label{fig:surveyDifferenceMaps} Survey difference maps at 30\,GHz. {\it Top}: difference between survey 1 and 2. {\it Middle}: difference between survey 1 and 3. Note that the 1-2 difference map has sharper features, due to the different scanning strategy between odd and even surveys that prevents optical systematics from being cancelled in the subtraction. {\it Bottom}: the mask used to distinguish between CMB-dominated regions and Galactic regions when computing the numbers reported in Table~\protect\ref{tab:accuracyBudgetResult} (30\,GHz case).}
\end{figure}

For all the reasons stated above, we believe that a fair upper limit of the calibration errors that are unrelated to optical effects can be deduced by studying odd-odd and even-even survey-difference null tests, as they are not affected by pointing errors, imbalance in frequency response, or ``fast'' systematics. Even-odd and odd-even differences are affected by systematics caused by optics as well. Therefore, these should be larger than the former and can also be used to produce upper limits for the calibration accuracy. Figure~\ref{fig:surveyDifferenceMaps} shows two examples. 

We have used odd-odd survey difference maps to produce the numbers reported in the row ``Unknown systematics'' in Table~\ref{tab:accuracyBudgetResult}. These numbers are the rms\ value of the pixels, after the maps have been degraded to a resolution which roughly coincides with the FWHM of the LFI beams. We subtracted from these values the part due to statistical noise, estimated from half-ring difference maps using the procedure described by \citet{planck2013-p02a}. As noted above, these rms\ values represent upper limits for calibration inaccuracy, because some of the systematics affecting survey difference maps are unrelated to calibration. Values under the ``Total'' group have been calculated from odd-even survey difference maps; as noted above, they include optical systematics as well.

\begin{figure*}
	\centering
	\begin{DIFnomarkup}
	\input{null_sims_halfrings.tex}
	\end{DIFnomarkup}
	\caption{\label{fig:nullTestSpectra} Spectrum of the half-difference between Sky Survey (SS) 1 and 2, $(\mathrm{SS1}-\mathrm{SS2})/2$ (odd-even number surveys, shown in frame A), and $(\mathrm{SS1}-\mathrm{SS3})/2$ (odd-odd, frame B), compared with the half ring spectra for the same time period at the three LFI bands. We also show the the simulated errors due to gain reconstruction (Sect.~\protect\ref{sec:noiseAndGain}). Spectra have been binned into five samples each, in order to remove high-frequency noise and highlight the general trend. At high multipoles ($\ell > 20$) there is a good match between survey differences and half-rings, which means that discrepancies between survey differences are mainly due to statistical noise. Note that the biggest discrepancy between survey differences and halfrings happens for 30\,GHz channels when comparing SS1 with SS2 (odd-even); this suggests that optical effects (sidelobes) are causing the discrepancy. Gain errors are calculated over the whole mission and are negligible except for the lowest multipoles of 44\,GHz and 70\,GHz spectra, where they probably account for the large-scale residuals found in survey difference maps.}
\end{figure*}
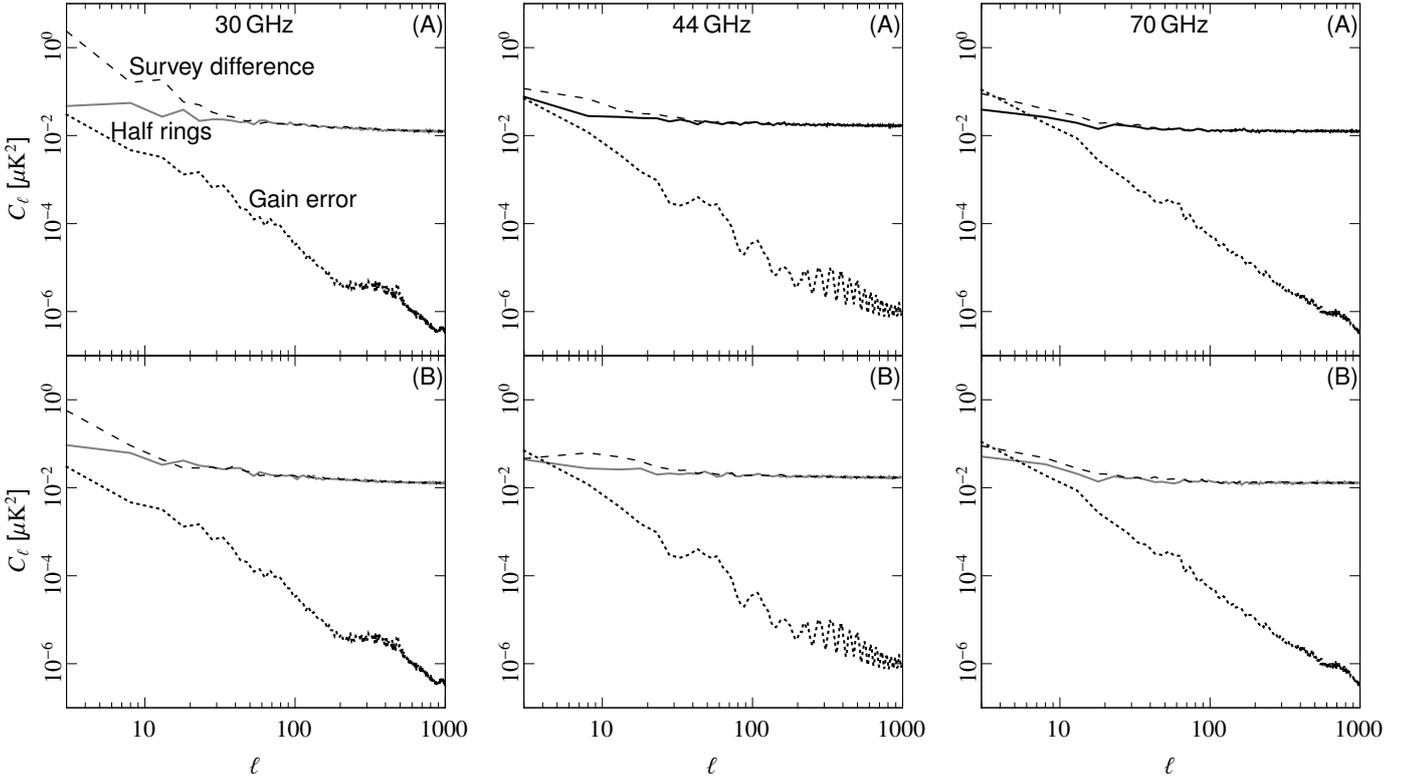

We show the power spectra of the even-odd and odd-odd null tests in Fig.~\ref{fig:nullTestSpectra}, where we compare them with the estimated statistical errors of the gains (Sect.~\ref{sec:noiseAndGain}) and the spectra of half-ring maps.  The latter provide an estimate of the level of statistical noise in the maps and generally agree with survey difference spectra for $\ell > 20$. This means that for a wide range of multipoles we do not expect any significant contribution of calibration systematics on the maps. For $\ell < 20$ there is some disagreement for the 30\,GHz maps. This might indicate the presence of residual optical systematics in the data. In fact, the chief reason that induced us to use a different calibration for 30\,GHz channels (using the 4\,K total-power output, see Sect.~\ref{sec:dVV}) was to minimize the impact of optical systematics on the calibration of these channels (which suffer from a relatively high level of sidelobes).

%% file: null_sims_halfrings.tex
% GNUPLOT: LaTeX picture with Postscript
\begingroup
  \fontfamily{phv}%
  \selectfont
  \makeatletter
  \providecommand\color[2][]{%
    \GenericError{(gnuplot) \space\space\space\@spaces}{%
      Package color not loaded in conjunction with
      terminal option `colourtext'%
    }{See the gnuplot documentation for explanation.%
    }{Either use 'blacktext' in gnuplot or load the package
      color.sty in LaTeX.}%
    \renewcommand\color[2][]{}%
  }%
  \providecommand\includegraphics[2][]{%
    \GenericError{(gnuplot) \space\space\space\@spaces}{%
      Package graphicx or graphics not loaded%
    }{See the gnuplot documentation for explanation.%
    }{The gnuplot epslatex terminal needs graphicx.sty or graphics.sty.}%
    \renewcommand\includegraphics[2][]{}%
  }%
  \providecommand\rotatebox[2]{#2}%
  \@ifundefined{ifGPcolor}{%
    \newif\ifGPcolor
    \GPcolorfalse
  }{}%
  \@ifundefined{ifGPblacktext}{%
    \newif\ifGPblacktext
    \GPblacktexttrue
  }{}%
  % define a \g@addto@macro without @ in the name:
  \let\gplgaddtomacro\g@addto@macro
  % define empty templates for all commands taking text:
  \gdef\gplbacktext{}%
  \gdef\gplfronttext{}%
  \makeatother
  \ifGPblacktext
    % no textcolor at all
    \def\colorrgb#1{}%
    \def\colorgray#1{}%
  \else
    % gray or color?
    \ifGPcolor
      \def\colorrgb#1{\color[rgb]{#1}}%
      \def\colorgray#1{\color[gray]{#1}}%
      \expandafter\def\csname LTw\endcsname{\color{white}}%
      \expandafter\def\csname LTb\endcsname{\color{black}}%
      \expandafter\def\csname LTa\endcsname{\color{black}}%
      \expandafter\def\csname LT0\endcsname{\color[rgb]{1,0,0}}%
      \expandafter\def\csname LT1\endcsname{\color[rgb]{0,1,0}}%
      \expandafter\def\csname LT2\endcsname{\color[rgb]{0,0,1}}%
      \expandafter\def\csname LT3\endcsname{\color[rgb]{1,0,1}}%
      \expandafter\def\csname LT4\endcsname{\color[rgb]{0,1,1}}%
      \expandafter\def\csname LT5\endcsname{\color[rgb]{1,1,0}}%
      \expandafter\def\csname LT6\endcsname{\color[rgb]{0,0,0}}%
      \expandafter\def\csname LT7\endcsname{\color[rgb]{1,0.3,0}}%
      \expandafter\def\csname LT8\endcsname{\color[rgb]{0.5,0.5,0.5}}%
    \else
      % gray
      \def\colorrgb#1{\color{black}}%
      \def\colorgray#1{\color[gray]{#1}}%
      \expandafter\def\csname LTw\endcsname{\color{white}}%
      \expandafter\def\csname LTb\endcsname{\color{black}}%
      \expandafter\def\csname LTa\endcsname{\color{black}}%
      \expandafter\def\csname LT0\endcsname{\color{black}}%
      \expandafter\def\csname LT1\endcsname{\color{black}}%
      \expandafter\def\csname LT2\endcsname{\color{black}}%
      \expandafter\def\csname LT3\endcsname{\color{black}}%
      \expandafter\def\csname LT4\endcsname{\color{black}}%
      \expandafter\def\csname LT5\endcsname{\color{black}}%
      \expandafter\def\csname LT6\endcsname{\color{black}}%
      \expandafter\def\csname LT7\endcsname{\color{black}}%
      \expandafter\def\csname LT8\endcsname{\color{black}}%
    \fi
  \fi
  \setlength{\unitlength}{0.0500bp}%
  \begin{picture}(10200.00,6220.00)%
    \gplgaddtomacro\gplbacktext{%
      \put(438,3904){\rotatebox{-270}{\makebox(0,0){\strut{}$10^{-6}$}}}%
      \put(438,4565){\rotatebox{-270}{\makebox(0,0){\strut{}$10^{-4}$}}}%
      \put(438,5227){\rotatebox{-270}{\makebox(0,0){\strut{}$10^{-2}$}}}%
      \put(438,5888){\rotatebox{-270}{\makebox(0,0){\strut{}$10^{0}$}}}%
      \put(1098,3387){\makebox(0,0){\strut{}}}%
      \put(2218,3387){\makebox(0,0){\strut{}}}%
      \put(3338,3387){\makebox(0,0){\strut{}}}%
      \csname LTb\endcsname%
      \put(168,4896){\rotatebox{-270}{\makebox(0,0){\strut{}$C_\ell$ [$\mu\mathrm{K}^2$]}}}%
      \put(850,5227){\makebox(0,0)[l]{\strut{}Half rings}}%
      \put(990,5715){\makebox(0,0)[l]{\strut{}Survey difference}}%
      \put(3338,6033){\makebox(0,0)[r]{\strut{}(A)}}%
      \put(1926,6033){\makebox(0,0){\strut{}30\,GHz}}%
      \put(1881,4723){\makebox(0,0)[l]{\strut{}Gain error}}%
    }%
    \gplgaddtomacro\gplfronttext{%
    }%
    \gplgaddtomacro\gplbacktext{%
      \colorrgb{0.50,0.50,0.50}%
      \put(438,1257){\rotatebox{-270}{\makebox(0,0){\strut{}$10^{-6}$}}}%
      \put(438,1918){\rotatebox{-270}{\makebox(0,0){\strut{}$10^{-4}$}}}%
      \put(438,2580){\rotatebox{-270}{\makebox(0,0){\strut{}$10^{-2}$}}}%
      \put(438,3241){\rotatebox{-270}{\makebox(0,0){\strut{}$10^{0}$}}}%
      \put(1098,740){\makebox(0,0){\strut{}$10$}}%
      \put(2218,740){\makebox(0,0){\strut{}$100$}}%
      \put(3338,740){\makebox(0,0){\strut{}$1000$}}%
      \csname LTb\endcsname%
      \put(168,2249){\rotatebox{-270}{\makebox(0,0){\strut{}$C_\ell$ [$\mu\mathrm{K}^2$]}}}%
      \put(1925,461){\makebox(0,0){\strut{}$\ell$}}%
      \put(3338,3386){\makebox(0,0)[r]{\strut{}(B)}}%
    }%
    \gplgaddtomacro\gplfronttext{%
    }%
    \gplgaddtomacro\gplbacktext{%
      \colorrgb{0.50,0.50,0.50}%
      \put(3850,3904){\rotatebox{-270}{\makebox(0,0){\strut{}$10^{-6}$}}}%
      \put(3850,4565){\rotatebox{-270}{\makebox(0,0){\strut{}$10^{-4}$}}}%
      \put(3850,5227){\rotatebox{-270}{\makebox(0,0){\strut{}$10^{-2}$}}}%
      \put(3850,5888){\rotatebox{-270}{\makebox(0,0){\strut{}$10^{0}$}}}%
      \put(4510,3387){\makebox(0,0){\strut{}}}%
      \put(5630,3387){\makebox(0,0){\strut{}}}%
      \put(6750,3387){\makebox(0,0){\strut{}}}%
      \put(6750,6033){\makebox(0,0)[r]{\strut{}(A)}}%
      \put(5338,6033){\makebox(0,0){\strut{}44\,GHz}}%
    }%
    \gplgaddtomacro\gplfronttext{%
    }%
    \gplgaddtomacro\gplbacktext{%
      \put(3850,1257){\rotatebox{-270}{\makebox(0,0){\strut{}$10^{-6}$}}}%
      \put(3850,1918){\rotatebox{-270}{\makebox(0,0){\strut{}$10^{-4}$}}}%
      \put(3850,2580){\rotatebox{-270}{\makebox(0,0){\strut{}$10^{-2}$}}}%
      \put(3850,3241){\rotatebox{-270}{\makebox(0,0){\strut{}$10^{0}$}}}%
      \put(4510,740){\makebox(0,0){\strut{}$10$}}%
      \put(5630,740){\makebox(0,0){\strut{}$100$}}%
      \put(6750,740){\makebox(0,0){\strut{}$1000$}}%
      \csname LTb\endcsname%
      \put(5337,461){\makebox(0,0){\strut{}$\ell$}}%
      \put(6750,3386){\makebox(0,0)[r]{\strut{}(B)}}%
    }%
    \gplgaddtomacro\gplfronttext{%
    }%
    \gplgaddtomacro\gplbacktext{%
      \colorrgb{0.50,0.50,0.50}%
      \put(7263,3904){\rotatebox{-270}{\makebox(0,0){\strut{}$10^{-6}$}}}%
      \put(7263,4565){\rotatebox{-270}{\makebox(0,0){\strut{}$10^{-4}$}}}%
      \put(7263,5227){\rotatebox{-270}{\makebox(0,0){\strut{}$10^{-2}$}}}%
      \put(7263,5888){\rotatebox{-270}{\makebox(0,0){\strut{}$10^{0}$}}}%
      \put(7923,3387){\makebox(0,0){\strut{}}}%
      \put(9043,3387){\makebox(0,0){\strut{}}}%
      \put(10162,3387){\makebox(0,0){\strut{}}}%
      \put(10162,6033){\makebox(0,0)[r]{\strut{}(A)}}%
      \put(8750,6033){\makebox(0,0){\strut{}70\,GHz}}%
    }%
    \gplgaddtomacro\gplfronttext{%
    }%
    \gplgaddtomacro\gplbacktext{%
      \put(7263,1257){\rotatebox{-270}{\makebox(0,0){\strut{}$10^{-6}$}}}%
      \put(7263,1918){\rotatebox{-270}{\makebox(0,0){\strut{}$10^{-4}$}}}%
      \put(7263,2580){\rotatebox{-270}{\makebox(0,0){\strut{}$10^{-2}$}}}%
      \put(7263,3241){\rotatebox{-270}{\makebox(0,0){\strut{}$10^{0}$}}}%
      \put(7923,740){\makebox(0,0){\strut{}$10$}}%
      \put(9043,740){\makebox(0,0){\strut{}$100$}}%
      \put(10162,740){\makebox(0,0){\strut{}$1000$}}%
      \csname LTb\endcsname%
      \put(8750,461){\makebox(0,0){\strut{}$\ell$}}%
      \put(10162,3386){\makebox(0,0)[r]{\strut{}(B)}}%
    }%
    \gplgaddtomacro\gplfronttext{%
    }%
    \gplbacktext
    \put(0,0){\includegraphics{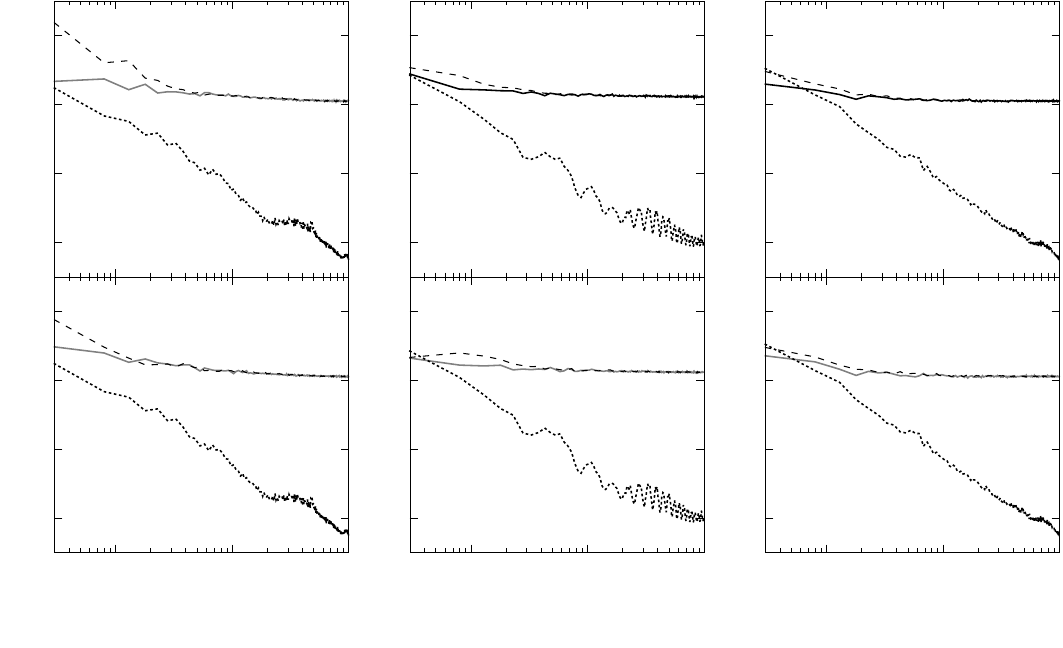}}%
    \gplfronttext
  \end{picture}%
\endgroup

%% file: P02b_5_4_1_source_fluxes.tex
% Figures
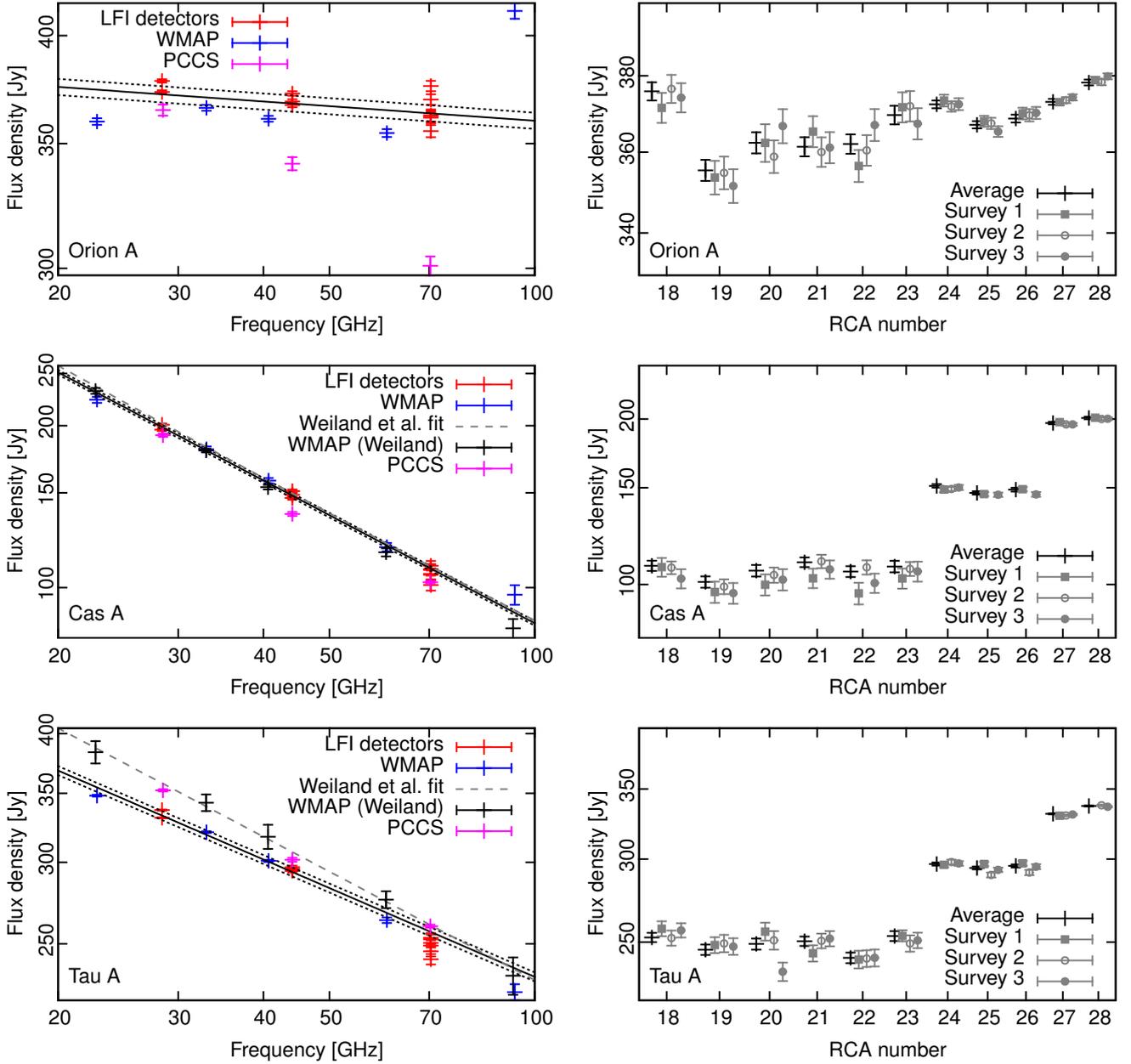
\begin{figure*}[tb]
\centering
\begin{DIFnomarkup}
\input{source_flux_plots.tex}
\end{DIFnomarkup}
\caption{Spectra (left) and per-survey and per-RCA measurements (right) for Orion A (top), Cas A (middle) and Tau A (bottom). For the spectra, the flux densities measured using aperture photometry are shown for LFI and {\it WMAP}, as well as the best-fitting power-law to the measurements  (solid line) and a $\pm$1\,\% range (dotted line). The fit and data points from \citet{weiland2010} are shown for Cas A and Orion A, and the Planck Catalogue of Compact Sources (PCCS) flux densities for all three sources are shown. For the per-RCA measurements, RCAs 18--23 are at 70\,GHz, 24--26 at 44\,GHz and 27--28 at 30\,GHz. For each RCA, the average from the three surveys is shown, followed by the measurements from each survey.}
\label{fig:sources}
\end{figure*}

In order to perform an independent check of the relative calibration between different RCAs and also to test both the colour corrections and central frequencies of each LFI band, we look at three bright radio sources in the LFI maps: Orion~A (M42); Taurus~A; and Cassiopeia~A.\footnote{Orion~A and Tau~A are not expected to vary significantly over the time scales relevant here. Cas~A has a secular decrease of $\sim 0.6\,\%$ at high frequencies \citep{1999osullivan}, which over the five year time difference between the averaged {\it WMAP} and \Planck{} datasets will add to 3\,\%. This has not been taken into account in this analysis; however, it has a negligible effect on the measured spectral index of the source that is well within the uncertainties.}. The emission in these three sources is due to synchrotron or free-free emission, such that they have very different spectra from the CMB. This means that colour corrections are important at the level of about 3--10\,\% depending on the RCA, which is in excess of the uncertainties on the flux densities of these sources.

Comparisons between ERCSC flux densities \citep{planck2011-1.10} at 30, 44, and 70\GHz\ and interpolated or extrapolated flux densities from the Karl G.~Jansky Very Large Array (VLA) follow-up observations have previously shown good statistical consistency at the 3\,\% level \citep{planck2011-6.2,Kurinsky2012,planck2013-p05}. A comparison has also been made between the \Planck-ATCA Coeval Observations (PACO) and PCCS catalogues in \citet{planck2013-p05}, which finds consistency at the level of around 4\,\%. However, we ideally want to test the calibration much more accurately than this, which can be accomplished comparing different LFI measurements and by comparing these results to {\it WMAP}. Additionally, the sources that we examine here are significantly less variable than the typical source in the VLA and PACO tests.

We use the \Planck{} single detector maps for each survey. We first average the main and side maps to create individual RCA maps per survey, which has the effect of cancelling out the polarization such that the maps are of total intensity only. We look at the individual survey maps, and the average of surveys 1--3. We also make use of the {\it WMAP}-9 deconvolved Stokes $I$ sky maps \citep{bennett2012}.

We measure source emission from the maps using aperture photometry, as described in \citet{planck2011-7.2}, with an aperture of 70\arcm\ radius and a background annulus of 80\arcm--100\arcm. By keeping the aperture and annuli sizes constant rather than changing them to reflect the resolution of the instrument, we ensure that we are measuring the same source extent and subtracting the same background emission at all frequencies, since the background may vary with distance due to the structure of the extended source, or background emission surrounding it. A correction factor for the source flux that falls outside of the aperture is applied as per \citet{planck2013-p05}; this factor is non-negligible for the {\it WMAP} data points. The measured emission, in $T_\mathrm{CMB}$, is converted to $T_\mathrm{RJ}$, and then into Jansky at the nominal frequency (28.4, 44.1 or 70.4\,GHz as appropriate). The uncertainty in the flux densities is calculated using the rms of the pixels in the background annulus.

% Table
\begin{table}[tmb]
\begingroup
\newdimen\tblskip \tblskip=5pt
\caption{Properties of three compact calibrators used to check the calibration: Orion A (M42); Cas A; and Tau A (M1)}
\label{tab:sources}
\nointerlineskip
\vskip -3mm
\footnotesize
\setbox\tablebox=\vbox{
   \newdimen\digitwidth 
   \setbox0=\hbox{\rm 0} 
   \digitwidth=\wd0 
   \catcode`*=\active 
   \def*{\kern\digitwidth}
   \newdimen\signwidth 
   \setbox0=\hbox{+} 
   \signwidth=\wd0 
   \catcode`!=\active 
   \def!{\kern\signwidth}
    \halign{\hfil#\hfil\tabskip 1.0em&
            \hfil#\hfil\tabskip 2pt&
            \hfil#\hfil\tabskip 2pt&
            \hfil#\hfil\tabskip 0pt\cr
    \noalign{\doubleline\vskip 2pt}
Property& Orion A& Cas A& Tau A\cr
\noalign{\vskip 4pt\hrule\vskip 6pt}
Position $(l,b)$ &  (209\pdeg01, $-$19\pdeg38)& (111\pdeg74, $-$02\pdeg14)& (184\pdeg56, $-$5\pdeg78)\cr
\noalign{\vskip 4pt\hrule\vskip 6pt}
$S_\mathrm{PCCS}^{28.4\,\mathrm{GHz}}$ [Jy]\tablefootmark{a}&  $365\pm2$ & $192.0\pm1.4$ & $352.4\pm0.9$\cr
$S_\mathrm{PCCS}^{44.1\,\mathrm{GHz}}$ [Jy]\tablefootmark{\hphantom{a}}&  $341\pm3$ & $137.0\pm1.2$ & $301.8\pm1.3$\cr
$S_\mathrm{PCCS}^{70.4\,\mathrm{GHz}}$ [Jy]\tablefootmark{\hphantom{a}}& $301\pm3$ & $102.3\pm0.5$  & $260.0\pm0.9$\cr
\noalign{\vskip 4pt\hrule\vskip 6pt}
$A$\tablefootmark{b} & $370.5\pm1.7$ & $179.9\pm0.8$ & $319.1\pm1.0$\cr
$\alpha$\tablefootmark{b} & $-0.026\pm0.014$ & $-0.668\pm0.014$ & $-0.287\pm0.013$\cr
\noalign{\vskip 4pt\hrule\vskip 6pt}
$S_\mathrm{aper}^{28.4\,\mathrm{GHz}}$ [Jy]\tablefootmark{c}& $384\pm*4$ (1.0\,\%) & $204\pm3$ (1.5\,\%) & $340\pm*4$ (1.3\,\%)\cr
$S_\mathrm{aper}^{44.1\,\mathrm{GHz}}$ [Jy]\tablefootmark{\hphantom{c}}& $373\pm*2$ (0.5\,\%) & $151\pm2$ (1.2\,\%) & $299\pm*2$ (0.5\,\%)\cr
$S_\mathrm{aper}^{70.4\,\mathrm{GHz}}$ [Jy]\tablefootmark{\hphantom{c}}& $372\pm10$ (2.7\,\%) & $110\pm4$ (3.9\,\%) & $255\pm11$ (4.2\,\%)\cr
\noalign{\vskip 4pt\hrule\vskip 6pt}
$S_\mathrm{aper,cc}^{28.4\,\mathrm{GHz}}$ [Jy]\tablefootmark{d}& $375\pm*4$ (1.0\,\%) & $198\pm3$ (1.5\,\%) & $334\pm4\phantom{.0}$ (1.2\,\%)\cr
$S_\mathrm{aper,cc}^{44.1\,\mathrm{GHz}}$ [Jy]\tablefootmark{\hphantom{d}}& $370\pm*3$ (0.7\,\%) & $149\pm2$ (1.5\,\%) & $296\pm1.2$ (0.4\,\%)\cr
$S_\mathrm{aper,cc}^{70.4\,\mathrm{GHz}}$ [Jy]\tablefootmark{\hphantom{d}}& $364\pm*7$ (2.0\,\%) & $106\pm3$ (2.8\,\%) & $249\pm4\phantom{.0}$ (1.8\,\%)\cr
\noalign{\vskip 3pt\hrule\vskip 4pt}
}}
\endPlancktable
\tablenote a Flux densities from PCCS, without colour correction.\par
\tablenote b Coefficients for the fitted power law, $S = A (\nu/33\mathrm{\,GHz})^\alpha$.\par
\tablenote c Average and standard deviation of the flux densities measured by photometry within a fixed aperture.\par
\tablenote d Average and standard deviation of the colour-corrected aperture photometry flux densities.\par
\endgroup
\end{table}

The properties of the sources are summarised in Table~\ref{tab:sources}. The flux densities for the sources in the first section of the table are the \texttt{APERFLUX} flux densities from PCCS. We only use these here for consistency checks, however, as the PCCS flux densities are the average across all radiometers of a given frequency rather than the flux densities as measured by each radiometer. We note that there is a difference in reported flux densities from {\it WMAP} between \citet{page2007} and \citet{weiland2010} of up to around 8\,\%, due to the method used: \citet{page2007} used aperture photometry, and \citet{weiland2010} fitted the beams to the sources. As such, care needs to be taken when comparing the numbers from the different analyses.

In order to calculate the necessary colour corrections, we iteratively calculate the amplitude (at a reference frequency of 33\,GHz) and spectral index of the source based on the \Planck{} and {\it WMAP} aperture photometry flux densities, the values for which are quoted in Table~\ref{tab:sources}. We then use this spectral index to colour-correct the flux densities per RCA using quadratic fits to the colour correction values given in Sect.~\ref{sec:colourCorrections}. The average flux densities, and the standard deviation calculated between the RCAs, both before and after colour correction, are given in Table~\ref{tab:sources}. The percentage of the standard deviation to the mean is also given. When the colour corrections are applied, there are a few significant improvements in the standard deviations and no relevant degradation, giving confidence that the colour corrections are improving the consistency between the RCAs.

Figure \ref{fig:sources} shows the spectra of the three sources (left-hand column) and the flux densities from individual survey maps for the different RCAs (right-hand column). For Orion A, two sets of points differ significantly from the spectrum from aperture photometry. At {\it WMAP} 93.5\,GHz, the measured flux density is much higher as it includes dust emission, which is not present at the LFI frequencies and as such is not considered here. Additionally, the PCCS aperflux is much steeper ($\alpha=-0.21$) than the best fit to the aperture photometry here; this is most likely due to the different sized apertures used, which depend on frequency for PCCS, meaning that it will systematically exclude the more diffuse emission from the source than is included in this analysis. This is particularly relevant for Orion A due to its more complex morphology. For Cas A, we find very good agreement between the results. For Tau A, we find a slightly flatter spectrum ($-0.287\pm0.013$) than the PCCS and \citet{weiland2010} ($-0.302\pm0.005$), and other analyses also report steeper spectra, e.g. \citet{Hafez2008} find $-0.32\pm0.01$ at 30\,GHz, and \citet{Perez2010} find $-0.296\pm0.006$; this is again likely due to the different aperture sizes considered here.

Although aperture size affects the comparison of the absolute flux densities, it does not affect the results between the different \Planck{} RCAs at the same frequencies, or between the different survey maps for the same RCAs. There is general consistency for all RCAs between the surveys, within the uncertainties. Note that some of the sources were not observed in every survey due to the survey definitions, for example Cas A was not observed by RCAs 25 and 26 in survey 2. However, we do find some systematic differences in the results from different RCAs. At 28.4\,GHz, RCA 27 is consistently lower than RCA 28 at the level of 2\,\%; this difference does not change when colour corrections are applied as the corrections for those two RCAs are so similar. The differences at 44\,GHz are 1.5\,\%, with RCA 24 systematically slightly higher than the 25 and 26. This might be due to the beams, as RCA 24 has a smaller beam size than RCAs 25 and 26; however, this difference is expected to be smaller than the effect seen here. At 70\,GHz, there are indications that RCA 18 is consistently high and RCA 19 is consistently low, with differences in the flux density between those two RCAs of up to 6\,\% (for Orion A). 

\begin{table}[tmb]
\begingroup
\newdimen\tblskip \tblskip=5pt
\caption{\label{tab:sourceFluxDiscrepancies} Average ratio of flux densities for RCA pairs.}
\nointerlineskip
\vskip -3mm
\footnotesize
\setbox\tablebox=\vbox{
   \newdimen\digitwidth 
   \setbox0=\hbox{\rm 0} 
   \digitwidth=\wd0 
   \catcode`*=\active 
   \def*{\kern\digitwidth}
   \newdimen\signwidth 
   \setbox0=\hbox{+} 
   \signwidth=\wd0 
   \catcode`!=\active 
   \def!{\kern\signwidth}
    \halign{\hbox to 2.5cm{#\leaderfil}\tabskip 0.5em&
            \hfil#\hfil\tabskip 5pt&
            \hfil#\hfil\tabskip 5pt&
            \hfil#\hfil\tabskip 5pt&
            \hfil#\hfil\tabskip=0pt\cr
    \noalign{\doubleline\vskip 0pt}
\omit&&&& Corrected\cr
\omit\hfil Ratio\hfil& Discrepancy\tablefootmark{a}& $\delta\cc_{\alpha=-1}$& $\delta\cc_{\alpha=0}$& discrepancy\tablefootmark{b}\cr
\noalign{\vskip 4pt\hrule\vskip 6pt}
\omit{\hspace{10pt} 70 GHz}\hfil\cr
\noalign{\vskip 2pt}
$S_\mathrm{RCA19} / S_\mathrm{RCA18}$& $-1$\,\%& *0.073& *0.049& 6\,\%\cr
$S_\mathrm{RCA20} / S_\mathrm{RCA18}$& *0\,\%& *0.047& *0.032& 4\,\%\cr
$S_\mathrm{RCA21} / S_\mathrm{RCA18}$& *0\,\%& *0.025& *0.017& 2\,\%\cr
$S_\mathrm{RCA22} / S_\mathrm{RCA18}$& *7\,\%& $-0.055$& $-0.035$& 1\,\%\cr
$S_\mathrm{RCA23} / S_\mathrm{RCA18}$& *2\,\%& $-0.024$& $-0.013$& 0\,\%\cr
\noalign{\vskip 6pt}
\omit{\hspace{10pt} 44 GHz}\hfil\cr
\noalign{\vskip 2pt}
$S_\mathrm{RCA24} / S_\mathrm{RCA26}$& *0\,\%& $-0.015$& $-0.011$& 1\,\%\cr
$S_\mathrm{RCA25} / S_\mathrm{RCA26}$& *1\,\%& $-0.007$& $-0.005$& 1\,\%\cr
\noalign{\vskip 6pt}
\omit{\hspace{10pt} 30 GHz}\hfil\cr
\noalign{\vskip 2pt}
$S_\mathrm{RCA27} / S_\mathrm{RCA28}$& *2\,\%& *0.001& *0.000& 2\,\%\cr
\noalign{\vskip 3pt\hrule\vskip 4pt}
}}
\endPlancktable
\tablenote a Value of $S_A/S_B - 1$, calculated over the whole sky.\par
\tablenote b Value of the column ``Discrepancy'' plus the difference of colour corrections for $\alpha=-1$ (see Table~\ref{tbl:colourCorrections}).\par
\endgroup
\end{table}

In order to better characterize these differences, we have studied how much the ratio $S_\mathrm{A} / S_\mathrm{B}$, the average ratio between compact source flux densities calculated using RCAs A and B, differs from unity. In addition to aperture photometry, we have estimated this ratio using two other methods:
\begin{enumerate}
\item We used the template fitting code from \citet{2006Davies}, run in a simplified way (i.e., without making use of a covariance matrix for noise or the CMB, such that it essentially does a least-squares fit), to compare ``template'' maps of RCA~28 (30\,GHz), RCA~26 (44\,GHz) and RCA~18 (70\,GHz) with the other maps at the same frequencies. The maps have been smoothed to $1\deg$, and a constant offset is used as a second template to remove any differences in the offsets between the template and the map. This provides an estimate of the ratio $S_\mathrm{A}/S_\mathrm{B}$ for the all-sky coefficients between RCAs. We have run this analysis on the full, unmasked sky and the Galactic plane ($\left|b\right| < 10^\circ$), which yields consistent results with those obtained through aperture photometry; in addition we have used the {\it WMAP} KQ75 mask on $3\deg$-smoothed maps and find consistent results at 30\,GHz, but better agreement at 44 and 70\,GHz, where the maps are more dominated by the CMB.
\item We combined pairs of RCA maps according to the formula $\vec{m}_\mathrm{A} - \gamma \vec{m}_\mathrm{B}$, with $\vec{m}_\mathrm{A}$ and $\vec{m}_\mathrm{B}$ the arrays of pixels for the two maps, and visually compared the three sources considered above. We found those values of $\gamma$ that either made asymmetries in the residual beam patterns around point sources disappear (for 30 and 44\,GHz), or that produced a source difference with the noise uncertainty (70\,GHz). We then assumed that $\gamma \approx S_A / S_B$. Although this method is rather qualitative, the values we obtained differ from aperture photometry estimates by only about $1\,\%$.
\end{enumerate}

Table~\ref{tab:sourceFluxDiscrepancies} shows a comparison of the average fractional difference between source densities calculated using two RCAs, e.g., RCA~18 vs.\ RCA~19. The table reports an estimate of the percentage discrepancy between the channels (calculated simply by averaging the discrepancies found by aperture photometry, template fitting and visual inspection of sources), as well as the discrepancy minus the expected difference in colour corrections between the RCAs.  We initially assume $\alpha = -1$ and tabulate the difference in colour corrections as $\delta\cc$ in Table~\ref{tab:sourceFluxDiscrepancies}. The ``corrected discrepancy'' in column~5 is the sum of $\delta\cc$ and the raw discrepancy. We also give $\delta\cc$ for $\alpha=0$ for comparison; these two values of $\alpha$ span the range of spectral indices of the sources considered. The expected uncertainties on the colour corrections for an RCA for $\alpha=-1$ are 0.42\,\%, 1.38\,\% and 1.53\,\% for 30, 44, and 70\,GHz, respectively, such that the effects shown in Table~~\ref{tab:sourceFluxDiscrepancies} are $4\sigma$ at 30 and 70\,GHz.

As the consistency between the RCAs is seen to be significantly better when comparing CMB emission between the RCAs (e.g., Sect.~\ref{sec:interchannelCalConsistency}), we conclude that the effects seen here are likely due to differences in bandpasses or the intermediate beams between the RCAs, although neither of these is expected to be sufficiently large to explain the differences seen here. These effects will be investigated further prior to the next release of \Planck{} data.

%% file: source_flux_plots.tex
% GNUPLOT: LaTeX picture with Postscript
\begingroup
  \fontfamily{phv}%
  \selectfont
  \makeatletter
  \providecommand\color[2][]{%
    \GenericError{(gnuplot) \space\space\space\@spaces}{%
      Package color not loaded in conjunction with
      terminal option `colourtext'%
    }{See the gnuplot documentation for explanation.%
    }{Either use 'blacktext' in gnuplot or load the package
      color.sty in LaTeX.}%
    \renewcommand\color[2][]{}%
  }%
  \providecommand\includegraphics[2][]{%
    \GenericError{(gnuplot) \space\space\space\@spaces}{%
      Package graphicx or graphics not loaded%
    }{See the gnuplot documentation for explanation.%
    }{The gnuplot epslatex terminal needs graphicx.sty or graphics.sty.}%
    \renewcommand\includegraphics[2][]{}%
  }%
  \providecommand\rotatebox[2]{#2}%
  \@ifundefined{ifGPcolor}{%
    \newif\ifGPcolor
    \GPcolorfalse
  }{}%
  \@ifundefined{ifGPblacktext}{%
    \newif\ifGPblacktext
    \GPblacktexttrue
  }{}%
  % define a \g@addto@macro without @ in the name:
  \let\gplgaddtomacro\g@addto@macro
  % define empty templates for all commands taking text:
  \gdef\gplbacktext{}%
  \gdef\gplfronttext{}%
  \makeatother
  \ifGPblacktext
    % no textcolor at all
    \def\colorrgb#1{}%
    \def\colorgray#1{}%
  \else
    % gray or color?
    \ifGPcolor
      \def\colorrgb#1{\color[rgb]{#1}}%
      \def\colorgray#1{\color[gray]{#1}}%
      \expandafter\def\csname LTw\endcsname{\color{white}}%
      \expandafter\def\csname LTb\endcsname{\color{black}}%
      \expandafter\def\csname LTa\endcsname{\color{black}}%
      \expandafter\def\csname LT0\endcsname{\color[rgb]{1,0,0}}%
      \expandafter\def\csname LT1\endcsname{\color[rgb]{0,1,0}}%
      \expandafter\def\csname LT2\endcsname{\color[rgb]{0,0,1}}%
      \expandafter\def\csname LT3\endcsname{\color[rgb]{1,0,1}}%
      \expandafter\def\csname LT4\endcsname{\color[rgb]{0,1,1}}%
      \expandafter\def\csname LT5\endcsname{\color[rgb]{1,1,0}}%
      \expandafter\def\csname LT6\endcsname{\color[rgb]{0,0,0}}%
      \expandafter\def\csname LT7\endcsname{\color[rgb]{1,0.3,0}}%
      \expandafter\def\csname LT8\endcsname{\color[rgb]{0.5,0.5,0.5}}%
    \else
      % gray
      \def\colorrgb#1{\color{black}}%
      \def\colorgray#1{\color[gray]{#1}}%
      \expandafter\def\csname LTw\endcsname{\color{white}}%
      \expandafter\def\csname LTb\endcsname{\color{black}}%
      \expandafter\def\csname LTa\endcsname{\color{black}}%
      \expandafter\def\csname LT0\endcsname{\color{black}}%
      \expandafter\def\csname LT1\endcsname{\color{black}}%
      \expandafter\def\csname LT2\endcsname{\color{black}}%
      \expandafter\def\csname LT3\endcsname{\color{black}}%
      \expandafter\def\csname LT4\endcsname{\color{black}}%
      \expandafter\def\csname LT5\endcsname{\color{black}}%
      \expandafter\def\csname LT6\endcsname{\color{black}}%
      \expandafter\def\csname LT7\endcsname{\color{black}}%
      \expandafter\def\csname LT8\endcsname{\color{black}}%
    \fi
  \fi
  \setlength{\unitlength}{0.0500bp}%
  \begin{picture}(10200.00,9620.00)%
    \gplgaddtomacro\gplbacktext{%
      \put(534,7070){\rotatebox{-270}{\makebox(0,0){\strut{} 300}}}%
      \put(534,8173){\rotatebox{-270}{\makebox(0,0){\strut{} 350}}}%
      \put(534,9128){\rotatebox{-270}{\makebox(0,0){\strut{} 400}}}%
      \put(609,6822){\makebox(0,0){\strut{} 20}}%
      \put(1663,6822){\makebox(0,0){\strut{} 30}}%
      \put(2411,6822){\makebox(0,0){\strut{} 40}}%
      \put(2991,6822){\makebox(0,0){\strut{} 50}}%
      \put(3866,6822){\makebox(0,0){\strut{} 70}}%
      \put(4793,6822){\makebox(0,0){\strut{} 100}}%
      \csname LTb\endcsname%
      \put(264,8211){\rotatebox{-270}{\makebox(0,0){\strut{}Flux density [Jy]}}}%
      \csname LTb\endcsname%
      \put(2803,6543){\makebox(0,0){\strut{}Frequency [GHz]}}%
      \put(711,7194){\makebox(0,0)[l]{\strut{}Orion A}}%
    }%
    \gplgaddtomacro\gplfronttext{%
      \csname LTb\endcsname%
      \put(2037,9248){\makebox(0,0)[r]{\strut{}LFI detectors}}%
      \csname LTb\endcsname%
      \put(2037,9062){\makebox(0,0)[r]{\strut{}WMAP}}%
      \csname LTb\endcsname%
      \put(2037,8876){\makebox(0,0)[r]{\strut{}PCCS}}%
    }%
    \gplgaddtomacro\gplbacktext{%
      \csname LT3\endcsname%
      \put(534,4246){\rotatebox{-270}{\makebox(0,0){\strut{} 100}}}%
      \put(534,5084){\rotatebox{-270}{\makebox(0,0){\strut{} 150}}}%
      \put(534,5678){\rotatebox{-270}{\makebox(0,0){\strut{} 200}}}%
      \put(534,6139){\rotatebox{-270}{\makebox(0,0){\strut{} 250}}}%
      \put(609,3615){\makebox(0,0){\strut{} 20}}%
      \put(1663,3615){\makebox(0,0){\strut{} 30}}%
      \put(2411,3615){\makebox(0,0){\strut{} 40}}%
      \put(2991,3615){\makebox(0,0){\strut{} 50}}%
      \put(3866,3615){\makebox(0,0){\strut{} 70}}%
      \put(4793,3615){\makebox(0,0){\strut{} 100}}%
      \csname LTb\endcsname%
      \put(264,5005){\rotatebox{-270}{\makebox(0,0){\strut{}Flux density [Jy]}}}%
      \csname LTb\endcsname%
      \put(2803,3336){\makebox(0,0){\strut{}Frequency [GHz]}}%
      \put(711,3987){\makebox(0,0)[l]{\strut{}Cas A}}%
    }%
    \gplgaddtomacro\gplfronttext{%
      \csname LTb\endcsname%
      \put(4005,6042){\makebox(0,0)[r]{\strut{}LFI detectors}}%
      \csname LTb\endcsname%
      \put(4005,5856){\makebox(0,0)[r]{\strut{}WMAP}}%
      \csname LTb\endcsname%
      \put(4005,5670){\makebox(0,0)[r]{\strut{}Weiland et al. fit}}%
      \csname LTb\endcsname%
      \put(4005,5484){\makebox(0,0)[r]{\strut{}WMAP (Weiland)}}%
      \csname LTb\endcsname%
      \put(4005,5298){\makebox(0,0)[r]{\strut{}PCCS}}%
    }%
    \gplgaddtomacro\gplbacktext{%
      \csname LT3\endcsname%
      \put(534,1095){\rotatebox{-270}{\makebox(0,0){\strut{} 250}}}%
      \put(534,1816){\rotatebox{-270}{\makebox(0,0){\strut{} 300}}}%
      \put(534,2426){\rotatebox{-270}{\makebox(0,0){\strut{} 350}}}%
      \put(534,2954){\rotatebox{-270}{\makebox(0,0){\strut{} 400}}}%
      \put(609,409){\makebox(0,0){\strut{} 20}}%
      \put(1663,409){\makebox(0,0){\strut{} 30}}%
      \put(2411,409){\makebox(0,0){\strut{} 40}}%
      \put(2991,409){\makebox(0,0){\strut{} 50}}%
      \put(3866,409){\makebox(0,0){\strut{} 70}}%
      \put(4793,409){\makebox(0,0){\strut{} 100}}%
      \csname LTb\endcsname%
      \put(264,1798){\rotatebox{-270}{\makebox(0,0){\strut{}Flux density [Jy]}}}%
      \csname LTb\endcsname%
      \put(2803,130){\makebox(0,0){\strut{}Frequency [GHz]}}%
      \put(711,781){\makebox(0,0)[l]{\strut{}Tau A}}%
    }%
    \gplgaddtomacro\gplfronttext{%
      \csname LTb\endcsname%
      \put(4005,2835){\makebox(0,0)[r]{\strut{}LFI detectors}}%
      \csname LTb\endcsname%
      \put(4005,2649){\makebox(0,0)[r]{\strut{}WMAP}}%
      \csname LTb\endcsname%
      \put(4005,2463){\makebox(0,0)[r]{\strut{}Weiland et al. fit}}%
      \csname LTb\endcsname%
      \put(4005,2277){\makebox(0,0)[r]{\strut{}WMAP (Weiland)}}%
      \csname LTb\endcsname%
      \put(4005,2091){\makebox(0,0)[r]{\strut{}PCCS}}%
    }%
    \gplgaddtomacro\gplbacktext{%
      \csname LT3\endcsname%
      \put(5634,7382){\rotatebox{-270}{\makebox(0,0){\strut{} 340}}}%
      \put(5634,8097){\rotatebox{-270}{\makebox(0,0){\strut{} 360}}}%
      \put(5634,8773){\rotatebox{-270}{\makebox(0,0){\strut{} 380}}}%
      \put(5951,6822){\makebox(0,0){\strut{} 18}}%
      \put(6415,6822){\makebox(0,0){\strut{} 19}}%
      \put(6855,6822){\makebox(0,0){\strut{} 20}}%
      \put(7273,6822){\makebox(0,0){\strut{} 21}}%
      \put(7672,6822){\makebox(0,0){\strut{} 22}}%
      \put(8054,6822){\makebox(0,0){\strut{} 23}}%
      \put(8419,6822){\makebox(0,0){\strut{} 24}}%
      \put(8769,6822){\makebox(0,0){\strut{} 25}}%
      \put(9105,6822){\makebox(0,0){\strut{} 26}}%
      \put(9429,6822){\makebox(0,0){\strut{} 27}}%
      \put(9741,6822){\makebox(0,0){\strut{} 28}}%
      \csname LTb\endcsname%
      \put(5364,8211){\rotatebox{-270}{\makebox(0,0){\strut{}Flux density [Jy]}}}%
      \csname LTb\endcsname%
      \put(7903,6543){\makebox(0,0){\strut{}RCA number}}%
      \put(5811,7194){\makebox(0,0)[l]{\strut{}Orion A}}%
    }%
    \gplgaddtomacro\gplfronttext{%
      \csname LTb\endcsname%
      \put(9105,7733){\makebox(0,0)[r]{\strut{}Average}}%
      \csname LTb\endcsname%
      \put(9105,7547){\makebox(0,0)[r]{\strut{}Survey 1}}%
      \csname LTb\endcsname%
      \put(9105,7361){\makebox(0,0)[r]{\strut{}Survey 2}}%
      \csname LTb\endcsname%
      \put(9105,7175){\makebox(0,0)[r]{\strut{}Survey 3}}%
    }%
    \gplgaddtomacro\gplbacktext{%
      \csname LT8\endcsname%
      \put(5634,4273){\rotatebox{-270}{\makebox(0,0){\strut{} 100}}}%
      \put(5634,5129){\rotatebox{-270}{\makebox(0,0){\strut{} 150}}}%
      \put(5634,5737){\rotatebox{-270}{\makebox(0,0){\strut{} 200}}}%
      \put(5951,3615){\makebox(0,0){\strut{} 18}}%
      \put(6415,3615){\makebox(0,0){\strut{} 19}}%
      \put(6855,3615){\makebox(0,0){\strut{} 20}}%
      \put(7273,3615){\makebox(0,0){\strut{} 21}}%
      \put(7672,3615){\makebox(0,0){\strut{} 22}}%
      \put(8054,3615){\makebox(0,0){\strut{} 23}}%
      \put(8419,3615){\makebox(0,0){\strut{} 24}}%
      \put(8769,3615){\makebox(0,0){\strut{} 25}}%
      \put(9105,3615){\makebox(0,0){\strut{} 26}}%
      \put(9429,3615){\makebox(0,0){\strut{} 27}}%
      \put(9741,3615){\makebox(0,0){\strut{} 28}}%
      \csname LTb\endcsname%
      \put(5364,5005){\rotatebox{-270}{\makebox(0,0){\strut{}Flux density [Jy]}}}%
      \csname LTb\endcsname%
      \put(7903,3336){\makebox(0,0){\strut{}RCA number}}%
      \put(5811,3987){\makebox(0,0)[l]{\strut{}Cas A}}%
    }%
    \gplgaddtomacro\gplfronttext{%
      \csname LTb\endcsname%
      \put(9105,4526){\makebox(0,0)[r]{\strut{}Average}}%
      \csname LTb\endcsname%
      \put(9105,4340){\makebox(0,0)[r]{\strut{}Survey 1}}%
      \csname LTb\endcsname%
      \put(9105,4154){\makebox(0,0)[r]{\strut{}Survey 2}}%
      \csname LTb\endcsname%
      \put(9105,3968){\makebox(0,0)[r]{\strut{}Survey 3}}%
    }%
    \gplgaddtomacro\gplbacktext{%
      \csname LT8\endcsname%
      \put(5634,1110){\rotatebox{-270}{\makebox(0,0){\strut{} 250}}}%
      \put(5634,1844){\rotatebox{-270}{\makebox(0,0){\strut{} 300}}}%
      \put(5634,2464){\rotatebox{-270}{\makebox(0,0){\strut{} 350}}}%
      \put(5951,409){\makebox(0,0){\strut{} 18}}%
      \put(6415,409){\makebox(0,0){\strut{} 19}}%
      \put(6855,409){\makebox(0,0){\strut{} 20}}%
      \put(7273,409){\makebox(0,0){\strut{} 21}}%
      \put(7672,409){\makebox(0,0){\strut{} 22}}%
      \put(8054,409){\makebox(0,0){\strut{} 23}}%
      \put(8419,409){\makebox(0,0){\strut{} 24}}%
      \put(8769,409){\makebox(0,0){\strut{} 25}}%
      \put(9105,409){\makebox(0,0){\strut{} 26}}%
      \put(9429,409){\makebox(0,0){\strut{} 27}}%
      \put(9741,409){\makebox(0,0){\strut{} 28}}%
      \csname LTb\endcsname%
      \put(5364,1798){\rotatebox{-270}{\makebox(0,0){\strut{}Flux density [Jy]}}}%
      \csname LTb\endcsname%
      \put(7903,130){\makebox(0,0){\strut{}RCA number}}%
      \put(5811,781){\makebox(0,0)[l]{\strut{}Tau A}}%
    }%
    \gplgaddtomacro\gplfronttext{%
      \csname LTb\endcsname%
      \put(9105,1320){\makebox(0,0)[r]{\strut{}Average}}%
      \csname LTb\endcsname%
      \put(9105,1134){\makebox(0,0)[r]{\strut{}Survey 1}}%
      \csname LTb\endcsname%
      \put(9105,948){\makebox(0,0)[r]{\strut{}Survey 2}}%
      \csname LTb\endcsname%
      \put(9105,762){\makebox(0,0)[r]{\strut{}Survey 3}}%
    }%
    \gplbacktext
    \put(0,0){\includegraphics{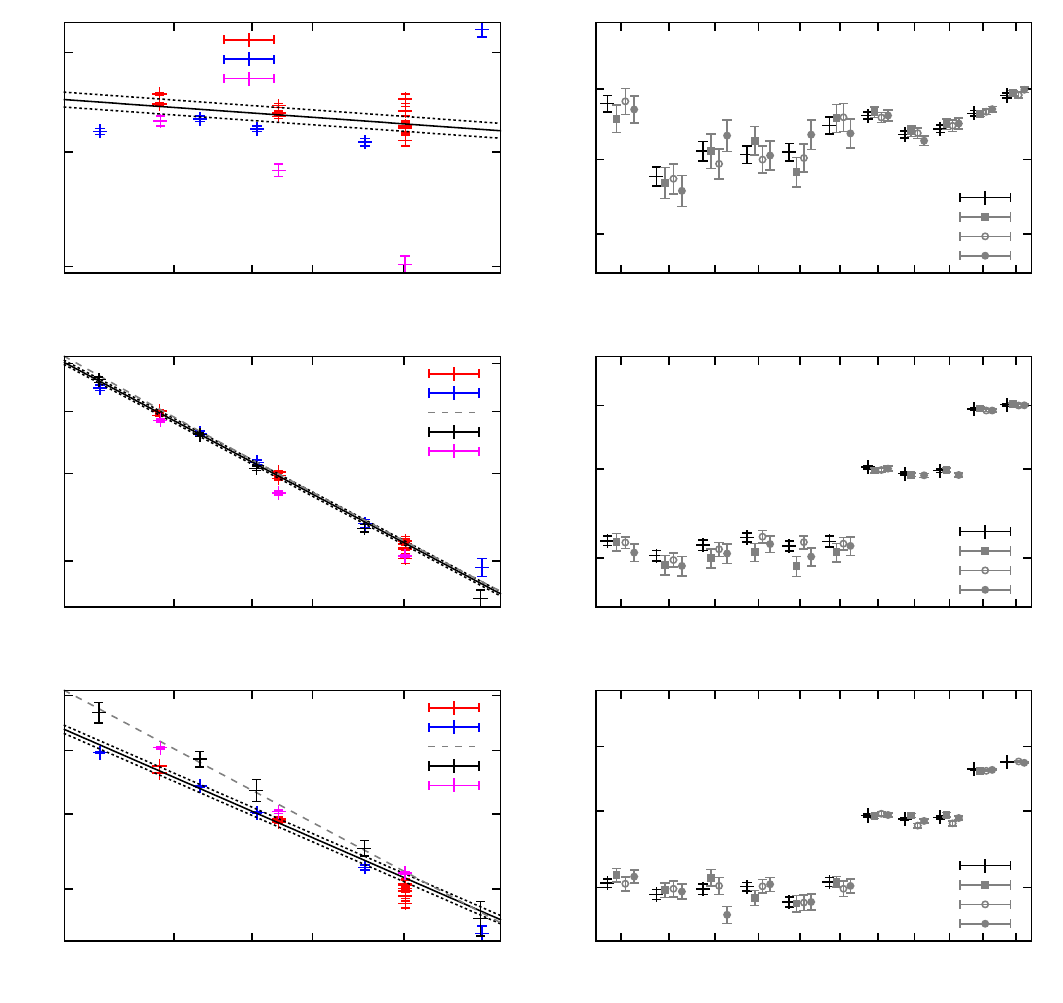}}%
    \gplfronttext
  \end{picture}%
\endgroup

%% file: P02b_5_4_2_planets.tex
\def\PlanetPointing{\vec{x}_{\mathrm{p}}}
\def\PlanetPointingBeam{\vec{y}_{\mathrm{p}}}
\def\PlanetI{I_{\mathrm{p}}}
\def\BackgroundI{I_{\mathrm{bg}}}
\def\PlanetIz{I_{\mathrm{p},0}}
\def\PlanetIzAver{\bar{I}_{\mathrm{p},0}}
\def\Geometry{\gamma}
\def\Beam{B_{t_i}}

\begin{table}
\begingroup
\newdimen\tblskip \tblskip=5pt
\caption{\label{tbl:planetCrossingTimes} Visibility times of planets$^a$.}
\nointerlineskip
\vskip -3mm
\footnotesize
\setbox\tablebox=\vbox{
   \newdimen\digitwidth 
   \setbox0=\hbox{\rm 0} 
   \digitwidth=\wd0 
   \catcode`*=\active 
   \def*{\kern\digitwidth}
   \newdimen\dotwidth % We need this to properly space the numbers below!
   \setbox0=\hbox{.} 
   \dotwidth=\wd0 
   \catcode`!=\active 
   \def!{\kern\dotwidth}
\halign{
% Template
#\tabskip 1.0em&
#\hfil&
#\hfil&
#\hfil\tabskip=0pt\cr
\noalign{\doubleline}
\omit&
\omit\hfil 30\,GHz\hfil&
\omit\hfil 44\,GHz\hfil&
\omit\hfil 70\,GHz\hfil\cr
\noalign{\vskip 3pt\hrule\vskip 5pt}
Mars& 09/10/18--21& 09/10/18--21&    09/10/20--24\cr
    & 10/04/17--20& 09/10/29--11/02& 10/04/15--18\cr
    &              & 10/04/09--12&             \cr
    &              & 10/04/18--20&             \cr
\noalign{\vskip 6pt}
Jupiter& 09/10/31--11/02& 09/10/24--27&    09/10/29--11/01\cr
       & 10/06/30--07/03& 09/10/31--11/02& 10/07/01--05\cr
       &                 & 10/06/30--07/02&          \cr
       &                 & 10/07/08--12&             \cr
\noalign{\vskip 6pt}
Saturn& 09/12/31--10/01/03& 09/12/31--10/01/03& 10/01/02--05\cr
      & 10/06/20--22&           10/01/08--11&           10/06/18--21\cr
      &                      & 10/06/10--13&                      \cr
      &                      & 10/06/20--22&                      \cr
\noalign{\vskip 6pt}
Uranus& 09/12/12--15& 09/12/13--15& 09/12/10--13\cr
      & 10/06/27--30& 10/06/27--29& 10/06/29--07/02\cr
      &              & 10/07/06--08&             \cr
\noalign{\vskip 6pt}
Neptune& 09/11/07&    09/10/31--11/02& 09/11/05--07\cr
       & 10/05/15--17& 09/11/07--09&        10/05/16--19\cr
       & 10/11/09--11& 10/05/15&   10/11/08--10\cr
       &              & 10/05/23--25&         \cr
       &              & 10/11/10&            \cr
\noalign{\vskip 5pt\hrule\vskip 3pt}}}
\endPlancktable
\tablenote a Dates are in the format YY/MM/DD. Only the times between 2009 Aug 13 and 2010 Nov 28 are reported here.\par
\endgroup
\end{table}

\begin{figure*}
	\centering
	\includegraphics{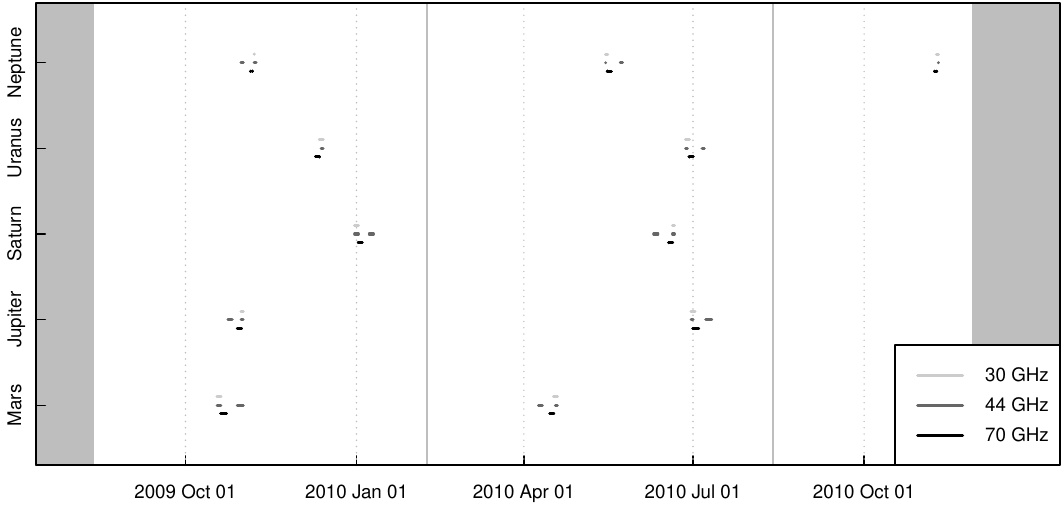}
	\caption{\label{fig:planetCrossingTimes} Timeline of planet crossings. Each line shows the times when the planet fell within the beam of at least one of the radiometers at 30, 44, or 70\,GHz. Gray bands at the edges mark the start of the survey and the end of the period considered in the 2013 data release. Vertical grey lines mark the separation between sky surveys.}
\end{figure*}

Planets can be used as a source for calibration cross-checks. In particular, Jupiter has several advantages: its signal is up to fifty times higher than the receiver noise per scan; it is visible to \Planck{} about twice a year; it has a
relatively well known spectrum; and at LFI's resolution it is point-like. Table~\ref{tbl:planetCrossingTimes} lists the dates when planets crossed LFI beams, and Fig.~\ref{fig:planetCrossingTimes} gives a visual timeline of these events.

The model for the time dependent signal produced by a planet in a timeline \citep{Cremonese:2002,Maris:MOBS:2007} is:
\begin{equation}\label{eq:planets:row}
 \PlanetI(t_i) = N^{-1} \int_{4\pi}\ud\Omega \left[ \BeamRd(\Pointing) 
  \PlanetIz\bigl(\Pointing-\RotBeamObs(t_i)\PlanetPointing(t_i)\bigr) + \BackgroundI(t_i)\right],
\end{equation}
where $\PlanetIz$ is the brightness distribution across the projected planet disk, $\PlanetPointing(t_i)$ represents the direction in which a planet is seen from the instrument at the sampling time $t_i$, $\RotBeamObs(t_i)$ is the matrix that converts pointing from the Ecliptic reference frame to the beam reference frame, and $\BackgroundI(t_i)$ is the observed background. We assume that the radiometers have an elliptical Gaussian beam pattern, and we neglect complications such as differential temperature distributions across the planetary disk, phase effects, and limb darkening. Therefore, Eq.~\ref{eq:planets:row} reduces to:
\begin{equation}
 \PlanetI  \approx 
\PlanetIzAver
\left(\frac{\Delta}{\Delta_0}
% \right)^2 \left(
\frac{b_0}{b}\right)^{2}
\left(1-f\cos\theta_{\mathrm{obs}} \right) 
% \frac{
\exp\left( 
-\frac{1}{2} \PlanetPointingBeam^\dagger \mathcal{C}^{-1} 
\PlanetPointingBeam
\right)
% }{N} 
+ \BackgroundI,
\end{equation}
where $\PlanetPointingBeam=\RotBeamObs \PlanetPointing$, $b$ is the beam FWHM, $f$ is the planetary disk flattening, $\theta_{\mathrm{obs}}$ is the observer planeto-centric latitude, and $\Delta$ is the distance between the planet and the observer. The diagonal matrix $\mathcal{C}=b^2/8\log2\,\mathrm{diag}(\epsilon, 1/\epsilon)$ represents the beam shape, whose ellipticity is $\epsilon$. The time dependency of $\PlanetI$, $\PlanetIzAver$, $\Delta$, $\theta_{\mathrm{obs}}$, $\RotBeamObs$, and $\PlanetPointing$ are omitted. The equation assumes a beam-averaged planet brightness $\PlanetIzAver$ calculated for a fiducial planet-observer distance $\Delta_0$ and beam FWHM $b_0$, in order to account for beam-to-beam variability.

The model is used to fit $\PlanetIzAver$ onto the timelines of each feed-horn, taken separately, by using a least-squares procedure. Samples for each timeline are accepted as valid data of the fit if their instantaneous angular distance from the planet is smaller than $5\deg$ at the epoch of observation. The background is assumed to be constant over the acceptance disc and is left as a free parameter. In order to properly destripe the data, we have used the baselines calculated by \texttt{Madam} \citep{keihanen2010}, the standard map-making tool used to produce the LFI maps.

The histogram of the data in the timelines can be approximated as the sum of two Gaussian distributions, one due to the signal and one due to the background. The histogram of differences of timelines and the fitted model overlap very well, one of the peaks assessing proper separation of the planetary signal. The width of the residual peak is an estimate of the confusion noise, dominated by white noise and CMB fluctuations.
Table~\ref{tab:planets:synoptic} reports the measurements of the brightness temperature for a number of planets. The two quoted uncertainties are the maximum intra-survey difference with respect to the  median of the measures and the confusion noise measured on the fitted data. The central wavelengths assumed for each of the three bands are 10.56\,mm, 6.80\,mm, and 4.26\,mm \citep[taken from the LFI instrument model, see][]{planck2013-p28}.

Jupiter is the only planet bright enough to be detected in the timeline without averaging. This is the reason why the random uncertainty is much smaller than the maximal survey-by-survey variation. Such variation can be ascribed to a number of small systematics not completely understood yet. One of them is the fact that planets have been extracted as soon as they were observed, so that different observations have had slightly different calibration procedures. Despite having a slightly worse signal-to-noise ratio (S/N), the same is true for Saturn. However, we have not introduced any correction for the effect of Saturn's rings yet. In the same manner the observations of Mars are not corrected for diurnal variability due to the rotation of the planet. Uranus and Neptune have quite low S/N, in particular at 30\,GHz, and therefore they are very sensitive to the confusion noise, background, uncertainties in beam orientations, and aperture corrections.

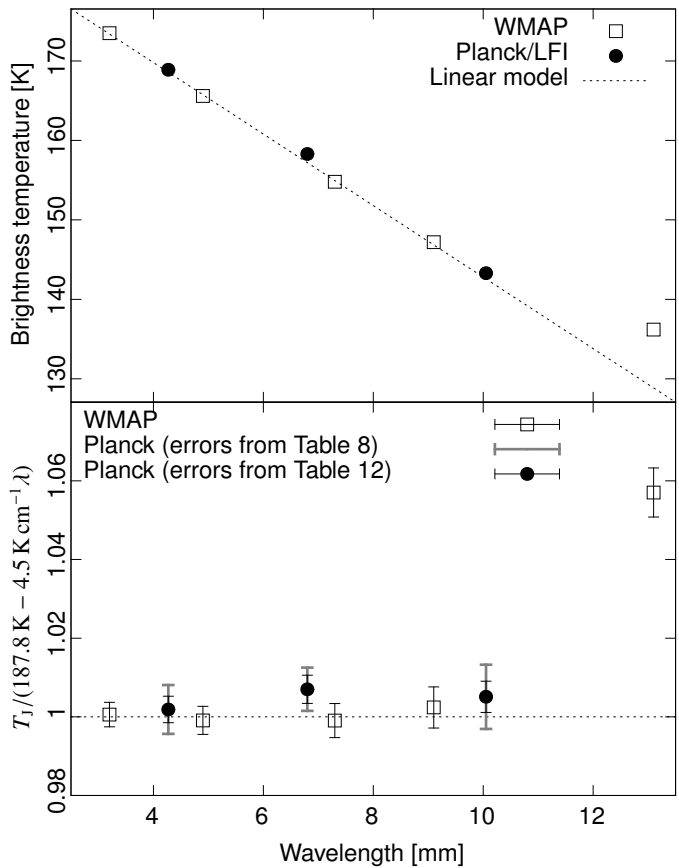
\begin{figure}
\centering
\begin{DIFnomarkup}
\input{jupiter_temperature.tex}
\end{DIFnomarkup}
\caption{\label{fig:jupiter} The Jupiter spectrum for LFI compared to the {\it WMAP} spectrum \citep{weiland2010}. \Planck's data are taken from Table~\ref{tab:planets:synoptic}. (The LFI points have been rescaled by a factor, 1.0693, which takes into account Jupiter's oblateness, in order to make them easier to compare with {\it WMAP}'s.) The upper frame shows the brightness temperatures as a function of the central wavelength, alongside a simple linear model $T_{\mathrm{J}}(\lambda) = 187.8\,\mathrm{K} - 4.5\,\mathrm{K}\,\mathrm{cm}^{-1} \lambda$, which by construction passes through the WMAP's points at $\lambda = 3.2\,\mathrm{mm}$ and $\lambda = 9.1\,\mathrm{mm}$. For both \Planck{} and {\it WMAP}, measurement errors are smaller than the size of the symbols. In the bottom frame, we rescaled our points to the linear model in order to better show deviations from it. The error bars for LFI data show the uncertainties quoted in Table~\protect\ref{tab:planets:synoptic}, as well as the overall calibration errors listed in Table~\protect\ref{tab:accuracyBudgetResult}.}
\end{figure}

Figure~\ref{fig:jupiter} compares the Jupiter spectrum from LFI averaged through the surveys with the {\it WMAP} measurements \citep{weiland2010}. It is possible to see the good agreement between the two datasets.

\begin{table}
\begingroup
\newdimen\tblskip \tblskip=5pt
\caption{\label{tab:planets:synoptic} Averaged brightness temperatures for planets\tablefootmark{a}. We have applied no correction for planetary disk oblateness.}
\nointerlineskip
\vskip 3mm
\footnotesize
\setbox\tablebox=\vbox{
   \newdimen\digitwidth 
   \setbox0=\hbox{\rm 0} 
   \digitwidth=\wd0 
   \catcode`*=\active 
   \def*{\kern\digitwidth}
   \newdimen\dotwidth % We need this to properly space the numbers below!
   \setbox0=\hbox{.} 
   \dotwidth=\wd0 
   \catcode`!=\active 
   \def!{\kern\dotwidth}
\halign{
% Template
#\tabskip 1.0em&
#\hfil&
#\hfil&
#\hfil\tabskip=0pt\cr
\noalign{\doubleline}
\omit&
\omit\hfil 30\,GHz\hfil&
\omit\hfil 44\,GHz\hfil&
\omit\hfil 70\,GHz\hfil\cr
\noalign{\vskip 3pt\hrule\vskip 5pt}
Mars&    $183 \pm **1!* \pm *4.1$& $187 \pm 10!* \pm *4.1$ & $183 \pm *3.5\pm *2.3$\cr
Jupiter& $134 \pm **0.5 \pm *0.2$& $148 \pm *0.5 \pm *0.2$ & $157 \pm *0.5\pm *0.2$\cr
Saturn&  $121 \pm **3.5 \pm *0.9$& $128 \pm *3.0 \pm *1.0$ & $131 \pm *2.5\pm *0.5$\cr
Uranus&  $190 \pm 133!* \pm 23!*$& $230 \pm 10.5 \pm 29!*$ & $138 \pm *7.5\pm 13!*$\cr
Neptune& $*79 \pm **5!* \pm 74!*$& $*74 \pm 22!* \pm 76!*$ & $101 \pm 11!*\pm 32!*$\cr
\noalign{\vskip 5pt\hrule\vskip 3pt}}}
\endPlancktable
\tablenote a Values are in Kelvin. The first error represents the intra-survey difference, the second one the confusion noise.\par
\endgroup
\end{table}

%% file: jupiter_temperature.tex
% GNUPLOT: LaTeX picture with Postscript
\begingroup
  \fontfamily{phv}%
  \selectfont
  \makeatletter
  \providecommand\color[2][]{%
    \GenericError{(gnuplot) \space\space\space\@spaces}{%
      Package color not loaded in conjunction with
      terminal option `colourtext'%
    }{See the gnuplot documentation for explanation.%
    }{Either use 'blacktext' in gnuplot or load the package
      color.sty in LaTeX.}%
    \renewcommand\color[2][]{}%
  }%
  \providecommand\includegraphics[2][]{%
    \GenericError{(gnuplot) \space\space\space\@spaces}{%
      Package graphicx or graphics not loaded%
    }{See the gnuplot documentation for explanation.%
    }{The gnuplot epslatex terminal needs graphicx.sty or graphics.sty.}%
    \renewcommand\includegraphics[2][]{}%
  }%
  \providecommand\rotatebox[2]{#2}%
  \@ifundefined{ifGPcolor}{%
    \newif\ifGPcolor
    \GPcolorfalse
  }{}%
  \@ifundefined{ifGPblacktext}{%
    \newif\ifGPblacktext
    \GPblacktexttrue
  }{}%
  % define a \g@addto@macro without @ in the name:
  \let\gplgaddtomacro\g@addto@macro
  % define empty templates for all commands taking text:
  \gdef\gplbacktext{}%
  \gdef\gplfronttext{}%
  \makeatother
  \ifGPblacktext
    % no textcolor at all
    \def\colorrgb#1{}%
    \def\colorgray#1{}%
  \else
    % gray or color?
    \ifGPcolor
      \def\colorrgb#1{\color[rgb]{#1}}%
      \def\colorgray#1{\color[gray]{#1}}%
      \expandafter\def\csname LTw\endcsname{\color{white}}%
      \expandafter\def\csname LTb\endcsname{\color{black}}%
      \expandafter\def\csname LTa\endcsname{\color{black}}%
      \expandafter\def\csname LT0\endcsname{\color[rgb]{1,0,0}}%
      \expandafter\def\csname LT1\endcsname{\color[rgb]{0,1,0}}%
      \expandafter\def\csname LT2\endcsname{\color[rgb]{0,0,1}}%
      \expandafter\def\csname LT3\endcsname{\color[rgb]{1,0,1}}%
      \expandafter\def\csname LT4\endcsname{\color[rgb]{0,1,1}}%
      \expandafter\def\csname LT5\endcsname{\color[rgb]{1,1,0}}%
      \expandafter\def\csname LT6\endcsname{\color[rgb]{0,0,0}}%
      \expandafter\def\csname LT7\endcsname{\color[rgb]{1,0.3,0}}%
      \expandafter\def\csname LT8\endcsname{\color[rgb]{0.5,0.5,0.5}}%
    \else
      % gray
      \def\colorrgb#1{\color{black}}%
      \def\colorgray#1{\color[gray]{#1}}%
      \expandafter\def\csname LTw\endcsname{\color{white}}%
      \expandafter\def\csname LTb\endcsname{\color{black}}%
      \expandafter\def\csname LTa\endcsname{\color{black}}%
      \expandafter\def\csname LT0\endcsname{\color{black}}%
      \expandafter\def\csname LT1\endcsname{\color{black}}%
      \expandafter\def\csname LT2\endcsname{\color{black}}%
      \expandafter\def\csname LT3\endcsname{\color{black}}%
      \expandafter\def\csname LT4\endcsname{\color{black}}%
      \expandafter\def\csname LT5\endcsname{\color{black}}%
      \expandafter\def\csname LT6\endcsname{\color{black}}%
      \expandafter\def\csname LT7\endcsname{\color{black}}%
      \expandafter\def\csname LT8\endcsname{\color{black}}%
    \fi
  \fi
  \setlength{\unitlength}{0.0500bp}%
  \begin{picture}(4980.00,6800.00)%
    \gplgaddtomacro\gplbacktext{%
      \put(335,3723){\rotatebox{-270}{\makebox(0,0){\strut{} 130}}}%
      \put(335,4320){\rotatebox{-270}{\makebox(0,0){\strut{} 140}}}%
      \put(335,4918){\rotatebox{-270}{\makebox(0,0){\strut{} 150}}}%
      \put(335,5515){\rotatebox{-270}{\makebox(0,0){\strut{} 160}}}%
      \put(335,6112){\rotatebox{-270}{\makebox(0,0){\strut{} 170}}}%
      \put(1025,3361){\makebox(0,0){\strut{}}}%
      \put(1845,3361){\makebox(0,0){\strut{}}}%
      \put(2665,3361){\makebox(0,0){\strut{}}}%
      \put(3485,3361){\makebox(0,0){\strut{}}}%
      \put(4305,3361){\makebox(0,0){\strut{}}}%
      \csname LTb\endcsname%
      \put(65,5025){\rotatebox{-270}{\makebox(0,0){\strut{}Brightness temperature [K]}}}%
      \csname LTb\endcsname%
      \put(2665,3119){\makebox(0,0){\strut{}}}%
    }%
    \gplgaddtomacro\gplfronttext{%
      \csname LTb\endcsname%
      \put(4132,6336){\makebox(0,0)[r]{\strut{}WMAP}}%
      \csname LTb\endcsname%
      \put(4132,6150){\makebox(0,0)[r]{\strut{}Planck/LFI}}%
      \csname LTb\endcsname%
      \put(4132,5964){\makebox(0,0)[r]{\strut{}Linear model}}%
    }%
    \gplgaddtomacro\gplbacktext{%
      \csname LTb\endcsname%
      \put(335,591){\rotatebox{-270}{\makebox(0,0){\strut{} 0.98}}}%
      \put(335,1182){\rotatebox{-270}{\makebox(0,0){\strut{} 1}}}%
      \put(335,1773){\rotatebox{-270}{\makebox(0,0){\strut{} 1.02}}}%
      \put(335,2365){\rotatebox{-270}{\makebox(0,0){\strut{} 1.04}}}%
      \put(335,2956){\rotatebox{-270}{\makebox(0,0){\strut{} 1.06}}}%
      \put(1025,405){\makebox(0,0){\strut{}4}}%
      \put(1845,405){\makebox(0,0){\strut{}6}}%
      \put(2665,405){\makebox(0,0){\strut{}8}}%
      \put(3485,405){\makebox(0,0){\strut{}10}}%
      \put(4305,405){\makebox(0,0){\strut{}12}}%
      \csname LTb\endcsname%
      \put(65,2069){\rotatebox{-270}{\makebox(0,0){\strut{}$T_{\mathrm{J}} / (187.8\,\mathrm{K} - 4.5\,\mathrm{K}\,\mathrm{cm}^{-1} \lambda)$}}}%
      \csname LTb\endcsname%
      \put(2665,126){\makebox(0,0){\strut{}Wavelength [mm]}}%
    }%
    \gplgaddtomacro\gplfronttext{%
      \csname LTb\endcsname%
      \put(512,3380){\makebox(0,0)[l]{\strut{}WMAP}}%
      \csname LTb\endcsname%
      \put(512,3194){\makebox(0,0)[l]{\strut{}Planck (errors from Table 8)}}%
      \csname LTb\endcsname%
      \put(512,3008){\makebox(0,0)[l]{\strut{}Planck (errors from Table 12)}}%
    }%
    \gplbacktext
    \put(0,0){\includegraphics{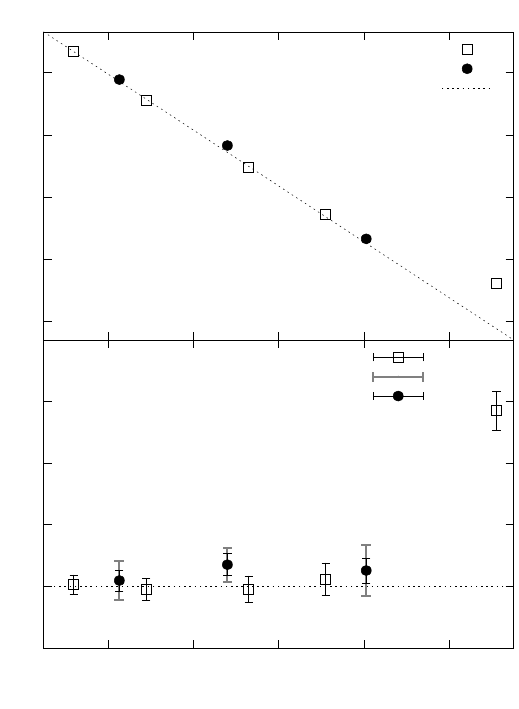}}%
    \gplfronttext
  \end{picture}%
\endgroup

%% file: P02b_5_5_consistency.tex
A key aspect of \Planck{}'s data analysis is the internal consistency of calibrated maps and power spectra obtained by LFI and HFI detectors. Because the two instruments employ highly independent technologies, inter-instrument comparisons provide useful indications on systematic effects of instrumental origin. An even more independent calibration cross-check can be made by comparing maps and power spectra obtained by LFI and WMAP in  similar frequency bands.  Because of the importance and complexity of these comparisons, a full paper of this 2013 release is dedicated to these issues, \citet{planck2013-p01a}.

The LFI 70 GHz and HFI 100 GHz power spectra in the region of the first CMB acoustic peak (multipole range 70--390) are consistent within $0.8\,\%$, corresponding to a map calibration consistency of $\sim 0.4\,\%$. These values, which include a small correction for unresolved sources, are within the LFI and HFI calibration uncertainties discussed in this paper and in \citet{planck2013-p03f}. In \citet{planck2013-p01a} we also show that an improved model of HFI near sidelobes, which were not fully accounted for initially, leads to a further improvement in the LFI-HFI consistency by a factor of 2. Such good agreement between LFI and HFI, while not a guarantee, is an indication of lack of major undetected systematics and of good calibration consistency.

A larger deviation, of order $1.8\,\%$ in the power spectrum ($\sim0.9\,\%$ in calibrated maps), is found between LFI 70\,GHz and {\it WMAP} V-band in the region of the first peak (the former measuring less power than the latter), a $1.4\sigma$ tension suggesting some undetected effect in LFI, in {\it WMAP}, or in both. These could be due to a combination of errors in the calibration process or beam window function, issues in the data processing, or imprecise estimation of the Solar dipole. Our estimates of the Solar dipole based on LFI 44\,GHz orbital dipole calibration (Table~\ref{tbl:prelim44GHzdipoles}) suggest good consistency with {\it WMAP}. Our results, however, are very preliminary, and more redundancy and cross-checks are needed before an assessment of inter-calibration at sub-percent level can be performed. We note that a $\sim 1\,\%$ calibration discrepancy would result in a noticeable offset in the Jupiter temperature as measured by LFI and by {\it WMAP} (Fig.~\ref{fig:jupiter}, bottom panel). Taken at face value, the combination of the good agreement of these two sets of measurements and the 1.8\% discrepancy of the power spectra at degree scales (first peak of the CMB) might suggest a beam or window function issue, either in LFI or {\it WMAP}. However, these deviations are (marginally) within uncertainties. The 2014 Planck release will greatly increase the LFI data redundancy compared to the present 2013 release (from 2.5 to 8 surveys, and from 2 to 7 Jupiter measurements), which we expect will help to settle the issue.

%% file: P02b_6_conclusions.tex
In this paper we have described the procedures used to calibrate \Planck{}/LFI's data and the ways we used to assess the accuracy of our calibration. We have shown that the level of consistency in the maps is quite good, since discrepancies are roughly $0.6\,\%$ at 44 and 70\,GHz, and $0.8\,\%$ at 30\,GHz.

However, we believe there are areas where we can significantly improve our understanding of the instrument and therefore produce a better calibration, e.g., modelling of far sidelobes (see sects.~\ref{sec:optics} and \ref{sec:fourPiConvolver}) and proper modeling of variations in the noise temperature of the radiometers (Sect.~\ref{sec:dVV}). We therefore need to concentrate our efforts in improving our algorithms and analysis methods for the next \Planck{} data delivery, in order also to provide a sound polarization calibration.

An important point which is still missing is the production of a full characterization of the dipole that is independent of {\it WMAP}. This would in turn allow us to calibrate the \Planck{}/LFI radiometers absolutely. As explained in Sect.~\ref{sec:orbitalDipole}, we were able to obtain some encouraging results, but not good enough to avoid using {\it WMAP}'s characterization of the dipole in our calibration pipeline. We aim to do so for the next release.

Another important point is to fully characterize the impact of optical systematics on the calibration. Our efficient dipole convolver (Sect.~\ref{sec:fourPiConvolver}) is a testimony to the work we have put in this task, but much more work is needed to fully understand the impact of far sidelobes on the calibration. We plan to run simulations to assess how much our uncertainty in the knowledge of the optical parameters of the beams impacts the estimation of gains and the production of calibrated timelines, and to further improve our model of the beams themselves.

Finally, we are determined to understand the origin of all the ``unknown systematics'' listed in the corresponding row in Table~\ref{tab:accuracyBudgetResult}, by means of simulations and more refined models of the radiometer.

%% file: P02b_appendix_A_beam_convolution_calculation.tex
In this section we provide the details of the model used to convolve the dipole signal used in the calibration with the beam response over the full sky (Sect.~\ref{sec:fourPiConvolver}).

\subsection{Analytical model}

Our starting point is Eq.~\ref{eq:dipole:fundamental}, which relates the dipole signal (as measured by a pencil beam) with the combined velocity of the spacecraft and of the Solar System with respect to the CMB rest frame. Following \cite{Peebles:Wilkinson:1968}, we rewrite Eq.~\ref{eq:dipole:fundamental} as
\begin{equation}\label{eq:dipole:extended}
\frac{\DeltaTDipoleDirac(t_i)}{\Tcmb} = \PointingE(t_i) \cdot \DipoleE + \left(\PointingE\cdot\DipoleE\right)^2 - \frac{1}{2}|\DipoleE|^2 + \mathcal{O}(|\DipoleE|^3),
\end{equation}
where $\DeltaTDipoleDirac$ is the cosmological kinematic signal as measured by a pencil beam, $\PointingE$ is the pointing direction in the Ecliptic reference frame and $\DipoleE = (\vsun + \vplanck)/c$ in the same reference frame. The equation accounts\footnote{Higher-order terms result in corrections at least a thousand times smaller and are therefore neglected.} for both the CMB dipole and the relativistic transverse Doppler effect, since the latter can be considered as a correction to the quadrupole proportional to $\Tcmb|\vsun/c|^2\approx3.8\,\muK$. (We can safely discard $- \frac{1}{2}|\DipoleE|^2$, since it is a constant and plays no role in the calibration.)
We denote with $\Pointing = \bigl(\PointingBx, \PointingBy, \PointingBz\bigr)$ the observing direction in the reference frame of the beam. Then the convolved dipole signal is
\begin{equation}
\begin{split}
\frac{\DeltaTDipole(t)}{\Tcmb} = &N \int_{4\pi} \ud\Omega\,\BeamRd(\Pointing)\,\Pointing \cdot \RotBeamObs(t)\DipoleE
\\
+ &N \int_{4\pi} \ud\Omega\,\BeamRd(\Pointing) \bigl(\Pointing \cdot \RotBeamObs(t)\DipoleE\bigr)^2,
\end{split}
\end{equation}
where $\RotBeamObs(t)$ is the matrix which transforms vectors from the Ecliptic reference frame to the beam reference frame, and $N$ is a normalization constant:
\begin{equation}
\label{eq:beamConvolveNormalizationConstant}
N^{-1} = \int_{4\pi} \ud\Omega\,B(\Pointing).
\end{equation}
The \emph{directional moments} of the beam are defined as
\begin{equation}\label{eq:directional:moments}
\RadiometerM{l,m,n}  =  N \int_{4\pi} \ud\Omega\,\BeamRd(\Pointing)\,\PointingBx^l\,\PointingBy^m\,\PointingBz^n.
\end{equation}
They depend only on the shape of the beam and their value does not change during the mission. If we set
\begin{equation}
\label{eq:radiometerS}
\Svector=
% (S_x,S_y,S_z) = 
\bigl(\RadiometerM{1,0,0},\RadiometerM{0,1,0},\RadiometerM{0,0,1}\bigr)
\end{equation}
(a vector in the beam reference frame) and
\begin{equation}
\label{eq:radiometerA}
\RadiometerA=\left(
\begin{array}{ccc}
\RadiometerM{2,0,0}, & \RadiometerM{1,1,0}, & \RadiometerM{1,0,1} \\ 
\RadiometerM{1,1,0}, & \RadiometerM{0,2,0}, & \RadiometerM{0,1,1} \\ 
\RadiometerM{1,0,1}, & \RadiometerM{0,1,1}, & \RadiometerM{0,0,2} \\ 
\end{array}
\right)
\end{equation}
(a symmetric matrix), then Eq.~\ref{eq:dipole:extended} reduces to the bilinear form
\begin{equation}
\frac{\Delta T(t)}{\Tcmb} = 
\vec{D}_E ^{\dagger}\RotBeamObs(t)^{\dagger} \Svector
+
\vec{D}_E^{\dagger}  \RotBeamObs(t)^{\dagger} \RadiometerA\RotBeamObs(t) \DipoleE.
\end{equation}

Because of a bug in the code which implemented Eq.~\eqref{eq:dipole:extended}, the value of $\left(\PointingE\cdot\DipoleE\right)^2$ was wrongly divided by two before the subtraction. Therefore, the LFI maps in the \Planck{} 2013 data release still have half of the kinematic quadrupole. This residual quadrupole has an amplitude of $\Tcmb|\vsun/c|^2/2\approx 1.9\,\muK$, and it will be properly removed in the next \Planck{} data release.

\subsection{Numerical calculation of the directional moments}

The computation of the components of $\Svector$ (Eq.~\ref{eq:radiometerS}) and $\RadiometerA$ (Eq.~\ref{eq:radiometerA}) was based on the integration of separate {\tt GRASP} maps for both sidelobes and main beam, plus intermediated beams computed for each radiometer at its nominal central frequency, i.e., 30\,GHz, 44~GHz, and 70~GHz. In this context, sidelobes are defined as the part of the beam pattern outside $5^\circ$ from the beam optical axis. The inner part of the beam pattern defines the intermediate beam and main beam maps.

Each map takes into account the small differences in the optical path for each radiometer, given its polarization and location in the focal-plane. For sidelobes, we used {\tt GRASP} maps with a resolution of $0.5\deg$ along the azimuthal angle and $0.5\deg$ along the zenith angle. For intermediate and main beams, the resolution is $0.5\deg$ (azimuth) and $1\arcm$ (zenith). We integrated Eq.~\ref{eq:directional:moments} separately for the intermediate beam and the sidelobes, and then we summed them together. (The same was done for the normalization constant $N$, Eq.~\ref{eq:beamConvolveNormalizationConstant}.) To compute the integral, we converted {\tt GRASP} maps into \HEALpix\ maps \citep{Gorski+2005} by bilinear interpolation and applied a simple midpoint algorithm. We chose the resolution of the \HEALpix\ maps in order to have a resolution much smaller than that of the corresponding {\tt GRASP} maps. Sidelobes were gridded onto an \HEALpix\ map with $\Nside = 1024$, and intermediate beams onto a map with resolution equivalent to $\Nside = 65\,536$.

In order to check the convergence, we computed the value of $\RadiometerM{l,m,n}$ a number of times for increasing values of $\Nside$. All the integrals showed a regular convergence curve. The largest $\Nside$ was fixed by having  $\RadiometerM{l,m,n}(\Nside)/\RadiometerM{l,m,n}(\Nside/2)-1 < 10^{-3}$ for the 
moment with the slower convergence (including $N$).

The beam pattern changes over the bandpass of each radiometer, causing the moments $\RadiometerM{l,m,n}$ to change slightly with frequency. However, the change mainly affects the sidelobes, which contribute
to less than $0.5\,\%$ of the full-sky average. For temperature maps it is therefore enough to use a monochromatic correction.